%% file: riuq_p2_map.tex
\documentclass[a4paper,fleqn,usenatbib]{mnras}

\usepackage{graphicx}

\usepackage{times}
\usepackage{aas_macros}
\usepackage{pdfpages}
\usepackage{grffile}
\usepackage{amsmath,amssymb}
\usepackage{amsfonts}
\usepackage{bm}
\usepackage{bbm}
\usepackage[ruled,linesnumbered]{algorithm2e}
\usepackage{bbold}
\usepackage{url}
\usepackage{arydshln}
\usepackage{booktabs}

\usepackage{xcolor,colortbl}

\usepackage{tikz}
\usetikzlibrary{positioning}
\usepackage{multirow}
\usepackage{pifont}
\newcommand{\cmark}{\ding{51}}%
\newcommand{\xmark}{\ding{55}}%

\newtheorem{theorem}{Theorem}[section]

\newtheorem{proposition}[theorem]{Proposition}
\newtheorem{lemma}[theorem]{Lemma}
\newtheorem{remark}[theorem]{Remark}
\newtheorem{example}[theorem]{Example}
\newtheorem{corollary}[theorem]{Corollary}

\def\argmin{\mathop{\rm argmin}}

\newcommand{\vect}[1]{\boldsymbol{#1}}

\newcommand{\tabL}{1.5mm}

\definecolor{Gray}{gray}{0.8}
\definecolor{Gray1}{gray}{0.9}
\newcolumntype{H}{>{\columncolor{Gray}}c}
\newcolumntype{I}{>{\columncolor{Gray1}}c}

\newcommand{\blue}[1]{\textcolor{blue}{#1}}

\newcommand{\yellow}[1]{\textcolor{yellow}{#1}}

\SetKwRepeat{Do}{do}{while}

\title[Uncertainty quantification for RI imaging II]{Uncertainty quantification for radio interferometric imaging:\\ II.~MAP estimation}
\author[Cai, Pereyra and McEwen]{Xiaohao Cai$^{1}$\thanks{E-mail:~x.cai@ucl.ac.uk~(XC);~m.pereyra@hw.ac.uk~(MP); \newline jason.mcewen@ucl.ac.uk (JDM)},
Marcelo Pereyra$^{2}$\blue{\footnotemark[1]} 
and Jason D. McEwen$^{1}$\blue{\footnotemark[1]}
\\
$^{1}$Mullard Space Science Laboratory,  University College London (UCL), Surrey RH5 6NT, United Kingdom  \\
$^{2}$Maxwell Institute for Mathematical Sciences, Heriot-Watt University, Edinburgh EH14 4AS, United Kingdom\\
} 

\begin{document}

\date{Accepted ---. Received ---; in original form ---}
\pagerange{\pageref{sec:intro}--\pageref{lastpage}}
\pubyear{2017}

\maketitle

\begin{abstract}
Uncertainty quantification is a critical missing component in radio
interferometric imaging that will only become increasingly important as the
big-data era of radio interferometry emerges.  
Statistical sampling approaches to perform Bayesian
inference, like Markov Chain Monte Carlo (MCMC) sampling, can in principle
recover the full posterior distribution of the image, from which uncertainties
can then be quantified.   However, for massive data sizes, like those
anticipated from the Square Kilometre Array (SKA), it will be difficult if not
impossible to apply any MCMC technique due to its inherent computational cost.
We formulate Bayesian inference problems with sparsity-promoting priors (motivated by compressive sensing), for
which we recover {\it maximum a posteriori} (MAP) point estimators of radio
interferometric images by convex optimisation. Exploiting recent developments in
the theory of probability concentration, we quantify
uncertainties by post-processing the recovered MAP estimate. Three strategies to
quantify uncertainties are developed:
(i) highest posterior density credible regions; (ii) local credible intervals
(\textit{cf.} error bars) for individual pixels and superpixels; and (iii)
hypothesis testing of image structure.   These forms of uncertainty
quantification provide rich information for analysing radio interferometric
observations in a statistically robust manner.  
Our MAP-based methods are
approximately $10^5$ times faster computationally than state-of-the-art 
MCMC methods and, in addition, support highly distributed and parallelised algorithmic structures.  For the first time, our MAP-based techniques
provide a means of quantifying uncertainties for radio interferometric imaging
for realistic data volumes and practical use, and scale to the emerging big-data
era of radio astronomy. 
\end{abstract}


\section{Introduction}\label{sec:intro}
Radio interferometric (RI) telescopes provide observations of the radio emission of the sky with high angular resolution and sensitivity, 
and provide a wealth of valuable information for astrophysics and cosmology \citep{RV46,ryl60,tho08}. 
Radio interferometers essentially acquire Fourier measurements of the sky image of interest. Imaging observations made 
by radio interferometers thus requires solving an ill-posed linear inverse problem \citep{tho08}, which is an important first step in many subsequent scientific analyses. 
Since the inverse problem is ill-posed (sometimes seriously), uncertainty information (\textit{e.g.} error estimates) regarding reconstructed images is critical.  
Nevertheless, uncertainty information is currently lacking in all RI imaging techniques used in practice.
In \cite{CPM17}, the first of these companion articles, we propose uncertainty quantification strategies 
for RI imaging based on state-of-the-art Markov chain Monte Carlo (MCMC) methods that sample the full posterior distribution of the image, 
with the sparsity-promoting priors that have been shown in practice 
to be highly effective \citep[\textit{e.g.}][]{PMdCOW16}.  Excellent results were achieved and a variety of different 
uncertainty quantification strategies were presented.  However, it is difficult to scale these strategies 
to big-data due to their high computational overhead.  We address this issue in the current article.

Over the coming decades radio astronomy will transition into the so-called big-data era.
Generally speaking, the new generation of radio telescopes, such as the LOw Frequency ARray (LO-FAR\footnote{\url{http://www.lofar.org}}),
the Extended Very Large Array (EVLA\footnote{\url{http://www.aoc.nrao.edu/evla}}),
the 
Australian Square Kilometre Array Pathfinder (ASKAP\footnote{\url{http://www.atnf.csiro.au/projects/askap}}), and the Murchison Widefield Array
(MWA\footnote{\url{http://www.mwatelescope.org/telescope}}), 
will achieve much higher dynamic range and angular resolution than previous instruments
and will acquire very large volumes of data.  The Square Kilometer Array (SKA\footnote{\url{http://www.skatelescope.org/}}) will provide a considerable step again in dynamic range (six or
seven orders of magnitude beyond prior telescopes) and angular resolution, and will acquire massive volumes of data, ushering in the big-data era of radio astronomy.
This emerging era of big-data, inevitably, will bring further challenges and so uncertainty quantification will be increasingly important.  As discussed in \cite{CPM17}, existing image reconstruction techniques,
such as CLEAN-based methods \citep{hog74,BC04,cor08,SFM11},
the maximum entropy method (MEM) \citep{A74,GD78,CE85},
and compressed sensing (CS) methods {\citep{wia09a,wia09b,S09,WMPBR10,mce11,li11a,li11b,CMW12,car14,wol13,dab15,DWPMW17,gar15,OCRMTPW16,ODW17,PMdCOW16,KCTW17}},
do not provide uncertainty information regarding their reconstructed images.
The approaches that do provide some form of uncertainty quantification \citep{SWMBKKTTZ14,JBSE16,GVJE17} cannot scale to big-data due to their high computational cost, are typically restricted to Gaussian or log-normal priors, and are not currently used in practice.  Please see our first article in this companion series \citep{CPM17} for a more thorough review of RI  imaging techniques and their properties.

The current state of the field thus triggers an urgent need to develop efficient uncertainty quantification methods for
RI imaging that scale to big-data.  Furthermore, we seek to support the sparsity-promoting priors that have been demonstrated in practice to be highly effective for RI imaging \citep[\textit{e.g.}][]{PMdCOW16}.  In \cite{CPM17} (the first part of this companion series), 
we proposed uncertainty quantification methods to address the RI imaging problem with sparse priors. 
In the current article (the second part of this companion series), we present fast uncertainty quantification methods that not only support sparse priors but also scale to big-data.  The techniques presented in this article are very different to those presented in \cite{CPM17} but support the same forms of uncertainty quantification.

The uncertainty quantification methods proposed in \cite{CPM17} are based on two proximal MCMC 
sampling methods, {\it i.e.} the
Moreau-Yoshida unadjusted Langevin algorithm (MYULA) \citep{DMP16} and the proximal Metropolis-adjusted Langevin algorithm (Px-MALA) \citep{M15}.
The main steps of the uncertainty quantification strategies presented in \cite{CPM17} can be briefly summarised as follows: 
firstly, the posterior distribution of the image is MCMC sampled; then, uncertainty quantification is performed by using the generated samples 
to compute local (pixel-wise) credible intervals,  highest posterior density (HPD) credible regions, 
and to perform hypothesis testing of image structure.  Two frameworks -- analysis and synthesis models -- are considered.
While excellent results were achieved in \cite{CPM17}, when it comes to big-data, the proposed approach would suffer due to the 
long computation time required to sample the posterior distribution (as would be the case for any MCMC sampling approach).

In this article we exploit an analytic method to approximate HPD credible regions from \textit{maximum a posteriori} (MAP) estimators, 
as derived in \cite{M16}, in order to develop very fast methods to perform uncertainty quantification for RI imaging.  
Our approach supports sparse priors and scales to massive data sizes, \textit{i.e.} to big-data.
We begin by formulating Bayesian MAP estimation for RI imaging as unconstrained convex optimisation problems, for analysis and synthesis forms. These are subsequently
solved efficiently by using convex minimisation algorithms \citep[\textit{e.g.}][]{CP10}. Recent advances in convex optimisation have resulted in techniques that achieve excellent reconstruction fidelity (with convergence guarantees), are flexible, and exhibit relatively low computational costs.
They also afford algorithmic structures that can be highly distributed and parallelised ({\it e.g.} \citealt{car14,OCRMTPW16}) and computed in an online manner \citep{CPraM17}.
Note, specifically, that only one point estimator is computed here for the analysis or synthesis form,
in contrast to sampling approaches that seek to explore the full posterior distribution as in \cite{CPM17},
which is very time consuming. MAP estimation is then followed by various strategies to quantify uncertainties.
Precisely, first the method of \cite{M16} is used to obtain approximate HPD credible regions for the recovered image. These HPD regions are then used, for the first time, to compute local credible intervals (\textit{cf.} error bars) that analyse uncertainty spatially and at different scales (pixles or superpixels). Finally, we also use the  HPD credible regions to perform hypothesis tests of image structure. We test our proposed approaches on simulated RI observations to demonstrate their effectiveness and compare with the MCMC methods presented in \cite{CPM17}.

The remainder of this article is organised as follows. In Section~\ref{sec:ri} we review the RI imaging inverse problem. 
In Section~\ref{sec:map-ri} we apply convex optimisation algorithms to solve the MAP estimation problem for RI imaging in the context of sparse priors.
{Note that Sections~\ref{sec:ri} and \ref{sec:map-ri} review background material for our specific problem to provide clarity and completeness (\textit{i.e.} so that all derivations are explicit and thus
one could implement our methods if one wanted).}
Uncertainty quantification techniques for RI imaging based on MAP estimation are formulated in Section~\ref{sec:uq}. The performance of the proposed methods is then evaluated numerically in Section~\ref{sec:exp}, where we compare uncertainties quantified by proximal MCMC methods and by MAP estimation.
Finally, we conclude in Section~\ref{sec:con} with a summary of our main contributions and a discussion of planned extensions.

\section{Radio interferometric imaging}\label{sec:ri}
In this section the inverse problem related to RI image reconstruction is introduced.
We briefly recall the use of proximal MCMC methods to solve this problem \citep{CPM17}, 
which we use as a benchmark in the experiments that follow. Finally, an introduction to Bayesian MAP estimation approaches for RI imaging is presented, which may be solved by efficient convex optimisation strategies.

\subsection{Radio interferometry}
{Here, we concisely recall the inverse problem of RI imaging (for further details see \citealt{CPM17} and references therein).}

In the discretised setting, let $\vect x \in \mathbb{R}^N$ represent the sampled intensity signal
(the sky brightness distribution). In particular, $\vect{x}$ can be represented by  
\begin{equation}\label{eqn:x}
{\vect x} = \bm{\mathsf{\Psi}} {\vect a} = \sum_{i} \bm{\mathsf{\Psi}}_i a_i,
\end{equation}
where $\bm{\mathsf{\bm{\mathsf{\Psi}}}} \in \mathbb{C}^{N\times L}$ is a basis or dictionary ({\it e.g.}, a wavelet basis or an over-complete frame)
and vector ${\vect a} = (a_1, \cdots, a_L)^\top$ represents the synthesis coefficients
of ${\vect x}$ under $\bm{\mathsf{\Psi}}$. In particular, ${\vect x}$ is said to be sparse if ${\vect a}$ contains only $K$ non-zero coefficients,
$K\ll N$, or compressible if many coefficients of $\vect{a}$ are nearly zero.  In practice, it is ubiquitous that natural images are sparse or compressible for approriate choices of $\bm{\mathsf{\Psi}}$.  Refer to \cite{CPM17} for more details about sparse representation.

{Let $\vect y \in \mathbb{C}^M$ be the $M$ visibilities acquired by a radio interferometric telescope observed under a linear measurement operator
$\bm{\mathsf{\Phi}} \in \mathbb{C}^{M\times N}$ modelling the acquisition of the sky brightness distribution. Then, we have
\begin{equation}\label{eqn:y}
{\vect y}=\bm{\mathsf{\Phi}} {\vect x} + {\vect n},
\end{equation}
where ${\vect n} \in \mathbb{C}^{M}$ is the instrumental noise. Without loss of generality, we subsequently consider independent and identically distributed (i.i.d.) Gaussian noise. 
In practice, $\vect y$ is only observed partially or with limited resolution.
Recovering the sky intensity signal $\vect x$ from 
the measured visibilities $\vect y$ acquired according to equation \eqref{eqn:y} then amounts to solving a linear inverse problem \citep{RBVC09}.}

\subsection{Bayesian inference}
The RI inverse problem \eqref{eqn:y} can be solved elegantly in the Bayesian statistical framework, 
which provides tools to estimate $\vect x$ {(or $\vect a$)} as well as to quantify the uncertainty in the estimated solutions. 
{After combining the observed and prior information, the posterior distribution $p(\vect x | \vect y)$ (or $p(\vect a | \vect y)$ )
can be obtained by using Bayes' theorem. Refer to \cite{CPM17} for more detailed discussion about Bayesian inference in the context of RI imaging.}

\subsection{Proximal MCMC methods}
To solve the ill-posed inverse problem in \eqref{eqn:y} with sparsity-promoting priors, which have been shown in practice to be highly effective \citep{PMdCOW16}, 
while also performing uncertainty quantification, two proximal MCMC methods to
perform Bayesian inference for RI imaging were developed in the companion article \citep{CPM17}. These proximal MCMC methods seek to 
sample the full posterior density $p(\vect x | \vect y)$ that models our understanding of the image $\vect{x}$ given data $\vect{y}$, in the context of prior information.  From the full posterior, summary estimators of $\vect{x}$ and other quantities of interest can be computed.  In particular, in \cite{CPM17} these methods are used to perform a range of uncertainty quantification analysis for RI images.

One of the proximal MCMC methods presented in \citet{CPM17}, MYULA, scales efficiently to high dimensions but suffers from some estimation bias  \citep{DMP16}. 
The other, Px-MALA, corrects this bias by using a Metropolis-Hastings correction step, at the expense of a higher computational cost 
and slower convergence \citep{M15}. Since Px-MALA can provide results with corrected bias and thus is more accurate,
we use it as a benchmark in the subsequent numerical tests presented in this work.
Nevertheless, the MCMC methods discussed in \cite{CPM17} will suffer when scaling to big-data (as will any MCMC method),
which motives us to explore alternative faster methods that can scale to big-data. 

In this article we develop methods for uncertainty quantification based on MAP estimation.
We emphasise that while MCMC methods such as Px-MALA are not as efficient as MAP estimation 
(the main focus in this article), and do not scale to large RI datasets, they are useful for smaller datasets and as a benchmark for the efficient alternative methods that we propose in Section~\ref{sec:uq}.

\subsection{Maximum a posteriori (MAP) estimation}
As discussed in the previous sections, sampling the full posterior $p(\vect x | \vect y)$ or $p(\vect a | \vect y)$ by MCMC methods is difficult because of the high dimensionality involved. Instead, Bayesian 
estimators that summarise $p(\vect x | \vect y)$ or $p(\vect a | \vect y)$ are often computed. 
In particular, one common approach is to compute MAP (maximum-a-posteriori) estimators given by
\begin{equation}\label{eqn:ir-un-af}
{\vect x}_{\rm map} = \argmin_{\vect x} \Big\{\mu \|\bm{\mathsf{\Psi}}^\dagger {\vect x}\|_1 
	+ \|{\vect y}-\bm{\mathsf{\Phi}} {\vect x}\|_2^2/2\sigma^2 \Big\},
\end{equation}
for the analysis model, and for the synthesis model by 
\begin{equation}\label{eqn:ir-un-sf}
{\vect x}_{\rm map} = \bm{\mathsf{\Psi}}\times\argmin_{{\vect a}} \Big\{ \mu \|{\vect a}\|_1 
	+ \|{\vect y}-\bm{\mathsf{\Phi}}\bm{\mathsf{\Psi}} {\vect a}\|_2^2/2\sigma^2 \Big\},
\end{equation}
{where the first term is a prior distribution to regularise the problem, reduce uncertainty, and improve estimation results,
and the second term is associated with the likelihood function of the model associated with \eqref{eqn:y}. }

As we discuss below, a main computational advantage of the MAP estimators \eqref{eqn:ir-un-af} and \eqref{eqn:ir-un-sf} 
is that they can be computed very efficiently, even in high dimensions, by using convex optimisation algorithms ({\it e.g.} \citealt{CP10,Green2015}). 
There is also abundant empirical evidence suggesting that these estimators deliver accurate reconstruction results (see \citealt{Pereyra:2016b} also
for a theoretical analysis of MAP estimation).  However, since MAP estimation results in a single point estimator, 
we typical lose uncertainty information that MCMC methods can provide  \citep{CPM17}.
On the contrary, however, as we show in this article it is possible to approximately quantify the uncertainties associated with MAP estimators by leveraging recent results in the theory of probability concentration \citep{M16}. Consequently, using the techniques presented later in this article MAP estimation can provide fast methods that scale to big-data and that quantify uncertainties.

\subsection{Convex optimisation methods for MAP estimation}
There are several convex optimisation methods that can be used to solve the MAP estimation problems \eqref{eqn:ir-un-af} and \eqref{eqn:ir-un-sf} efficiently, such as forward-backward  splitting, Douglas-Rachford splitting, or alternating direction method of multipliers (ADMM) (see \citealt{CP10}).
In our experiments \eqref{eqn:ir-un-af} and \eqref{eqn:ir-un-sf} are solved by adopting the simple forward-backward 
algorithm, which we detail {in Appendix~\ref{sec:appendix}}.

\section{Sparse MAP estimation for RI imaging} \label{sec:map-ri}
In this section we present the algorithmic details of implementing the forward-backward splitting algorithm to solve the sparse MAP estimation problems 
for both the analysis setting \eqref{eqn:ir-un-af} and synthesis setting \eqref{eqn:ir-un-sf}.
For the sake of brevity, henceforth the labels \ $\bar{}$ \ and \ $\hat{}$ \ denote 
symbols related to the analysis and synthesis models, respectively. 

\subsection{Analysis}
For the analysis setting \eqref{eqn:ir-un-af}, set ${\bar f}({\vect x}) = \mu \|\bm{\mathsf{\Psi}}^\dagger {\vect x}\|_1$ 
and ${\bar g}({\vect x}) = \|{\vect y}-\bm{\mathsf{\Phi}} {\vect x}\|_2^2/2\sigma^2$. Then 
\begin{equation} 
\argmin_{\vect x} \Big\{{\bar f}({\vect x}) + {\bar g}({\vect x})  \Big\}
\end{equation}
can be solved using the forward-backward iteration formula ({shown in Appendix~\ref{sec:appendix}}), leading to the iterations
\begin{equation} \label{eqn:fb-i-ana}
{\vect x}^{(i+1)} = {\rm prox}_{\lambda^{(i)} {\bar f}} ({\vect x}^{(i)} - \lambda^{(i)} \nabla {\bar g}({\vect x}^{(i)})). 
\end{equation}
Assume for now $\bm{\mathsf{\Psi}}^\dagger \bm{\mathsf{\Psi}} = \bm{\mathsf{ I}}$, where $\bm{\mathsf{ I}}$ is identity matrix (although this assumption is not essential and relaxed later). 
We have, $\forall \bar{\vect z} \in \mathbb{R}^N$,
\begin{equation} \label{eqn:prox-a}
{\rm prox}_{\lambda {\bar f}} (\bar{\vect z}) = \bar{\vect z} + \bm{\mathsf{\Psi}} 
	\left ( {\rm soft}_{\lambda \mu}(\bm{\mathsf{\Psi}}^\dagger \bar{\vect z}) - \bm{\mathsf{\Psi}}^\dagger \bar{\vect z} \right ),
\end{equation}
and
\begin{equation} \label{eqn:grad-a}
\nabla {\bar g}({\vect x}) =  \bm{\mathsf{\Phi}}^\dagger (\bm{\mathsf{\Phi}} {\vect x} - {\vect y})/\sigma^2,
\end{equation}
where {${\rm soft}_{\lambda\mu}({\vect z})$  is the pointwise soft-thresholding operator of vector ${\vect z}$ defined in \eqref{eqn:soft-t}}.
{See Remark 4.1 in \cite{CPM17} when $\bm{\mathsf{\Psi}}^\dagger \bm{\mathsf{\Psi}} \neq \bm{\mathsf{ I}}$ for computing
${\rm prox}_{\lambda {\bar f}} (\bar{\vect z})$.
}
Substituting \eqref{eqn:prox-a} and \eqref{eqn:grad-a} into \eqref{eqn:fb-i-ana},
the analysis problem \eqref{eqn:ir-un-af} can be solved iteratively by
\begin{align} 
{\vect v}^{(i+1)} &= {\vect x}^{(i)} -  \lambda^{(i)} \bm{\mathsf{\Phi}}^\dagger (\bm{\mathsf{\Phi}} {\vect x}^{(i)} - {\vect y})/\sigma^2, \label{eqn:ir-un-af-fb-i-1}  \\
{\vect x}^{(i+1)} &= {\vect v}^{(i+1)} \! + \! \bm{\mathsf{\Psi}} \left ( {\rm soft}_{\lambda^{(i)} \mu}(\bm{\mathsf{\Psi}}^\dagger {\vect v}^{(i+1)}) \! -\! \bm{\mathsf{\Psi}}^\dagger {\vect v}^{(i+1)} \right ).
	\label{eqn:ir-un-af-fb-i-2}
\end{align}
As initialisation use, {\it e.g.}, ${\vect x}^{(0)} = \bm{\mathsf{\Phi}}^{\dagger} \vect y$, {\it i.e.} the dirty image.

\subsection{Synthesis}
For the synthesis setting \eqref{eqn:ir-un-sf}, set $\hat{f}(\vect a) = \mu \|{\vect a}\|_1$ and $\hat{g}(\vect a) = \|{\vect y}-\bm{\mathsf{\Phi}}\bm{\mathsf{\Psi}} {\vect a}\|_2^2/2\sigma^2$. Then  
\begin{equation} 
\argmin_{\vect x} \Big\{{\hat f}({\vect a}) + {\hat g}({\vect a})  \Big\}
\end{equation}
can be solved using the forward-backward iteration formula ({shown in Appendix~\ref{sec:appendix}}), leading to the iterations
\begin{equation} \label{eqn:fb-i-syn}
{\vect a}^{(i+1)} = {\rm prox}_{\lambda^{(i)} {\hat f}} ({\vect a}^{(i)} - \lambda^{(i)} \nabla {\hat g}({\vect a}^{(i)})). 
\end{equation}
We have, $\forall \hat{\vect z} = (\hat{z}_1, \cdots, \hat{z}_L) \in \mathbb{R}^L$,
\begin{equation} \label{eqn:prox}
\begin{split}
{\rm prox}_{\lambda {\hat f}} (\hat{\vect z})  
	&=  \argmin_{{\vect u} \in \mathbb{R}^L}  \lambda \mu \|{\vect u}\|_1 + \|{\vect u} - \hat{\vect z}\|^2/2   \\
	&= {\rm soft}_{\lambda \mu}(\hat{\vect z})
\end{split}
\end{equation}
and 
\begin{equation} \label{eqn:grad}
\nabla {\hat g}({\vect a}) =  \bm{\mathsf{\Psi}}^\dagger \bm{\mathsf{\Phi}}^\dagger (\bm{\mathsf{\Phi}}\bm{\mathsf{\Psi}} {\vect a} - {\vect y})/\sigma^2.
\end{equation}
Finally, substituting \eqref{eqn:prox} and \eqref{eqn:grad} into \eqref{eqn:fb-i-syn}, the synthesis problem \eqref{eqn:ir-un-sf} can be solved iteratively by
\begin{equation} \label{eqn:ir-un-sf-fb-i}
{\vect a}^{(i+1)} = {\rm soft}_{\lambda^{(i)} \mu} \left ({\vect a}^{(i)} - \lambda^{(i)}  \bm{\mathsf{\Psi}}^\dagger \bm{\mathsf{\Phi}}^\dagger (\bm{\mathsf{\Phi}}\bm{\mathsf{\Psi}} {\vect a}^{(i)} - {\vect y})/\sigma^2 \right).  
\end{equation}

\begin{remark}
Note that in both the analysis and synthesis settings various terms can be precomputed.  For example, in \eqref{eqn:ir-un-af-fb-i-1} and \eqref{eqn:grad} the operators $\bm{\mathsf{\Phi}}^\dagger\bm{\mathsf{\Phi}}$ and 
$\bm{\mathsf{\Psi}}^\dagger \bm{\mathsf{\Phi}}^\dagger \bm{\mathsf{\Phi}}\bm{\mathsf{\Psi}}$ can be precomputed offline. 
Similarly, the terms of $\bm{\mathsf{\Phi}}^\dagger \vect y$ (the so-called dirty map) and $\bm{\mathsf{\Psi}}^\dagger \bm{\mathsf{\Phi}}^\dagger \vect y$
respectively in \eqref{eqn:ir-un-af-fb-i-1} and \eqref{eqn:grad} can also be precomputed to improve computation efficiency.
\end{remark}

We summarise the  forward-backward splitting algorithms for the analysis and synthesis reconstruction forms 
in Algorithms \ref{alg:fb-a} and \ref{alg:fb-s}.
We consider stopping criteria based on a maximum iteration number and when the relative difference between solutions at two consecutive iterations is within some tolerance, 
{\it i.e.}, $\|{\vect x}^{(i+1)} - {\vect x}^{(i)}\|_2/\|{\vect x}^{(i)}\|_2$
(for Algorithm \ref{alg:fb-a}) and \mbox{$\|\bm{\mathsf{\Psi}}{\vect a}^{(i+1)} - \bm{\mathsf{\Psi}}{\vect a}^{(i)}\|_2/\|\bm{\mathsf{\Psi}}{\vect a}^{(i)}\|_2$} (for Algorithm \ref{alg:fb-s}).  The iteration is terminated when either of the stopping criteria are reached.  
{The complexity of the algorithms is simply given by the complexity of application of the measurement operator $\bm{\mathsf{\Phi}}$. 
However, the measurement operator (and its adjoint) needs to be applied multiple times, hence the pre-factor associated with the complexity is significant.  In general fast, optimised algorithms are applied for realistic measurement operators (essentially based on non-uniform fast Fourier transforms), resulting in a complexity of $\mathcal{O}(MJ + N \log N) $, where $J$ denotes the support of the kernel used to perform convolutional degridding (see, \textit{e.g.}, \citealt{PMdCOW16} for further details).}

\begin{algorithm} 
\caption{Forward-backward  algorithm for analysis}
\label{alg:fb-a}
 \textbf{Input:}  ${\vect y} \in \mathbb{R}^M$, ${\vect x}^{(0)} \in \mathbb{R}^N$, $\sigma$ and $\lambda^{(i)} \in (0, \infty)$\\
 \textbf{Output:} ${\vect x}^\prime$ \vspace{0.05in} \\ 
\Do{Stopping criterion is not reached}{
   	update ${\vect v}^{(i+1)} = {\vect x}^{(i)} -  \lambda^{(i)} \bm{\mathsf{\Phi}}^\dagger (\bm{\mathsf{\Phi}} {\vect x}^{(i)} - {\vect y})/\sigma^2 $ \\
	compute $\vect u = \bm{\mathsf{\Psi}}^\dagger {\vect v}^{(i+1)}$ \\
	update ${\vect x}^{(i+1)} = {\vect v}^{(i+1)} + \bm{\mathsf{\Psi}} \left ( {\rm soft}_{\lambda^{(i)} \mu}(\vect u) - \vect u\right )$  \\
        $i=i+1$
}   \vspace{0.05in}

set ${\vect x}^\prime = {\vect x}^{(i)}$
\end{algorithm}
\begin{algorithm} 
\caption{Forward-backward  algorithm for synthesis}
\label{alg:fb-s}
 \textbf{Input:}  ${\vect y} \in \mathbb{R}^M$, ${\vect a}^{(0)} \in \mathbb{R}^L$, $\sigma$ and $\lambda^{(i)} \in (0, \infty)$\\
 \textbf{Output:} ${\vect a}^\prime$  \vspace{0.05in}  \\ 
\Do{Stopping criterion is not reached}{
	compute $\vect u =  {\vect a}^{(i)} - \lambda^{(i)}  
		\bm{\mathsf{\Psi}}^\dagger \bm{\mathsf{\Phi}}^\dagger (\bm{\mathsf{\Phi}}\bm{\mathsf{\Psi}} {\vect a}^{(i)} - {\vect y})/\sigma^2$ \\
   	update ${\vect a}^{(i+1)} = {\rm soft}_{\lambda^{(i)} \mu} (\vect u ) $  \\
        $i=i+1$
} \vspace{0.05in} 

set ${\vect a}^\prime = {\vect a}^{(i)}$
\end{algorithm}

\section{Bayesian uncertainty quantification:\quad MAP Estimation}\label{sec:uq}
The analysis and synthesis reconstruction models address inverse problems
which are generally ill-conditioned or ill-posed (especially when the measurements are only observed partially or
with limited resolution). Consequently, the corresponding estimators have significant intrinsic uncertainty that is very challenging
to analyse and quantify. In \cite{M16} a general methodology was proposed to use MAP estimators to accurately approximate Bayesian credible regions 
for $p(\vect x | \vect y)$. These credible regions indicate the regions of the parameter space where most of the posterior probability mass lies. 
A remarkable property of the approximation is that it only requires knowledge of ${\vect x}_{\rm map}$ and therefore 
it can be computed very efficiently, even in very large-scale problems.

\begin{figure}
  \begin{center}
    \begin{tabular}{c}
  	 \input{fig_uq_diag_part2}
    \end{tabular}
  \end{center}
  \caption{Our proposed uncertainty quantification procedure for RI imaging based on MAP estimation. 
  	The light green areas on the right show the types of uncertainty quantification developed.
    Firstly, an image is reconstructed by MAP estimation using convex optimisation techniques, which scale to big-data.
   Then, various forms of uncertainty quantification are performed.  Global approximate Bayesian credible regions are computed. 
   These are then used to compute 
   local credible intervals ({\it cf.} error bars) corresponding to individual pixels and 
   superpixels and to perform hypothesis testing of image structure 
   to test whether a structure is physical or an artefact.
    }
  \label{fig:uq_diag}
\end{figure}
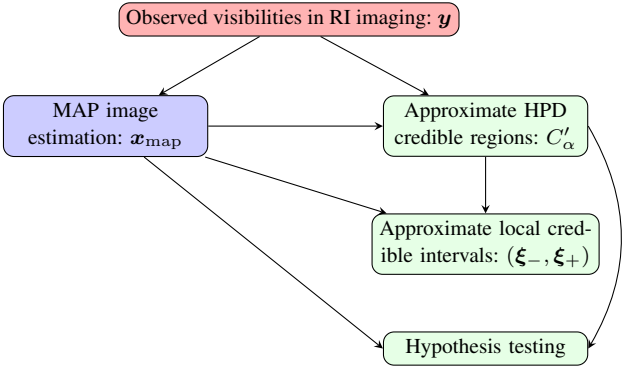

The diagram in Figure \ref{fig:uq_diag} shows the main components of our proposed uncertainty quantification 
methodology based on MAP estimation.
As is shown, firstly, an image is reconstructed by MAP estimation. 
MAP estimation can be computed extremely rapidly and is therefore ideal for application to big-data. 
Then, various forms of uncertainty quantification are performed.  Firstly, global approximate Bayesian credible regions are computed. 
These are then used to compute 
local credible intervals ({\it cf.} error bars) corresponding to individual pixels and 
superpixels. Finally, again using the global approximate Bayesian credible regions, hypothesis testing of image structure can be performed  
to test whether a structure is physical or an artefact. For consistency, we adopt the same notation as in the companion article \citep{CPM17}.

\subsection{Approximate highest posterior density (HPD) credible regions}
The first step in our uncertainty quantification methodology is to compute a credible region for $p(\vect x|\vect y)$. A posterior credible region 
with credible level $100(1-\alpha)\%$ is a set $C_{\alpha} \in \mathbb{R}^N$  that satisfies
\begin{equation}\label{eqn:cr}
p (\vect x \in C_{\alpha} | \vect y) = \int_{\vect x \in \mathbb{R}^N} p(\vect x | \vect y) \mathbb{1}_{C_{\alpha}} {\rm d} \vect x = 1-\alpha,
\end{equation}
where $\mathbb{1}_{C_{\alpha}}$ is the indicator function for $C_{\alpha}$, defined by
\mbox{$\mathbb{1}_{C_{\alpha}}({\vect u}) = 1$} if $\vect u \in C_{\alpha}$ and 0 otherwise. Many regions satisfy the above property. 
We focus on the HPD (Highest Posterior Density) region defined by
\begin{equation}\label{eqn:cr-hpd}
C_{\alpha} := \{\vect x: f(\vect x) + g(\vect x) \le \gamma_{\alpha}\},
\end{equation}
where the threshold $\gamma_{\alpha}$ which defines an isocontour or level-set of the log-posterior
 is set such that \eqref{eqn:cr} holds, and we recall that $p(\vect x | \vect y) \propto \exp\{-f(\vect x) - g(\vect x)\}$. 
This region is decision-theoretically optimal in the sense of minimum volume \citep{R01}.

Computing HPD credible regions in \eqref{eqn:cr-hpd} is difficult because of the high-dimensional integral in \eqref{eqn:cr}. 
For RI models that are not too high dimensional, $C_{\alpha}$ can be computed efficiently by using 
proximal MCMC method as described in \citealt{CPM17}. However, this is not possible in big-data settings. 

Here we use an approximation of $C_{\alpha}$ proposed recently in \cite{M16} for convex inverse problems solved by MAP estimation. 
The approximation is given by
\begin{equation}\label{eqn:cr-e}
{C}^\prime_{\alpha} := \{\vect x: f(\vect x) + g(\vect x) \le \gamma^{\prime}_{\alpha}\},
\end{equation}
where $\gamma^{\prime}_{\alpha}$ is an approximation of the HPD threshold ${\gamma}_{\alpha}$ given by 
\begin{equation}\label{eqn:gamma}
{\gamma}^{\prime}_{\alpha} = f({\vect x}_{\rm map}) +  g({\vect x}_{\rm map}) + \tau_{\alpha}\sqrt{N} + N,
\end{equation}
with universal constant $\tau_{\alpha} = \sqrt{16\log(3/\alpha)}$. Recall that $N$ is the dimension of $\vect x$ and $100(1-\alpha)$\% the credible level considered.
After computing ${\vect x}_{\rm map}$ by using modern convex optimisation algorithms,
$\gamma^\prime_{\alpha}$ can be calculated straightforwardly using \eqref{eqn:gamma}, even in very high dimensions. 
The approximation given in \eqref{eqn:gamma} was motivated from recent results in information theory in terms of 
a probability concentration inequality (refer to \citealt{M16} for more details).

For any $\alpha \in (4 {\rm exp} (-N/3), 1)$, the error between ${\gamma}^{\prime}_{\alpha}$ and $\gamma_{\alpha}$ is bounded by the following inequality 
\begin{equation}
0 \le {\gamma}^{\prime}_{\alpha} - \gamma_{\alpha} \le \eta_{\alpha}\sqrt{N} + N,
\end{equation}
where $\eta_{\alpha} = \sqrt{16\log(3/\alpha)} + \sqrt{1/\alpha}$. Since the error ${\gamma}^{\prime}_{\alpha} - \gamma_{\alpha}$ grows 
at most linearly with respect to $N$ when $N$ is large, the credible region ${C}^{\prime}_{\alpha}$ associated with ${\gamma}^{\prime}_{\alpha}$ is a stable approximation of $C_{\alpha}$. 
Moreover, since ${\gamma}^{\prime}_{\alpha} - \gamma_{\alpha} \geq 0$ the approximation is theoretically
 conservative in the sense that ${C}^{\prime}_{\alpha}$ overestimates ${C}_{\alpha}$.
Precisely, in the analysis formulation, we first compute the reconstructed image ${\vect x}_{\rm map}$ by using Algorithm \ref{alg:fb-a}, 
and then obtain an approximate HPD credible region 
\begin{equation}
\bar{C}^{\prime, {\rm map}}_{\alpha} := \{\vect x: {\bar f}(\vect x) + {\bar g}(\vect x) \le \bar{\gamma}_{\alpha}^{\prime}\}
\end{equation}
with 
\begin{equation}\label{eqn:gamma-a}
\bar{\gamma}_{\alpha}^{\prime} = {\bar f}({\vect x}_{\rm map}) + {\bar g}({\vect x}_{\rm map}) + \tau_{\alpha}\sqrt{N} + N.
\end{equation}
Similarly, in the synthesis setting we compute ${\vect a}_{\rm map}$ via Algorithm \ref{alg:fb-s}, and then construct 
\begin{equation}
\hat{C}^{\prime, {\rm map}}_{\alpha}:= \{\bm{\mathsf{\Psi}}\vect a : {\hat f}({\vect a}) + {\hat g}({\vect a}) \le \hat{\gamma}_{\alpha}^{\prime} \}
\end{equation}
with 
\begin{equation}\label{eqn:gamma-s}
\hat{\gamma}_{\alpha}^{\prime} = {\hat f}({\vect a}_{\rm map}) + {\hat g}({\vect a}_{\rm map}) + \tau_{\alpha}\sqrt{N} + N.
\end{equation}
Note that $\bar{\gamma}_{\alpha}^{\prime}$ and $\hat{\gamma}_{\alpha}^{\prime}$ define the HPD credible regions implicitly.

The HPD credible regions can be used to quantify uncertainties in a variety of manners.  In the reminder of this section we describe two such strategies. 

\subsection{Local credible intervals}
The first strategy we propose is a novel approach to compute local credible intervals corresponding to pixels and superpixels, as a means for 
quantifying uncertainty spatially at different scales. This presents a new form 
of Bayesian uncertainty quantification tailored for image data and is easy to visualise and interpret. The method is based on the HPD credible regions discussed above and is applicable for any method for which 
HPD credible regions can be computed. Here we promote the MAP-based approach, based on the approximations \eqref{eqn:gamma-a} and \eqref{eqn:gamma-s}, 
and benchmark our results against the MCMC approach Px-MALA, introduced in \cite{CPM17}.

Let $\Omega = \cup_{i}\Omega_i $ be a partition of the image domain $\Omega$ into subsets or \emph{superpixels} $\Omega_i$ 
such that $\Omega_i\cap\Omega_j = \emptyset, i\neq j$. The image domain can be partitioned at different scales, from a single pixel to larger scales involving 
blocks of several pixels. To index superpixels we define the index operator 
$\vect \zeta_{\Omega_i} = (\zeta_1, \cdots, \zeta_N) \in \mathbb{R}^N$ on $\Omega_i$, which satisfies
\begin{equation}
\zeta_k = 
\begin{cases}
1, \ {\rm if} \ k \in \Omega_i, \\
0, \ {\rm otherwise}.
\end{cases}
\end{equation}
To quantify the uncertainty associated with the region $\Omega_i$ we calculate the points ${\xi}_{-, \Omega_i}$ and ${\xi}_{+,  \Omega_i}$ 
that saturate the HPD credible region ${C}^{\prime, {\rm map}}_{\alpha}$ from above and from below at $\Omega_i$, given by
\begin{align}
&{\xi}_{-,  \Omega_i} = \min_{\xi}\left \{ \xi |  f({\vect x}_{i,\xi}) + g({\vect x}_{i,\xi}) \le {\gamma}^\prime_{\alpha}, \forall \xi \in [0, +\infty )  \right \},  \label{eqn:cr-local-sp-i-l}\\
&{\xi}_{+,  \Omega_i} = \max_{\xi}\left \{ \xi |  f({\vect x}_{i,\xi}) + g({\vect x}_{i,\xi}) \le {\gamma}^\prime_{\alpha}, \forall \xi \in [0, +\infty )  \right \},  \label{eqn:cr-local-sp-i-h}
\end{align}
where ${\vect x}_{i,\xi} = {\vect x}^* (\bm{\mathsf{ I}} - \vect \zeta_{\Omega_i}) + \xi \vect \zeta_{\Omega_i}$ represents a point estimator generated 
by replacing the intensity of ${\vect x}^*$ in $\Omega_i$ by $\xi$.  We recall that ${\gamma}^{\prime}_{\alpha}$ 
is the threshold or isocontour level defining ${C}^{\prime, {\rm map}}_{\alpha}$. We then construct the  interval $({\xi}_{-, \Omega_i}, {\xi}_{+,  \Omega_i})$ 
that represents the range of intensity values $\xi$ of $\Omega_i$ for which ${\vect x}_{i,\xi} \in {C}^{\prime, {\rm map}}_{\alpha}$. 

Finally, for visualisation, we gather all the lower and upper bounds ${\xi}_{-,  \Omega_i}$, ${\xi}_{+,  \Omega_i}$, $\forall i$, into the following two images:
\begin{equation}\label{eqn:cr-local-sp}
{\vect \xi}_- = \sum_i {\xi}_{-, \Omega_i} \vect \zeta_{\Omega_i}, \quad {\vect \xi}_+ = \sum_i {\xi}_{+, \Omega_i} \vect \zeta_{\Omega_i}.
\end{equation}
We typically consider the difference image $({\vect \xi}_+ - {\vect \xi}_-)$ that shows the length of the local credible intervals (\textit{cf.} error bars).
These images can be constructed at different scales to analyse structure of different sizes. In our experiments, as examples,
we consider superpixels of sizes $10\times 10$, $20\times 20$, and $30\times 30$ pixels.

{To conclude, notice that visualising uncertainty in high dimensional problems is fundamentally difficult. For example, even the simple case of $N$-dimensional Gaussian models involves covariance matrices of size $N \times N$;  the models considered here are significantly more complex. As a result, uncertainty information could potentially structure along directions of the parameter space that the visual uncertainty plots described above fail to capture.  However, we believe that correlations in images are predominantly local, albeit at potentially different scales. What our analyses seek to capture and visually display are precisely these local correlations at superpixel scales of different levels.}

\subsection{Hypothesis testing of image structure}
In a manner akin to the companion article \cite{CPM17}, we use \emph{knock-out} posterior tests to assess specific areas or 
structures of interest in the reconstructed images. These tests proceed by constructing a surrogate test image ${\vect x}^{*, {\rm sgt}}$ 
by carefully replacing the structure of interest in an point estimator ${\vect x}^{*}$ (or $\bm{\mathsf{\Psi}} {\vect a}^{*}$) 
with background information. If removing the structure has pushed ${\vect x}^{*, {\rm sgt}}$ outside of the HPD credible region (\textit{i.e.} ${\vect x}^{*, {\rm sgt}} \notin {C}^{\prime, {\rm map}}_{\alpha}$), 
this indicates that the data strongly supports the structure under consideration. 
Conversely, if ${\vect x}^{*, {\rm sgt}}$ remains inside of the HPD credible region (\text{i.e.} ${\vect x}^{*, {\rm sgt}} \in {C}^{\prime, {\rm map}}_{\alpha}$), then the likelihood is insensitive to the modification, 
indicating lack of strong evidence for the scrutinised structure.

Algorithmically, a surrogate ${\vect x}^{*, {\rm sgt}}$ for a test area $\Omega_D \subset \Omega$ is generated by performing 
segmentation-inpainting of ${\vect x}^{*}$, for example by applying a wavelet filter $\bm{\mathsf{\Lambda}}$ iteratively by using
\begin{equation} \label{eqn:inpaint}
{\vect x}^{(m+1), {\rm sgt}} = {\vect x}^{*}  \mathbb{1}_{\Omega - \Omega_D} + \bm{\mathsf{\Lambda}}^{\dagger}{\rm soft}_{\lambda_{\rm thd}} 
	(\bm{\mathsf{\Lambda}}{\vect x}^{(m), {\rm sgt}}) \mathbb{1}_{\Omega_D},
\end{equation}
with ${\vect x}^{(0), {\rm sgt}} = {\vect x}^{*}$ or ${\vect x}^{(0), {\rm sgt}} = \bm{\mathsf{\Psi}} {\vect a}^{*}$ 
for the synthesis formulation (usually 100 iterations are sufficient for convergence). 
To determine if ${\vect x}^{*, {\rm sgt}} \in {C}^{\prime, {\rm map}}_{\alpha}$, 
it suffices to check if 
\begin{equation}
f({\vect x}^{*, {\rm sgt}}) + g({\vect x}^{*, {\rm sgt}}) \le {\gamma}^\prime_{\alpha}.
\end{equation}

{In addition to the approach presented above to assess the existence of specific areas or structures of interest, we also propose the following approach 
to focus on assessing sub-structure within areas of interest.   
Briefly speaking, we create surrogate test images with the sub-structure in question effectively removed by smoothing 
the corresponding region. Algorithmically, a surrogate ${\vect x}^{*, {\rm sgt}}$ for a test area $\Omega_D \subset \Omega$ is then generated by 
\begin{equation} \label{eqn:smooth}
{\vect x}^{*, {\rm sgt}} = {\vect x}^{*}  \mathbb{1}_{\Omega - \Omega_D} +  
	(\bm{\mathsf{S}}{\vect x}^*) \mathbb{1}_{\Omega_D},
\end{equation}
where $\bm{\mathsf{S}}$ is a smoothing operator applied to remove sub-structure within the test area $\Omega_D$.
}

\section{Experimental results}\label{sec:exp}
We now investigate the performance of the proposed uncertainty quantification methodology for the three strategies
discussed in Section~\ref{sec:uq}. We also report a detailed comparison with the proximal MCMC method Px-MALA, 
which is one of the MCMC methods introduced in the companion article \citep{CPM17} and that can also support sparsity-promoting priors. 
Px-MALA produces (asymptotically) exact inferences and therefore we use it here as an accurate benchmark for the methods
proposed in this article.

\subsection{Simulations}
\begin{figure*}
	\centering
	\begin{tabular}{cccc}
		\includegraphics[trim={{.15\linewidth} {.07\linewidth} {.02\linewidth} {.07\linewidth}}, clip, width=0.24\linewidth, height = 0.21\linewidth]
		{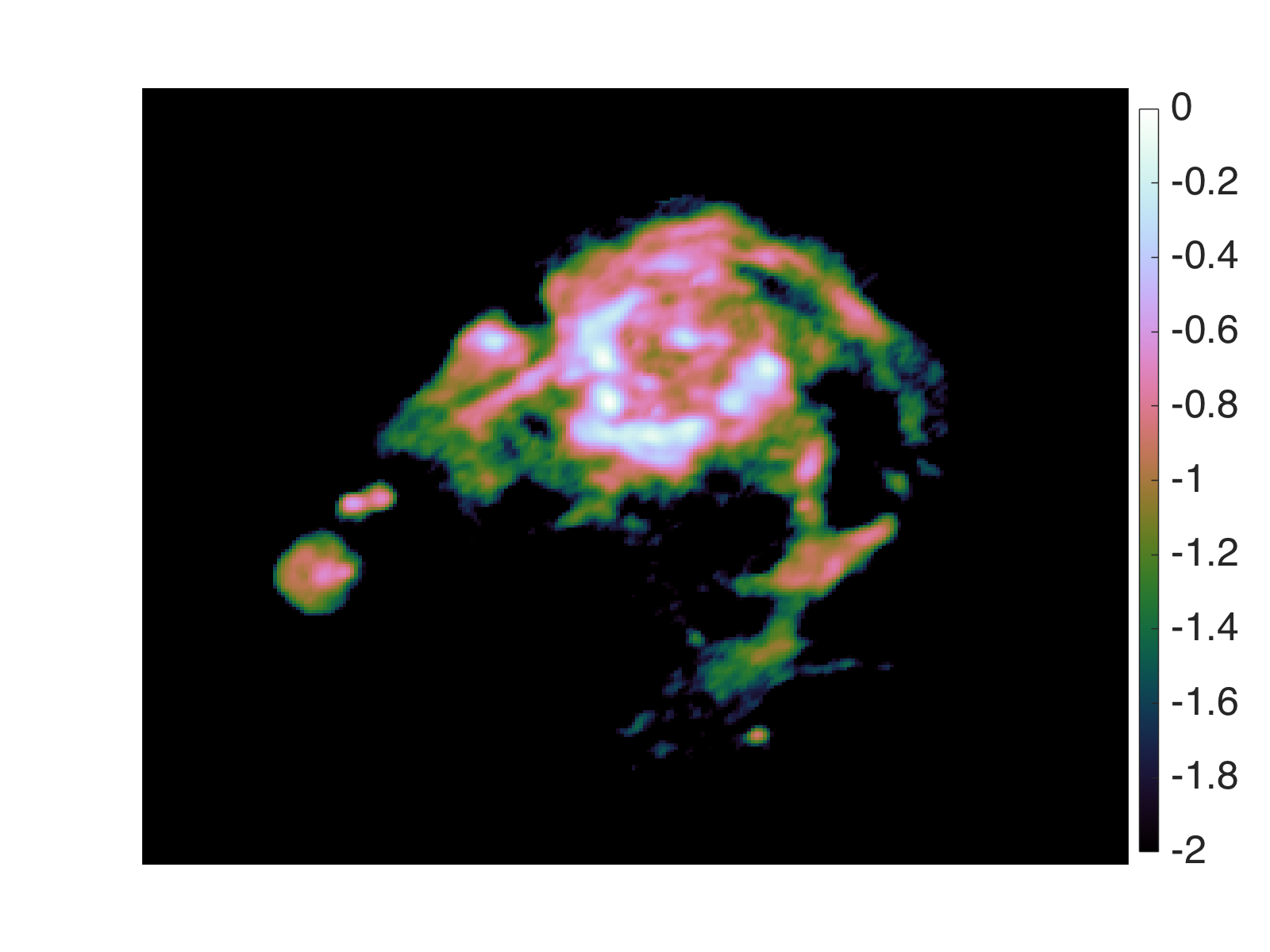} &
		\includegraphics[trim={{.15\linewidth} {.07\linewidth} {.03\linewidth} {.073\linewidth}}, clip, width=0.24\linewidth, height = 0.21\linewidth]
		{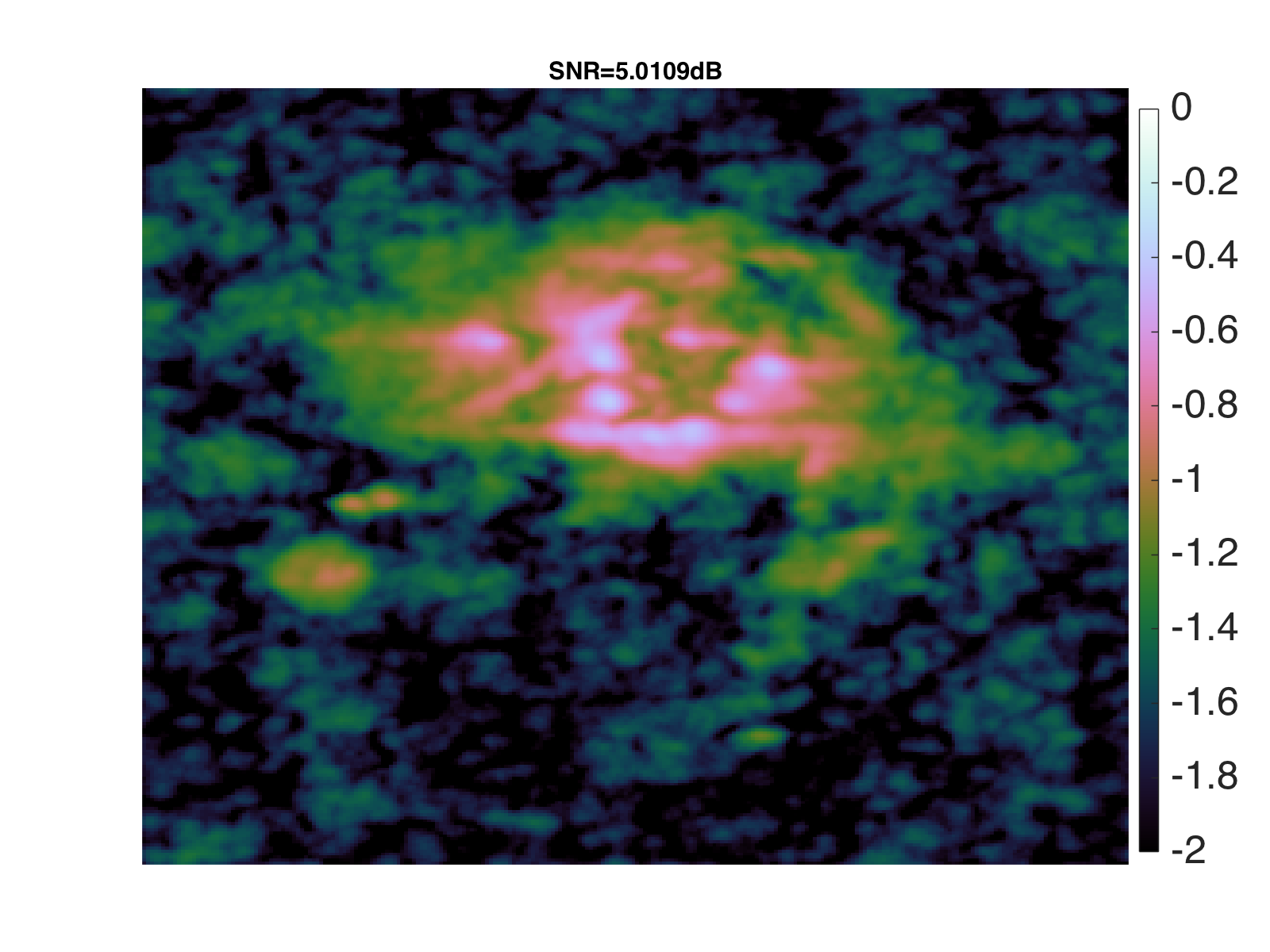} &
		\includegraphics[trim={{.15\linewidth} {.07\linewidth} {.02\linewidth} {.073\linewidth}}, clip, width=0.24\linewidth, height = 0.21\linewidth]
		{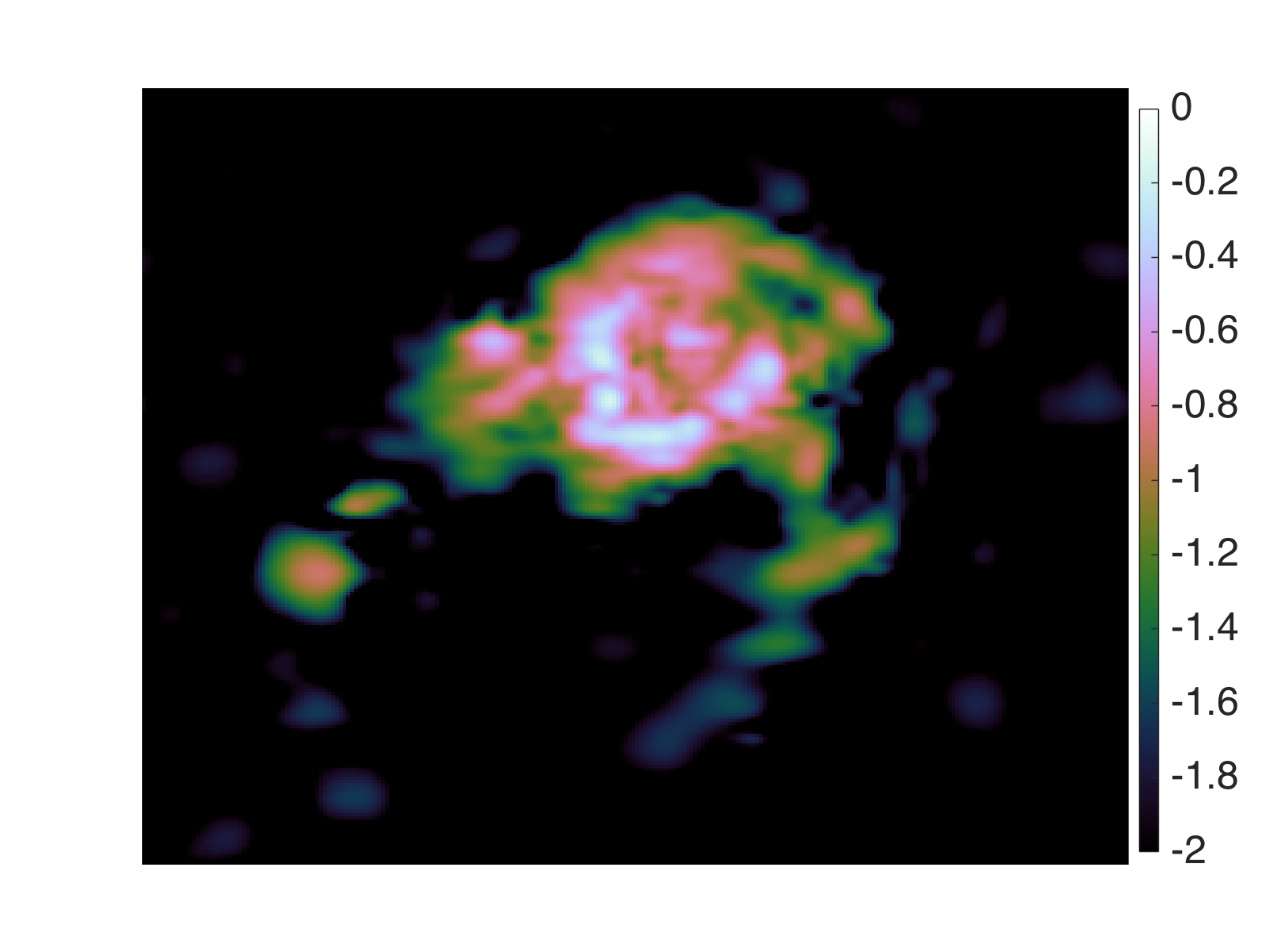} &
		\includegraphics[trim={{.15\linewidth} {.07\linewidth} {.02\linewidth} {.073\linewidth}}, clip, width=0.24\linewidth, height = 0.21\linewidth]
		{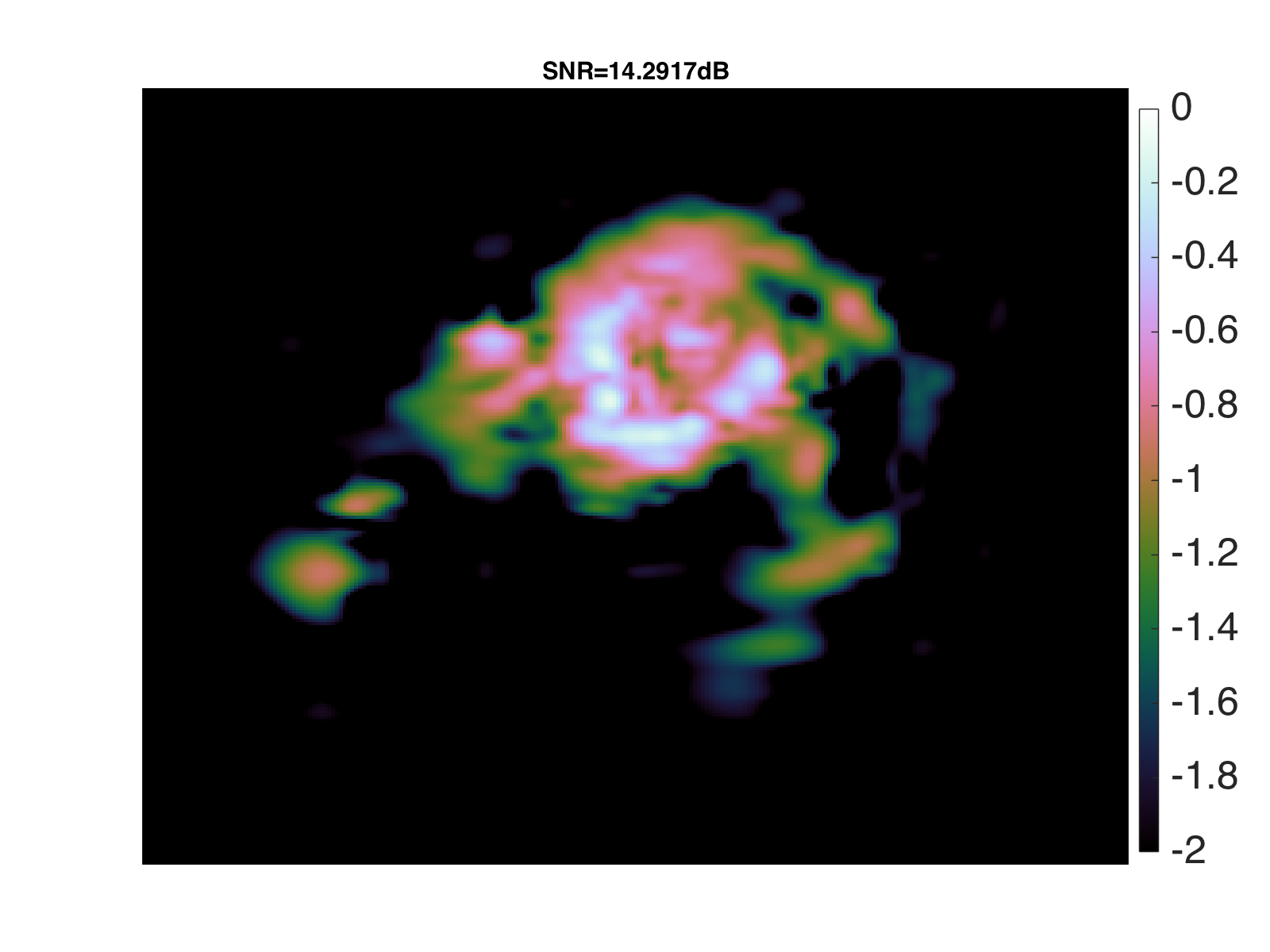} 
		\\
		{\small (a) ground truth} & {\small (b) dirty map} & {\small (c)  Px-MALA for analysis model} & {\small (d) MAP for analysis model }
		\\
		&
		&
		\includegraphics[trim={{.15\linewidth} {.07\linewidth} {.02\linewidth} {.073\linewidth}}, clip, width=0.24\linewidth, height = 0.21\linewidth]
		{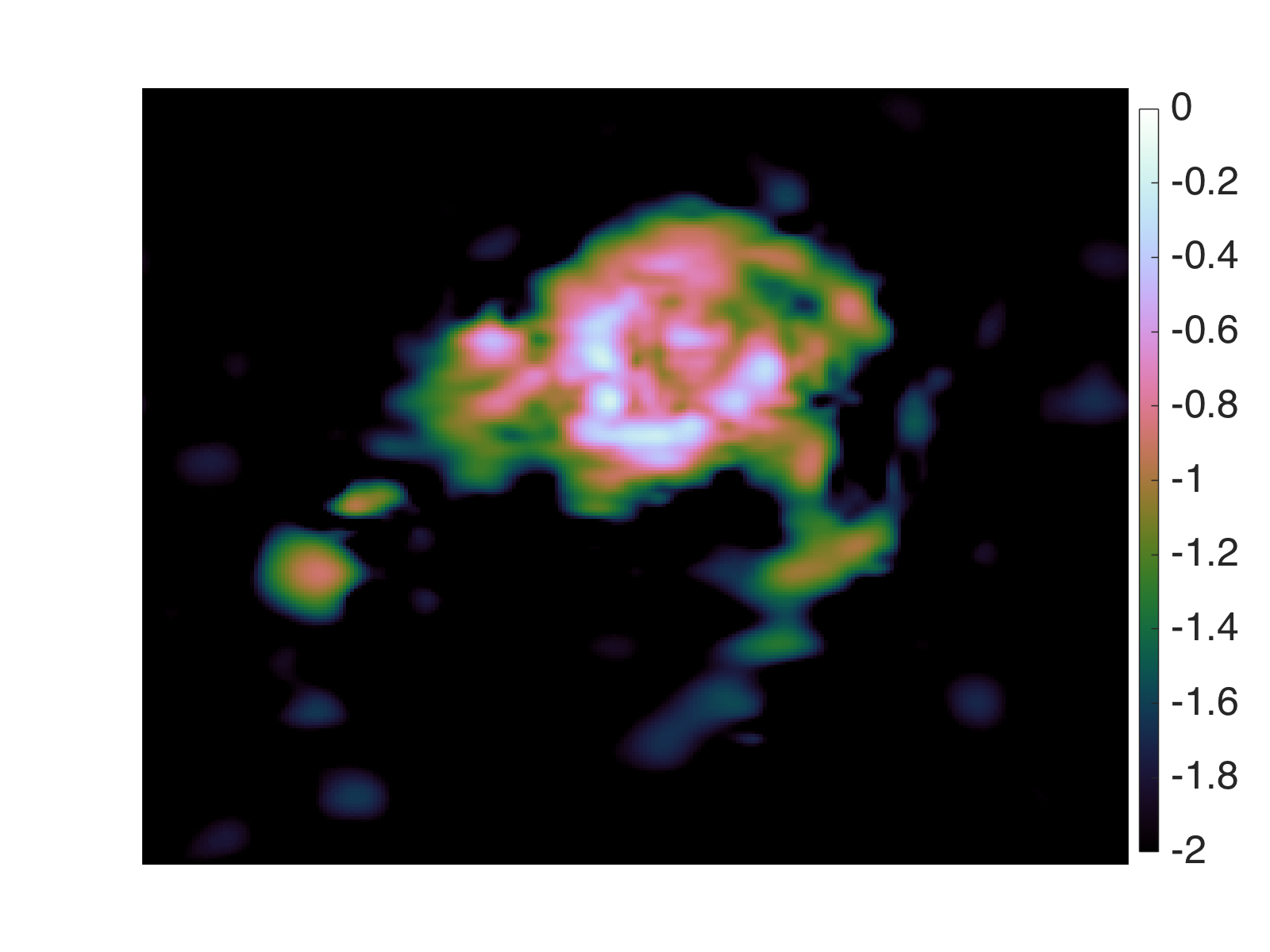} & 
		\includegraphics[trim={{.15\linewidth} {.07\linewidth} {.02\linewidth} {.073\linewidth}}, clip, width=0.24\linewidth, height = 0.21\linewidth]
		{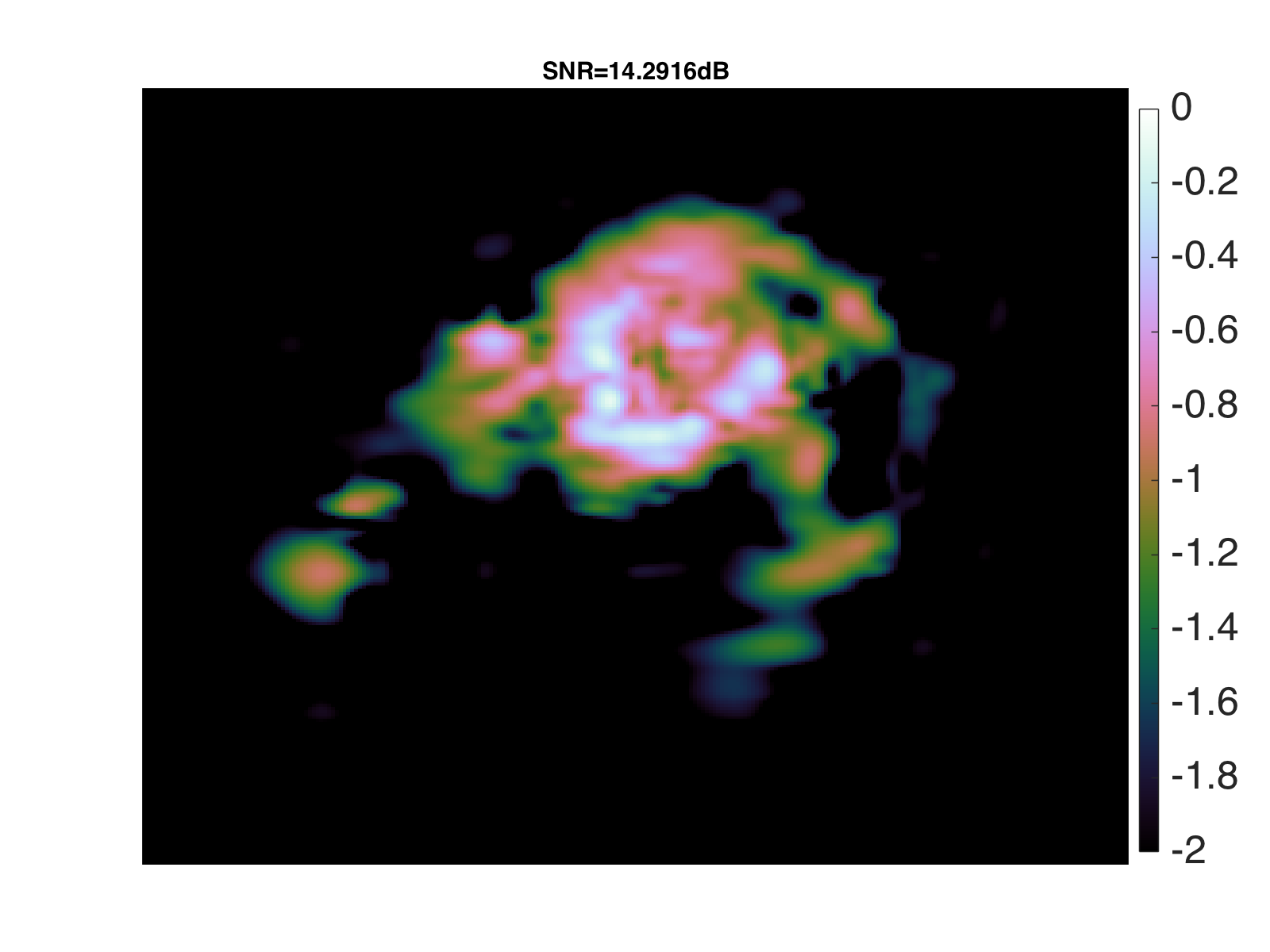} 
		\\
		& & {\small (e) Px-MALA for synthesis model  }  & {\small (f) MAP for synthesis model }  	 
        \end{tabular}
	\caption{Image reconstructions for M31 (size $256\times 256$). All images are shown in ${\tt log}_{10}$ scale 
	{(\textit{i.e.} the numeric labels on the colour bar are the logarithms of the image intensity).}. (a): ground truth; 
	(b): dirty image (reconstructed by inverse Fourier transform);
	(c) and (d): point estimators for the analysis model \eqref{eqn:ir-un-af} computed by Px-MALA and MAP estimation, respectively;
	(e) and (f): the same as (c) and (d) but for the synthesis model \eqref{eqn:ir-un-sf}.
	In particular, the point estimators of Px-MALA are the sample mean.
	Clearly, consistent results between Px-MALA and MAP estimation and between the analysis and synthesis models are obtained.
	}
	\label{fig-m31}
\end{figure*}

In a manner akin to \cite{CPM17}, we perform our experiments with the following four RI images: M31 galaxy (size $256\times 256$), 
Cygnus A galaxy (size $256\times 512$), W28 supernova remnant (size $256\times 256$), and 3C288 (size $256\times 256$). 
These images are depicted in Figure \ref{fig-m31} (a) and Figure \ref{fig-others} (a).   Radio interferometric observations are simulated for these ground truth images in a similar manner as in \citet{CPM17}.

The numerical experiments performed in this article for MAP estimation were run on a Macbook laptop with an i7 Intel CPU and memory of 16 GB,
running MATLAB R2015b. The Px-MALA algorithm used as a benchmark is significantly more computationally expensive and required 
a high-performance workstation  (see \citealt{CPM17}). For further details about the experiment setup and the implementation of Px-MALA please 
see \cite{CPM17}.

Regarding the models used for the experiments, the $\ell_1$ regularisation parameter $\mu$ in the analysis and synthesis 
models is set to $10^{4}$
and the dictionary $\bm{\mathsf{\Psi}}$ in the analysis and synthesis models
is set to Daubechies 8 wavelets.
In Algorithms \ref{alg:fb-a} and \ref{alg:fb-s}, 
we use $\lambda^{(i)} = 0.5$, {with stopping criteria set by a maximum iteration number of 500 or relative difference between solutions of $10^{-4}$}. In formulas \eqref{eqn:gamma-a} and \eqref{eqn:gamma-s}, the range of values for $\alpha$ is $[0.01, 0.99]$. 
In particular, credible regions and intervals are reported at $\alpha = 0.01$, corresponding to the 99\% credible level.  
The maximum number of iterations for segmented-inpainting in \eqref{eqn:inpaint} is set to 200.

\addtolength{\tabcolsep}{-\tabL}
\addtolength{\tabcolsep}{\tabL}

\addtolength{\tabcolsep}{-\tabL}
{ \renewcommand{\arraystretch}{0.0}
\begin{figure*}
	\centering
	\begin{tabular}{cccc}
		\includegraphics[trim={{.15\linewidth} {.07\linewidth} {.025\linewidth} {.07\linewidth}}, clip, width=0.24\linewidth, height = 0.13\linewidth]
		{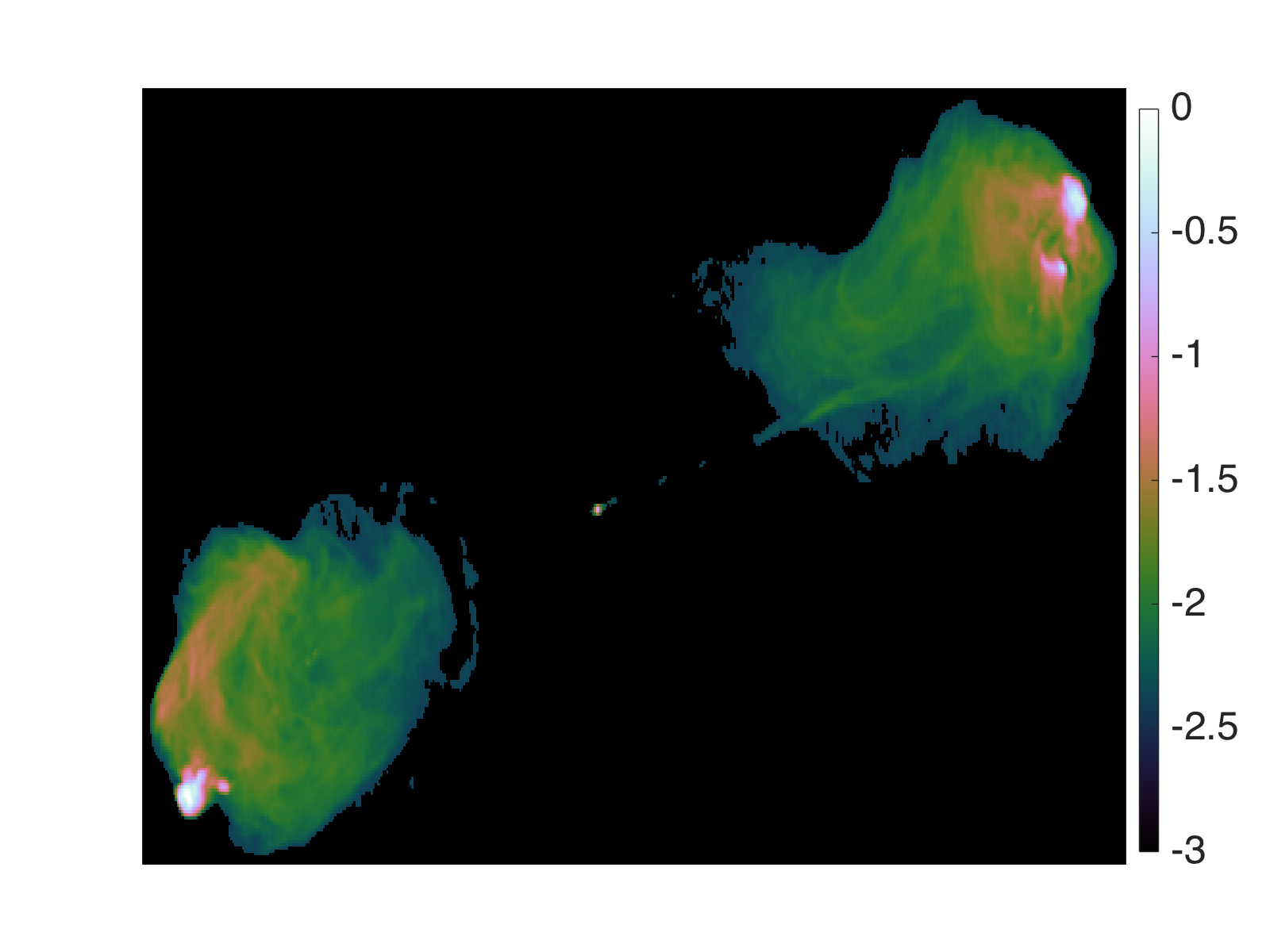} &
		\includegraphics[trim={{.15\linewidth} {.07\linewidth} {.025\linewidth} {.072\linewidth}}, clip, width=0.24\linewidth, height = 0.13\linewidth]
		{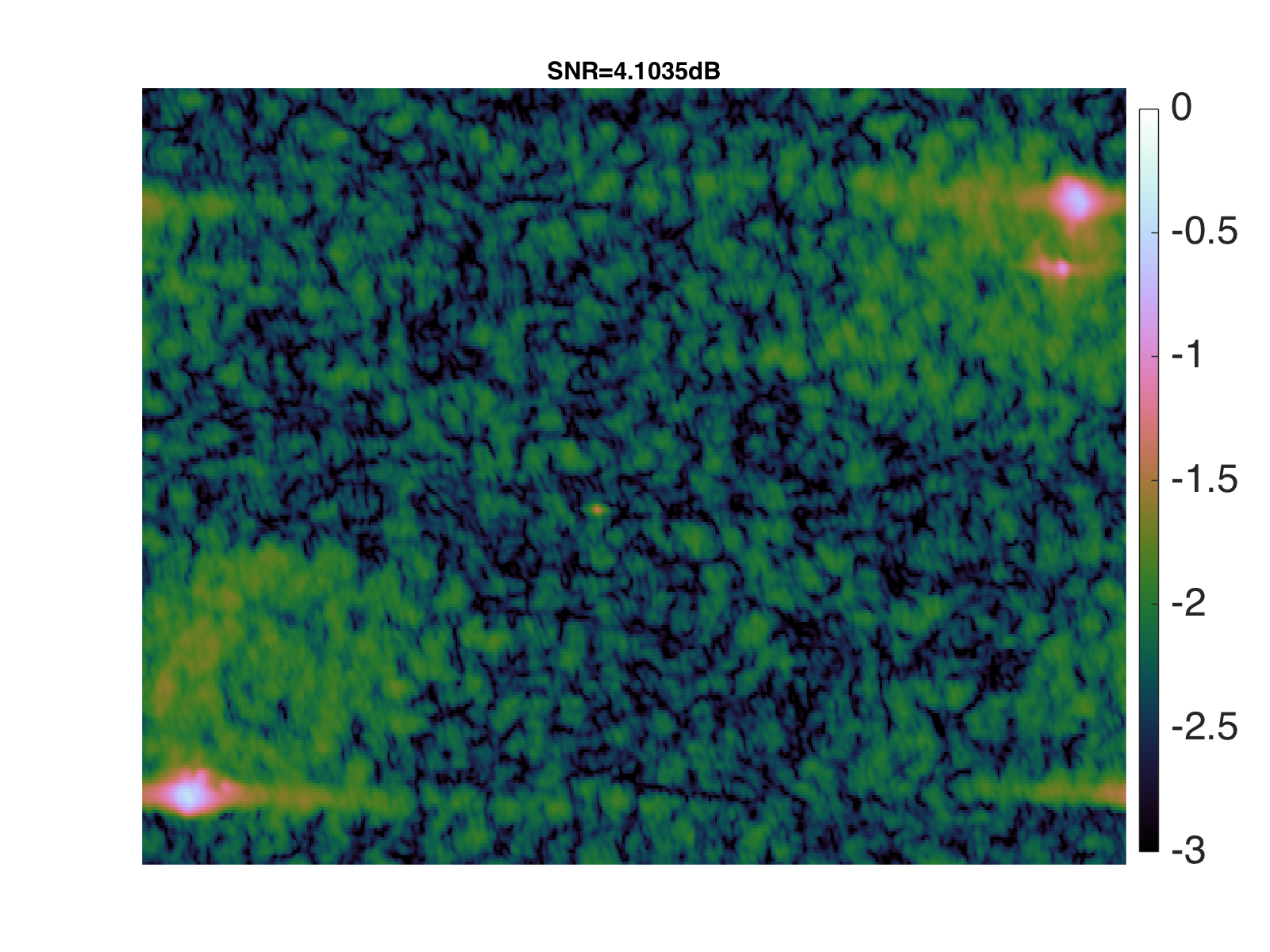} &
		\includegraphics[trim={{.15\linewidth} {.07\linewidth} {.025\linewidth} {.072\linewidth}}, clip, width=0.24\linewidth, height = 0.13\linewidth]
		{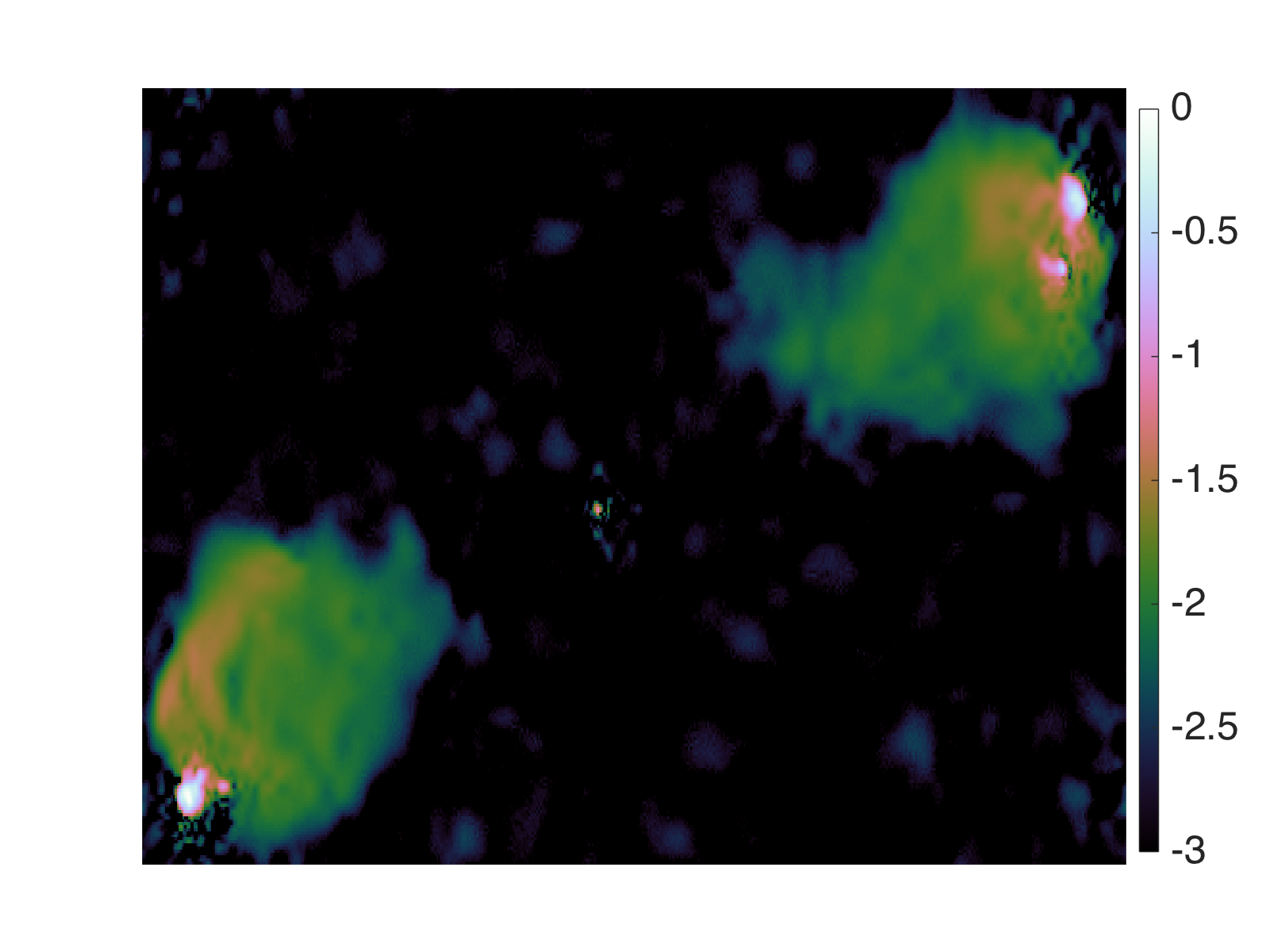} &
		\includegraphics[trim={{.15\linewidth} {.07\linewidth} {.025\linewidth} {.073\linewidth}}, clip, width=0.24\linewidth, height = 0.13\linewidth]
		{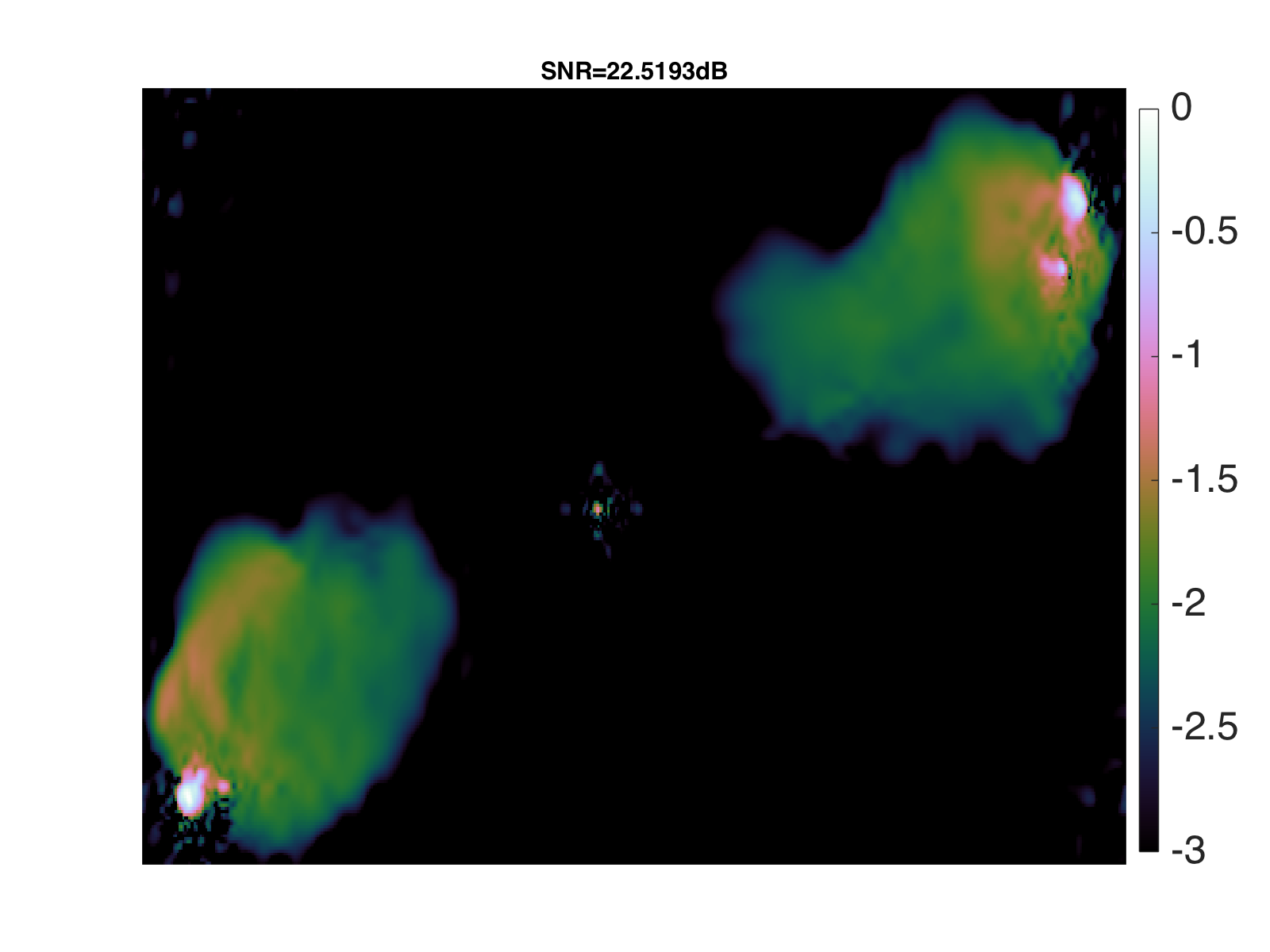}  
		\\
		\includegraphics[trim={{.15\linewidth} {.07\linewidth} {.02\linewidth} {.07\linewidth}}, clip, width=0.24\linewidth, height = 0.21\linewidth]
		{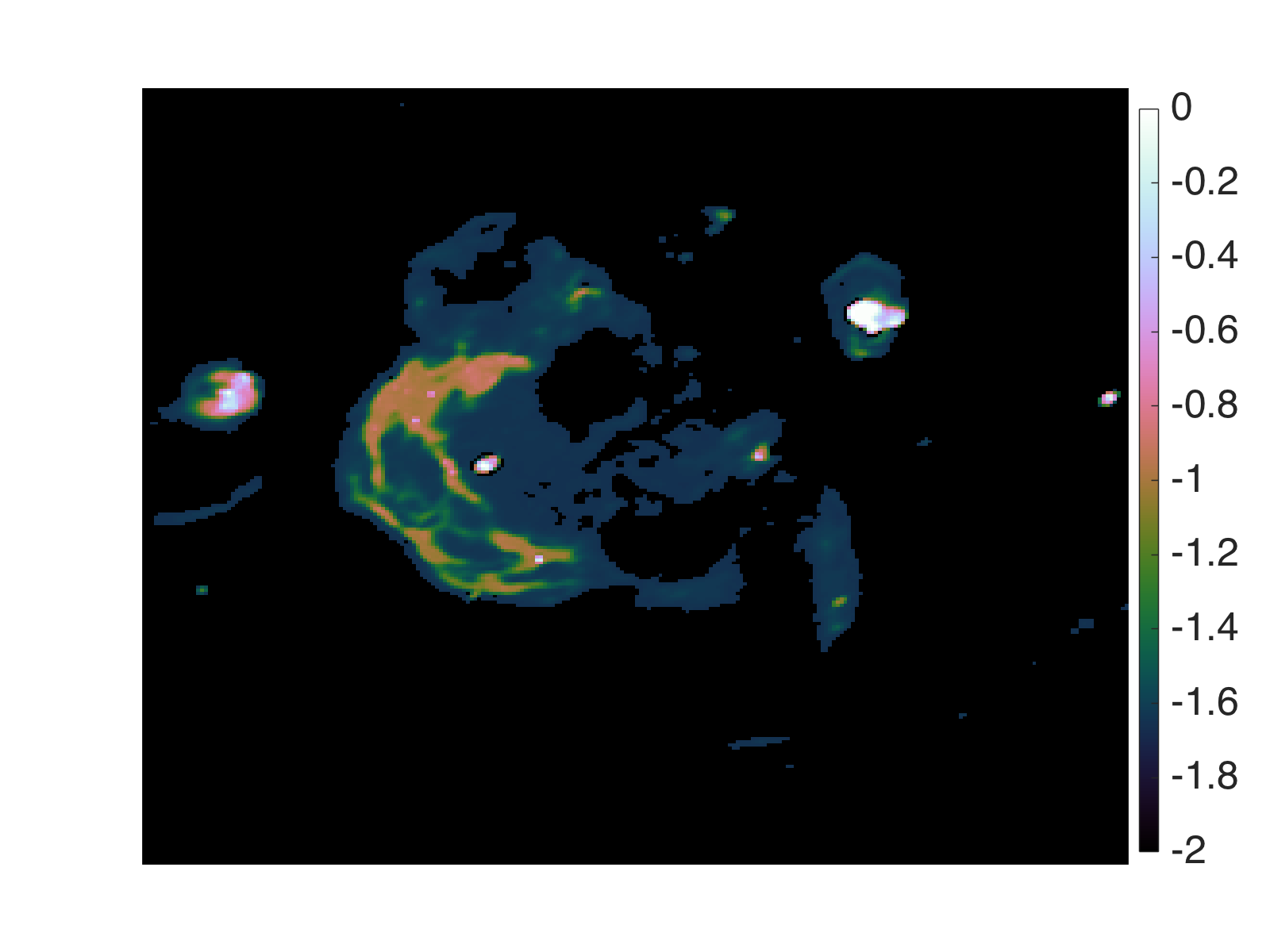} &
		\includegraphics[trim={{.15\linewidth} {.07\linewidth} {.03\linewidth} {.073\linewidth}}, clip, width=0.24\linewidth, height = 0.21\linewidth]
		{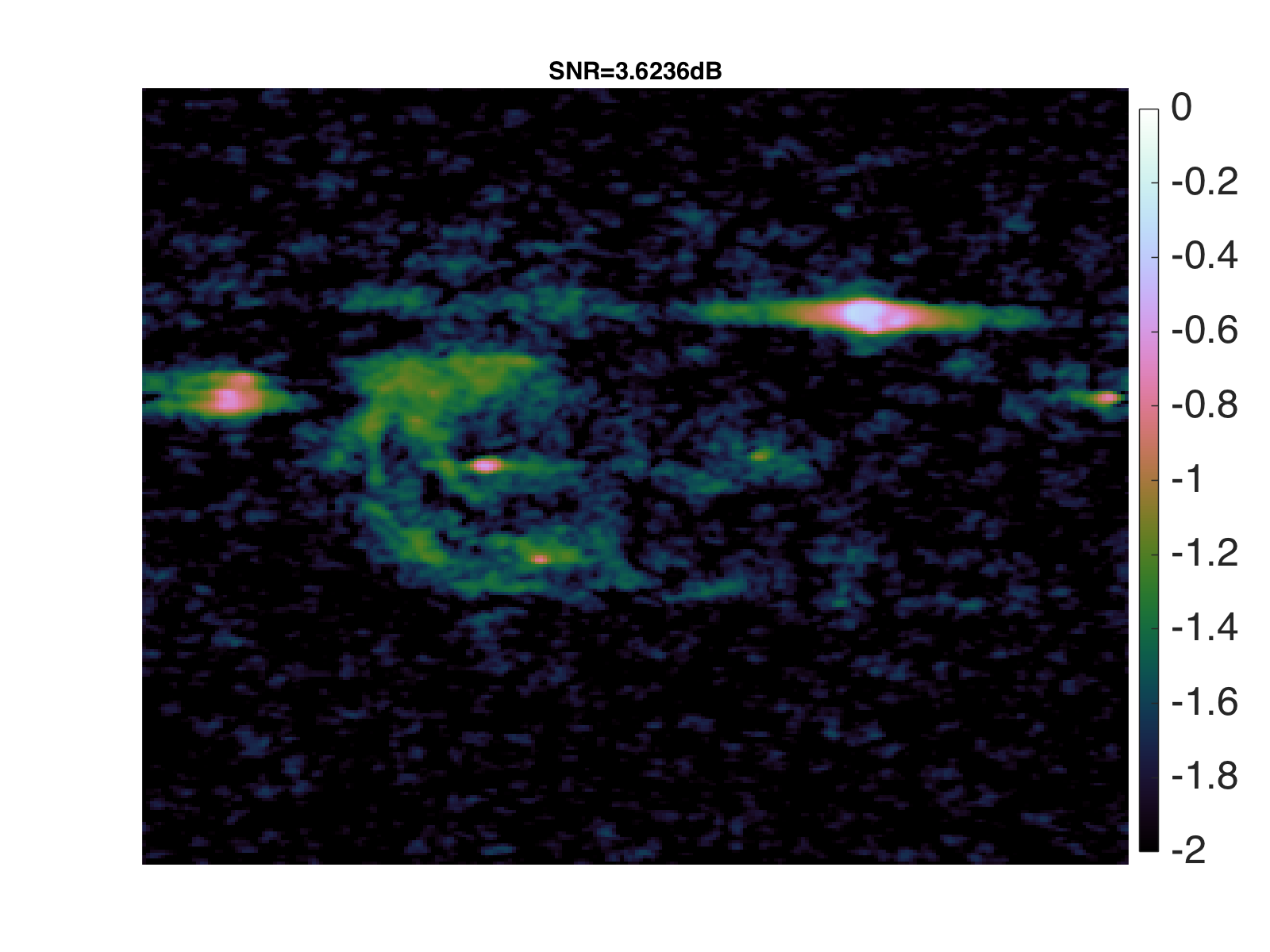} &
		\includegraphics[trim={{.15\linewidth} {.07\linewidth} {.02\linewidth} {.073\linewidth}}, clip, width=0.24\linewidth, height = 0.21\linewidth]
		{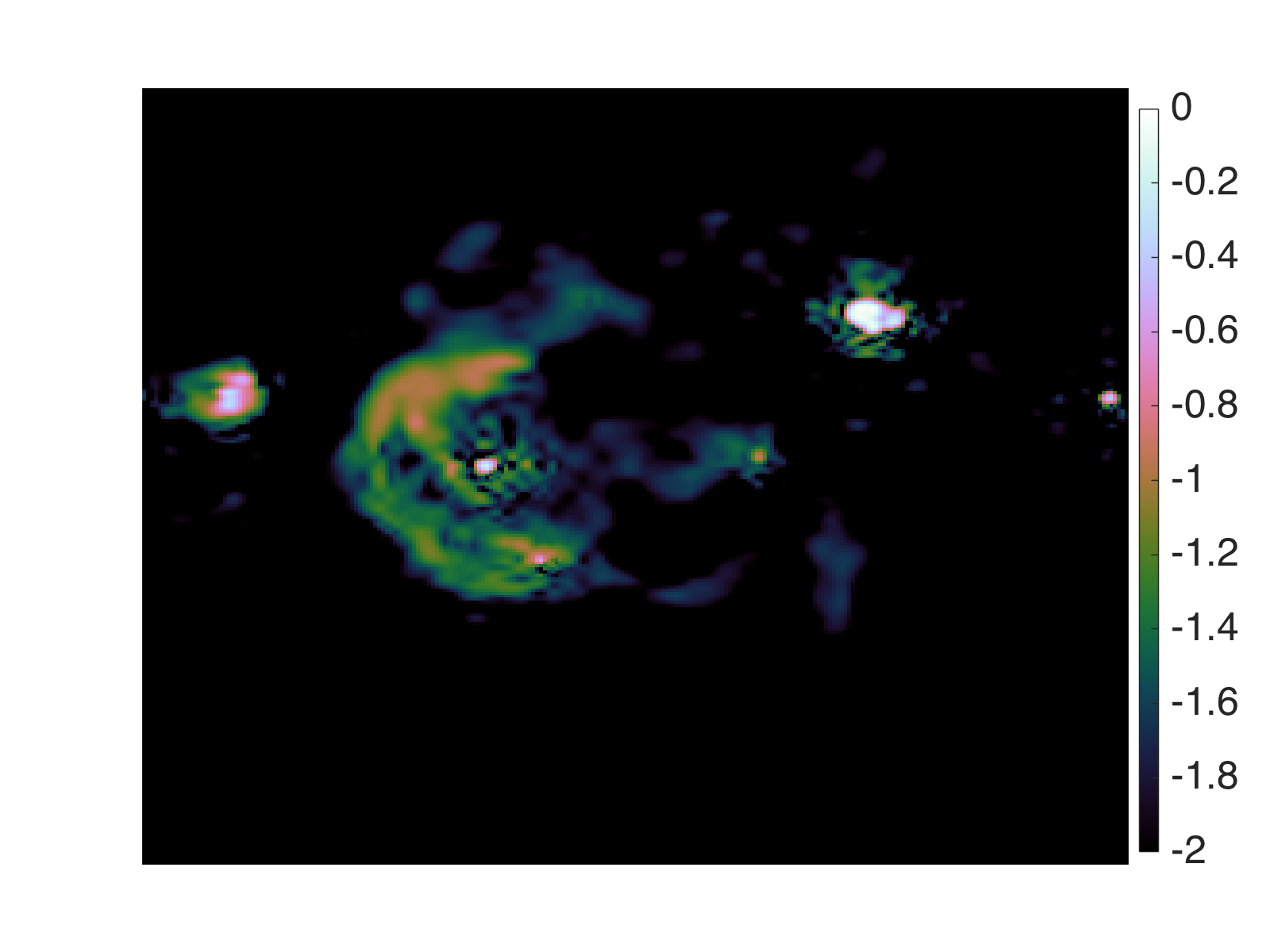} &
		\includegraphics[trim={{.15\linewidth} {.07\linewidth} {.02\linewidth} {.073\linewidth}}, clip, width=0.24\linewidth, height = 0.21\linewidth]
		{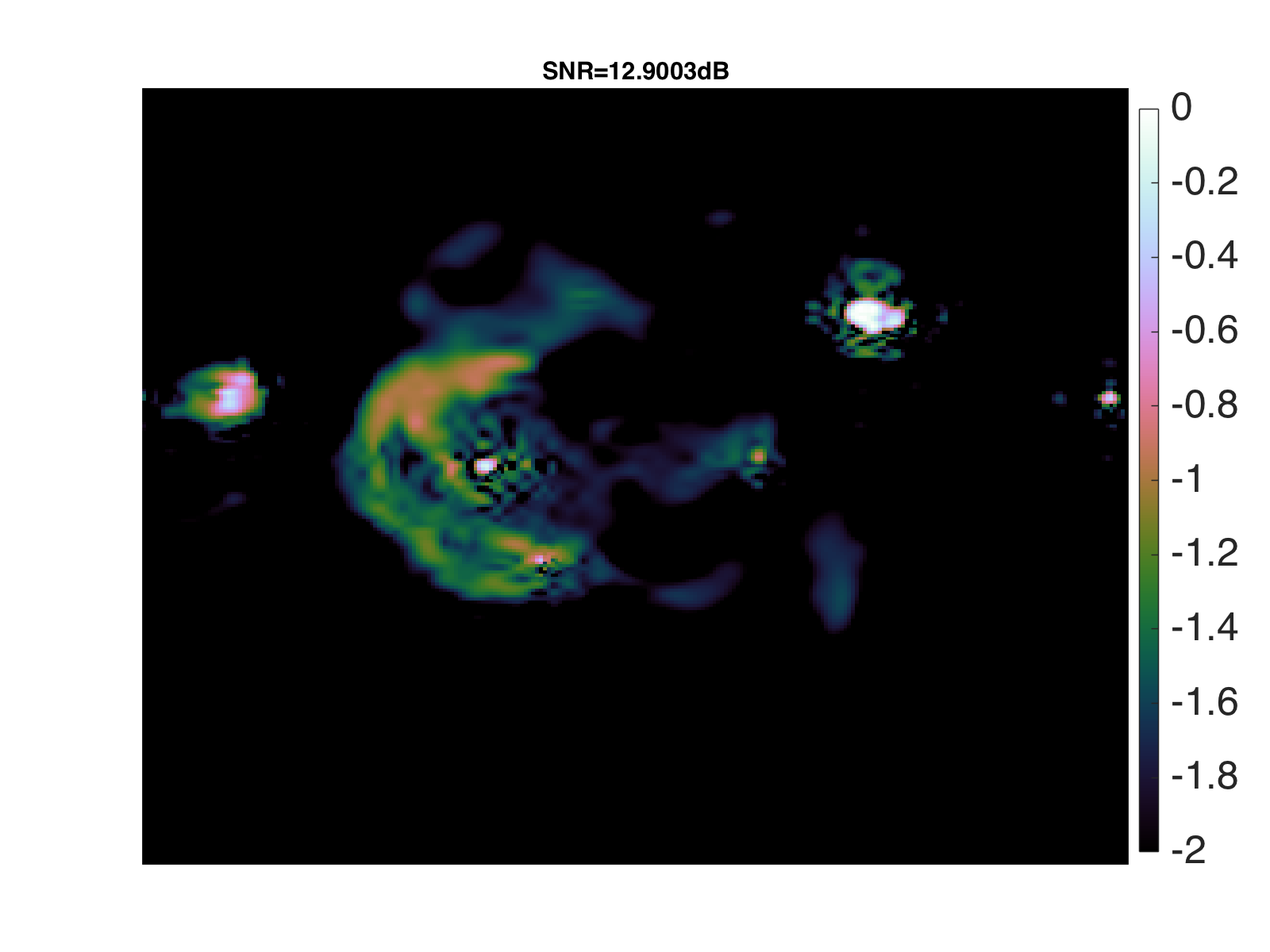} 
		\\
		\includegraphics[trim={{.15\linewidth} {.07\linewidth} {.02\linewidth} {.07\linewidth}}, clip, width=0.24\linewidth, height = 0.21\linewidth]
		{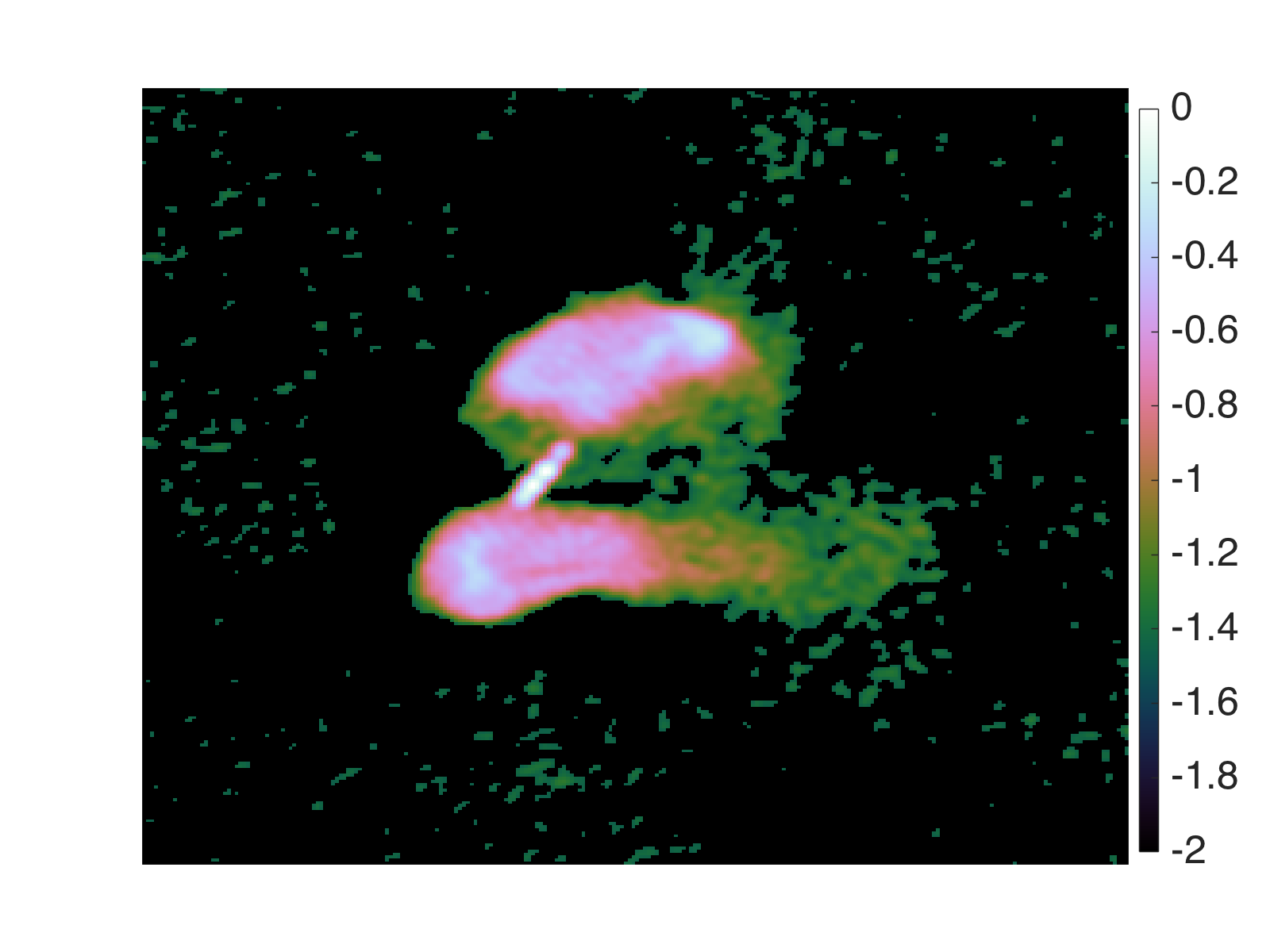} &
		\includegraphics[trim={{.15\linewidth} {.07\linewidth} {.03\linewidth} {.073\linewidth}}, clip, width=0.24\linewidth, height = 0.21\linewidth]
		{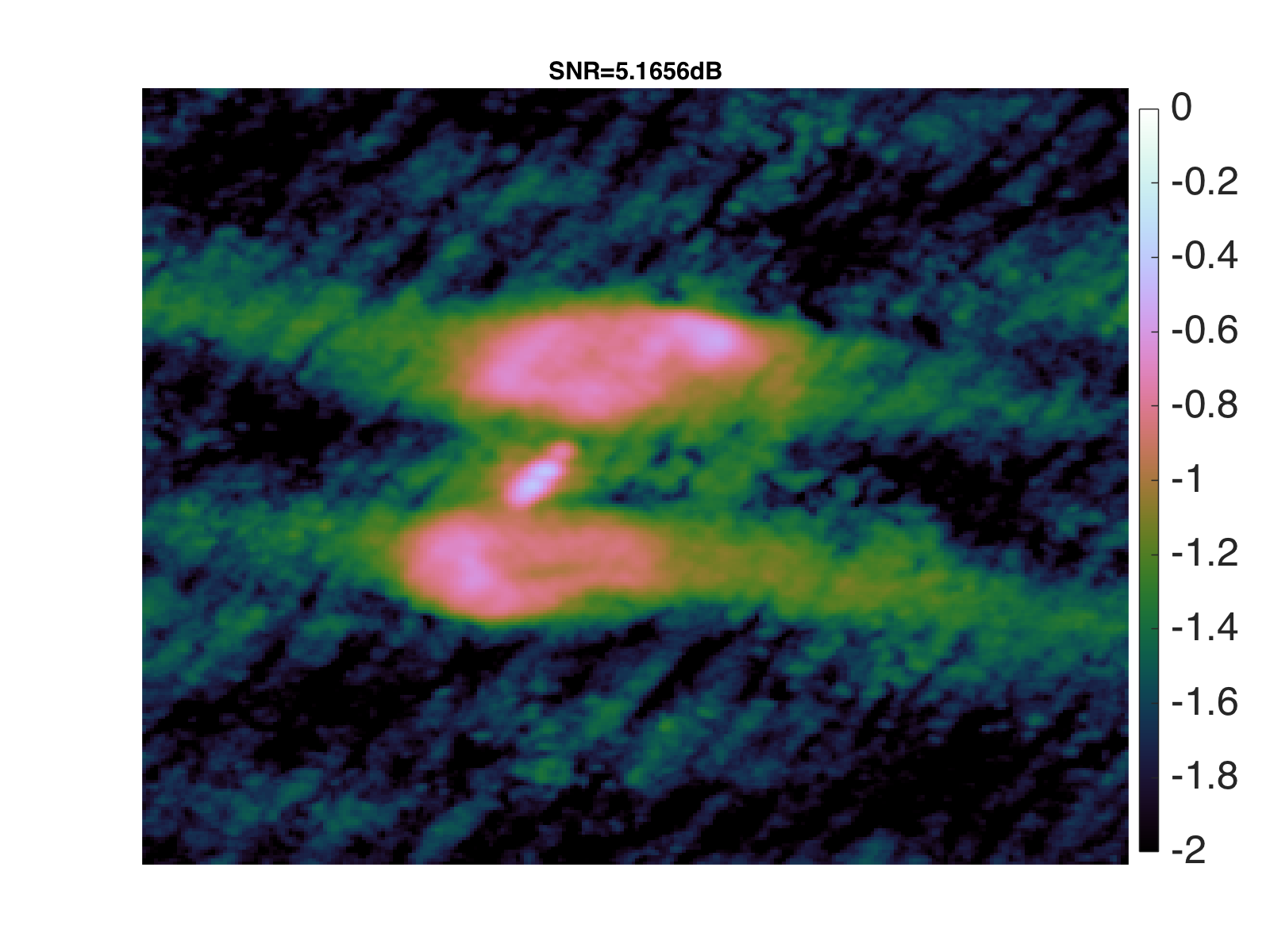} &
		\includegraphics[trim={{.15\linewidth} {.07\linewidth} {.02\linewidth} {.073\linewidth}}, clip, width=0.24\linewidth, height = 0.21\linewidth]
		{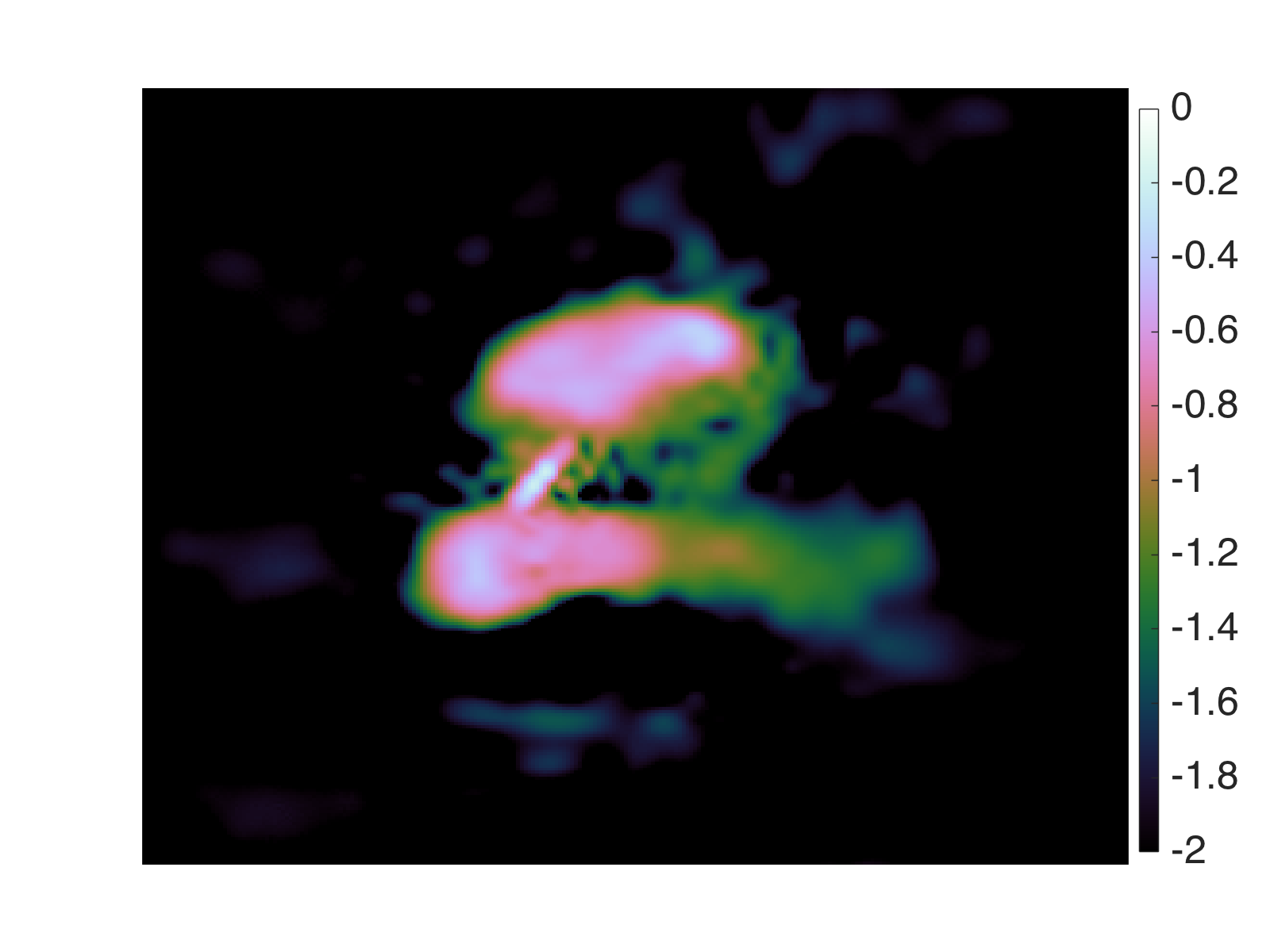} &		
		\includegraphics[trim={{.15\linewidth} {.07\linewidth} {.02\linewidth} {.073\linewidth}}, clip, width=0.24\linewidth, height = 0.21\linewidth]
		{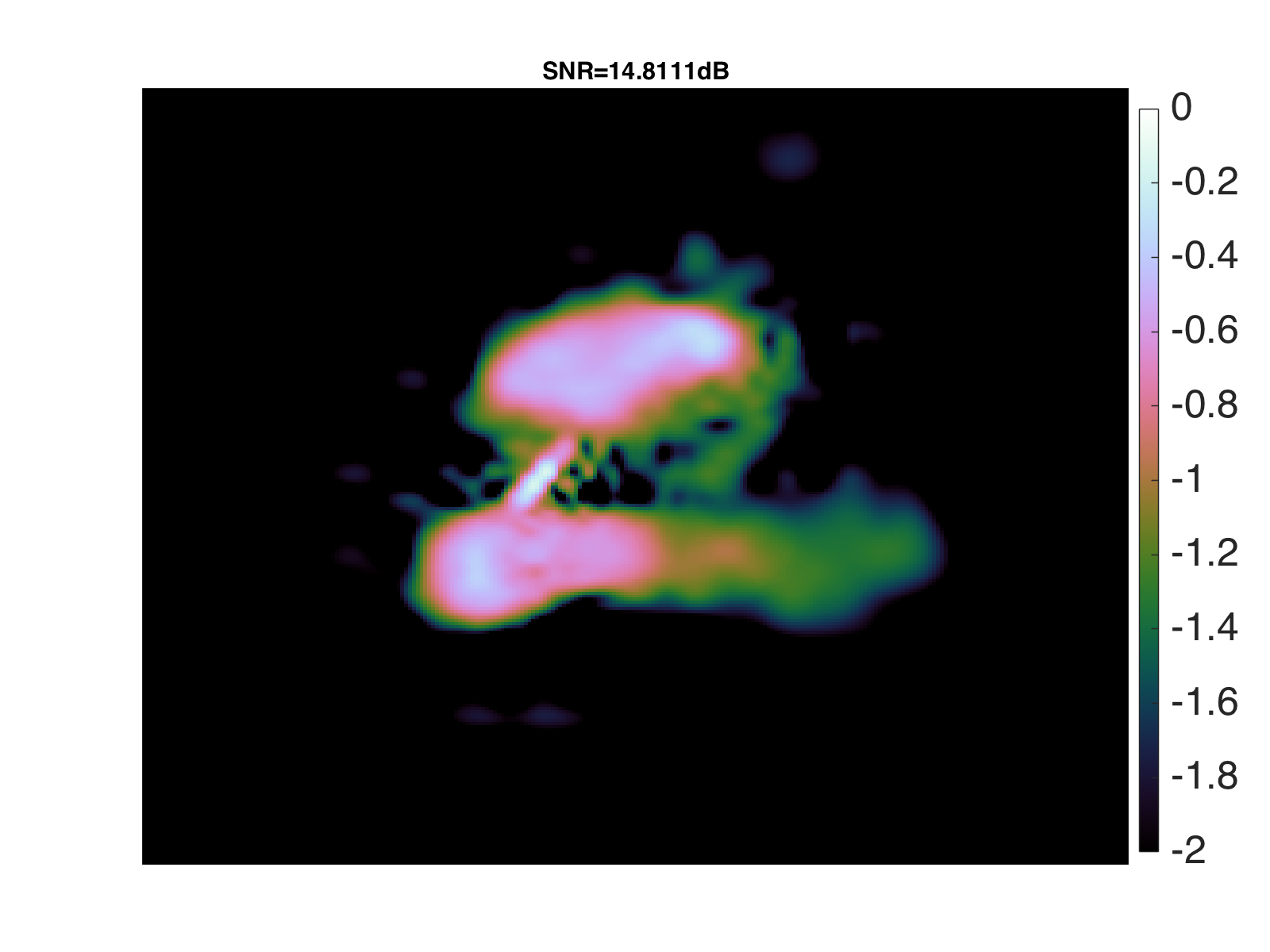} 			
		\\ \vspace{0.05in}\\
		{\small (a) ground truth} & {\small (b) dirty map} & {\small (c) Px-MALA for analysis model }  & {\small (d) MAP for analysis model }
        \end{tabular}
	\caption{Image reconstructions for Cygnus A (size $256\times 512$), W28 (size $256\times 256$), and 3C288 (size $256\times 256$) radio galaxies (first to third rows). 
	All images are shown in ${\tt log}_{10}$ scale.
	First column: (a) ground truth. 
	Second to forth columns: (b) dirty images; (c) and (d) point estimators for the analysis model \eqref{eqn:ir-un-af} computed by Px-MALA and MAP estimation, respectively.
	Clearly, consistent results between Px-MALA and MAP estimation are obtained.
	}
	\label{fig-others}
\end{figure*}
}
\addtolength{\tabcolsep}{\tabL}

\subsection{Image reconstruction}
As the first step in our analysis we perform Bayesian image reconstruction for the four images considered. 
Precisely, for each image we compute two Bayesian estimators, the MAP estimator computed by convex optimisation 
and the sample mean estimator computed with Px-MALA. For completeness, we consider both the analysis 
and the synthesis models \eqref{eqn:ir-un-af} and \eqref{eqn:ir-un-sf}.

The Bayesian estimators related to the analysis model are shown in Figures \ref{fig-m31} and \ref{fig-others}.  
Observe that both estimators produce similar, excellent reconstruction results. For comparison, dirty maps 
(reconstructed by applying the inverse Fourier transform directly to the visibilities) of the test images are shown in
Figure \ref{fig-m31} (b) and Figure \ref{fig-others} (b). As expected, 
the results of the analysis and synthesis models \eqref{eqn:ir-un-af} and \eqref{eqn:ir-un-sf} under 
an orthogonal basis $\bm{\mathsf{\Psi}}$ are nearly undistinguishable\footnote{Note that, 
when $\bm{\mathsf{\Psi}}^\dagger \bm{\mathsf{\Psi}} = \bm{\mathsf{ I}}$, as considered here, 
the analysis and synthesis models are identical. However, when $\bm{\mathsf{\Psi}}^\dagger \bm{\mathsf{\Psi}} \neq \bm{\mathsf{ I}}$,
they are very different and we expect different reconstructed images.} (see results for M31 in Figure \ref{fig-m31};
to avoid redundancy the results for the other images are not reported here). For this reason, in the reminder 
of this article only the results for the analysis model are presented.

We emphasise again that MAP estimators computed by convex optimisation are significantly faster to compute than 
the estimators that require MCMC methods. In particular, in our experiments there is a gain of order $10^5$ 
in terms of computation time (see Table \ref{tab:time} for the computation time comparisons with Px-MALA).
Furthermore, MAP estimation based on convex optimisation supports algorithmic structures that can be highly distributed  \citep[\textit{e.g.}][]{car14,OCRMTPW16} to further assist in scaling to big-data.
MCMC algorithms cannot typically be distributed to such a high degree.
We have not yet considered distributed MAP algorithms here; our MAP-based methods therefore provide additional performance improvements over MCMC beyond the already dramatic improvements shown in Table \ref{tab:time}.

\begin{figure*}
	\centering
	\begin{tabular}{cccc}
		\includegraphics[trim={{.01\linewidth} {.05\linewidth} {.1\linewidth} {.01\linewidth}}, clip, width=0.22\linewidth, height = 0.2\linewidth]
		{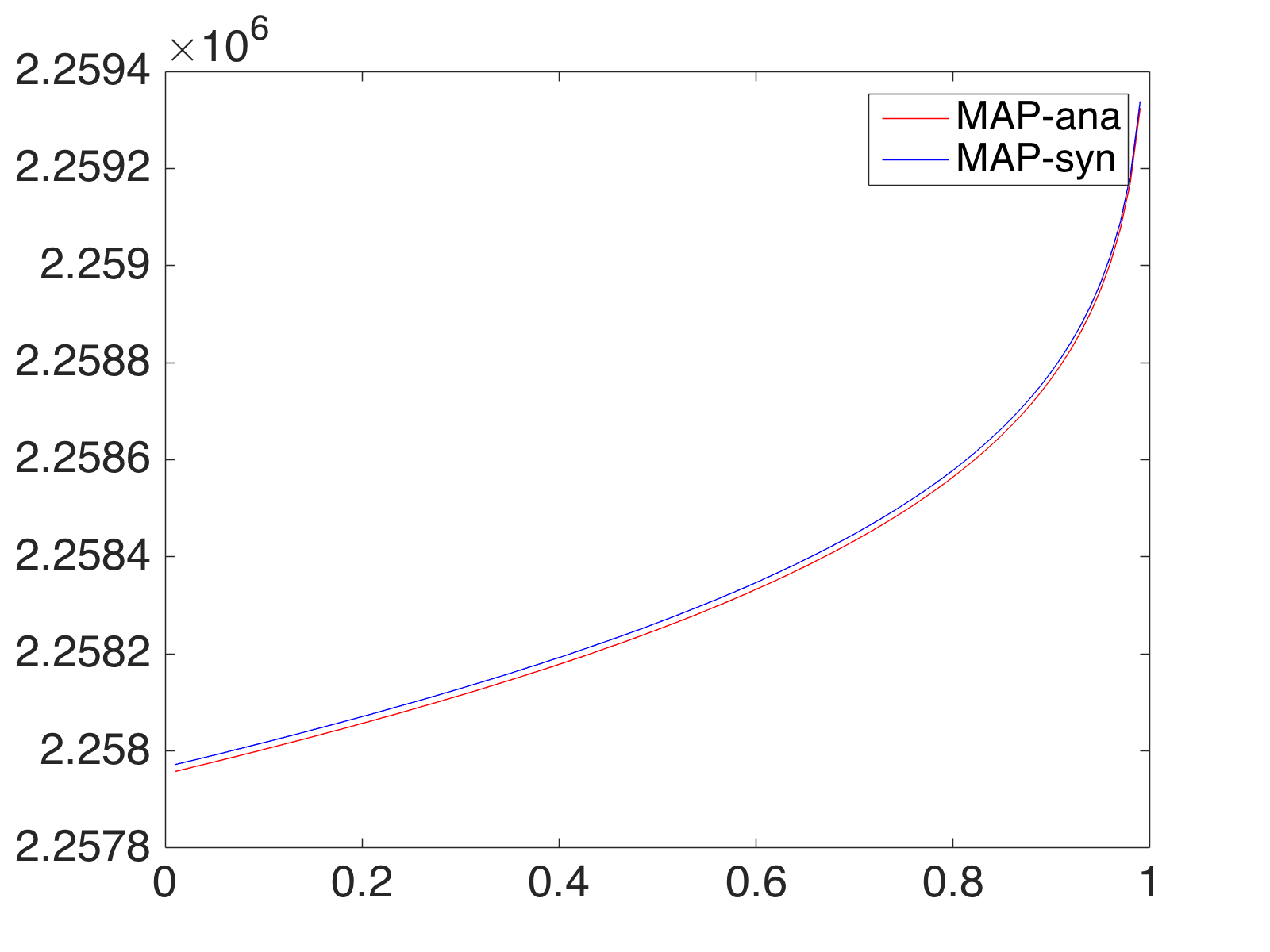}  \put(-60,-4){\tiny $1-\alpha$}&
		\includegraphics[trim={{.01\linewidth} {.05\linewidth} {.1\linewidth} {.01\linewidth}}, clip, width=0.22\linewidth, height = 0.2\linewidth]
		{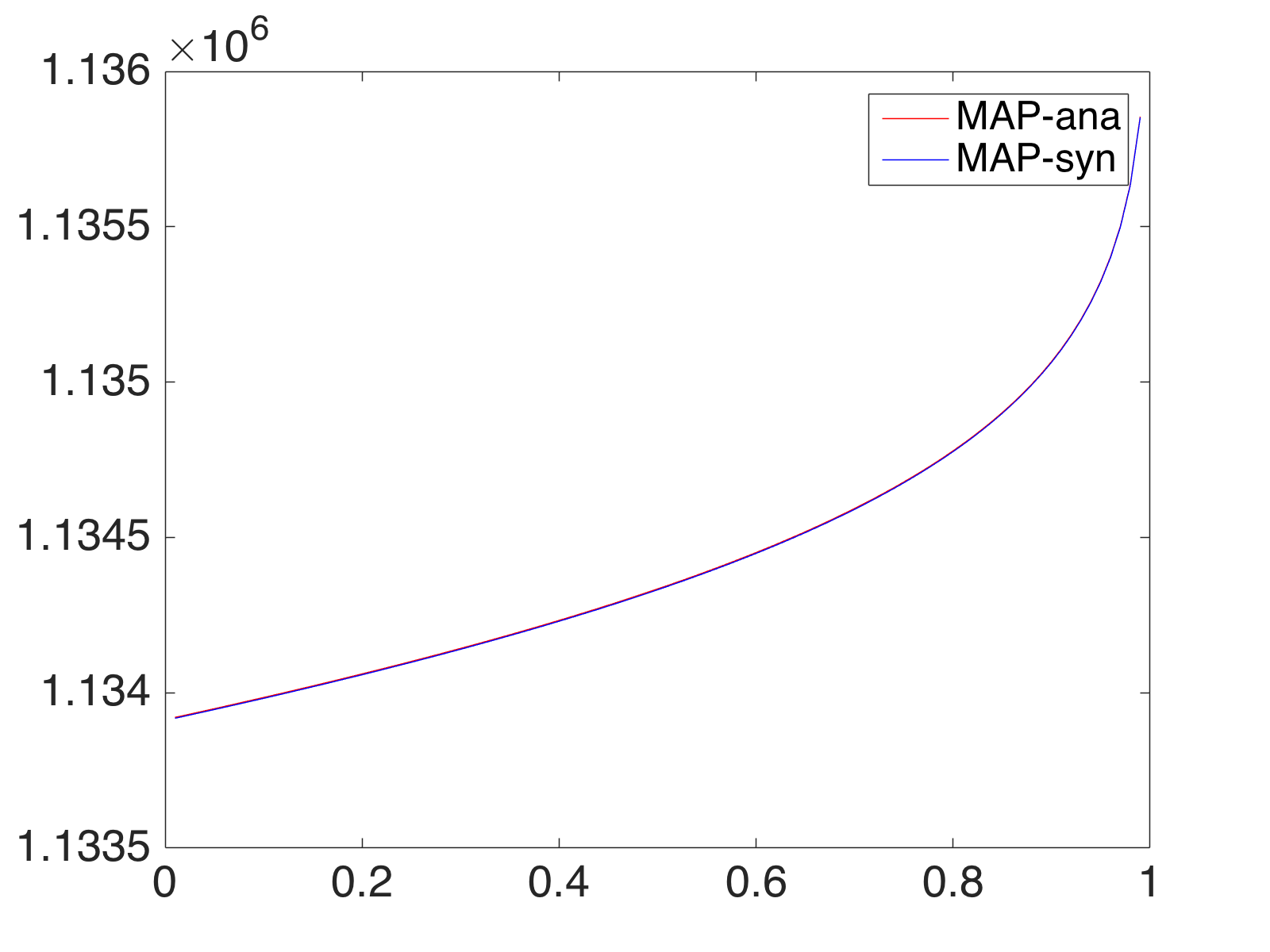} \put(-60,-4){\tiny $1-\alpha$} &
		\includegraphics[trim={{.01\linewidth} {.05\linewidth} {.1\linewidth} {.01\linewidth}}, clip, width=0.22\linewidth, height = 0.2\linewidth]
		{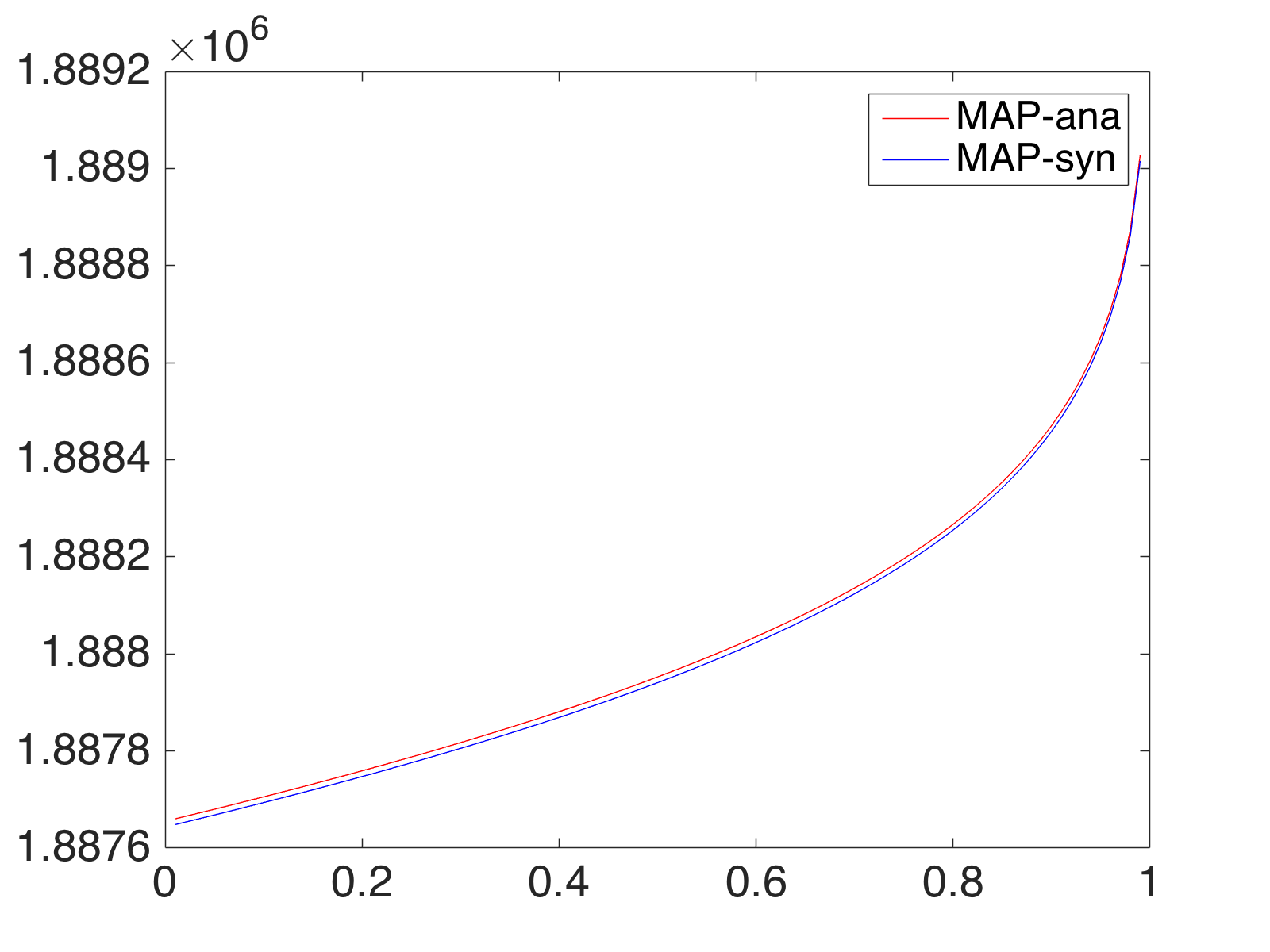} \put(-60,-4){\tiny $1-\alpha$} &
		\includegraphics[trim={{.01\linewidth} {.05\linewidth} {.1\linewidth} {.01\linewidth}}, clip, width=0.22\linewidth, height = 0.2\linewidth]
		{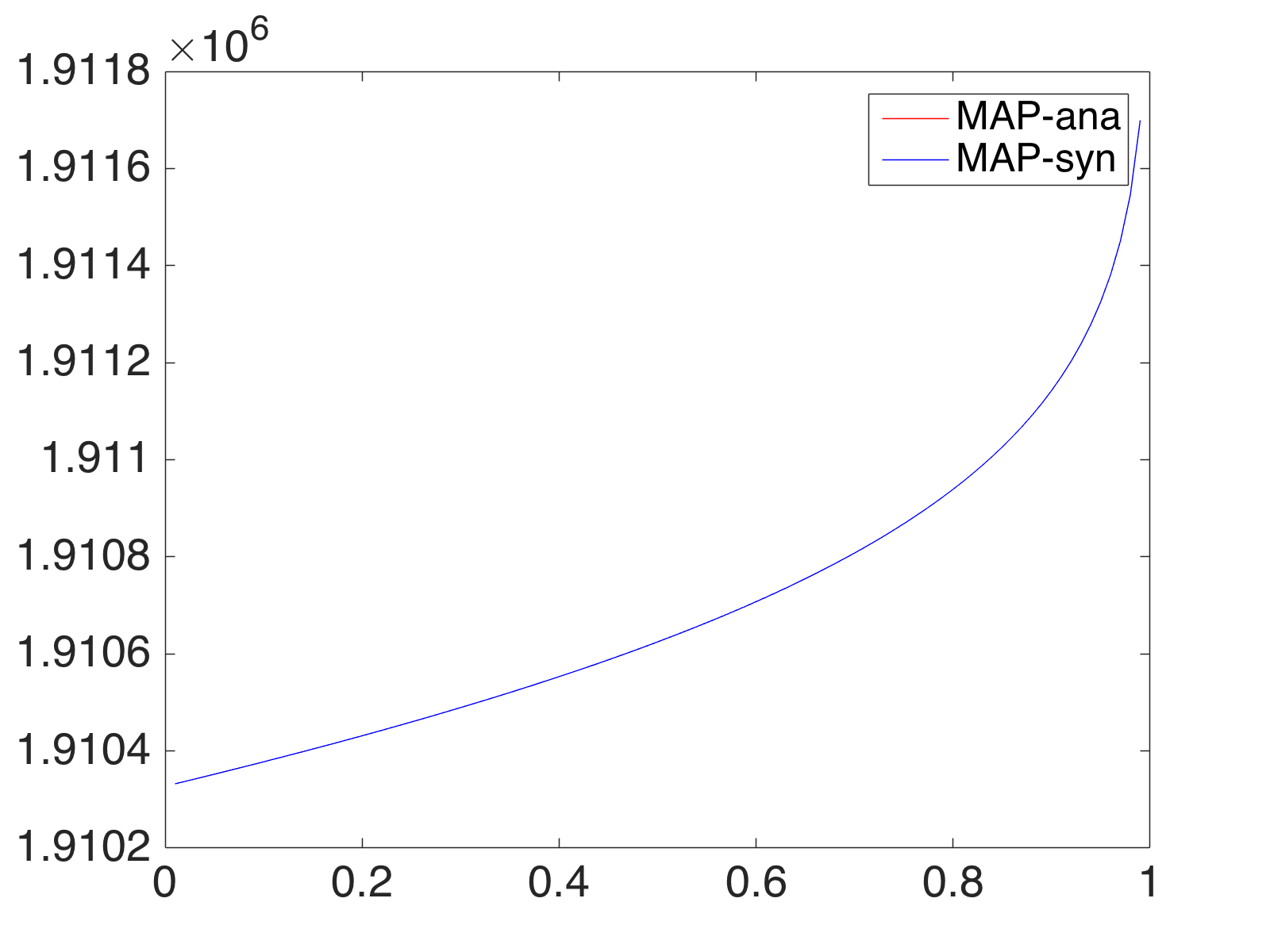} \put(-60,-4){\tiny $1-\alpha$} 
		\\
		 {\small  (a) M31 } & {\small (b) Cygnus A }  & {\small (c) W28}  & {\small (d) 3C288} 
        \end{tabular}
	\caption{HPD credible region isocontour levels $\bar{\gamma}^\prime_{\alpha}$ and $\hat{\gamma}^\prime_{\alpha}$
	 computed using MAP-based methods, for test images 
	(a) M31, (b) Cygnus A, (c) W28, and (d) 3C288. In particular, MAP-ana ({\it resp.} MAP-syn) represents the results by MAP estimation for 
	the analysis ({\it resp.} synthesis) model. Note that the red line in plot (d) is overlaid by the blue line and thus may not be visible, 
	due to the high degree of similarity between the two results.  In all cases the results of the analysis and synthesis models are in close agreement.  
	}
	\label{fig-hpd-cr}
\end{figure*}

\begin{table}
\begin{center}
\caption{CPU time in minutes for the proximal MCMC method Px-MALA (generating full posterior samples) and MAP-based methods (computing a point estimator), 
for the analysis and synthesis models and for test images of M31, Cygnus A, W28 and 3C288. 
MAP estimation is approximately $10^{5}$ times faster than Px-MALA and can be scaled to big-data.  
	 } \label{tab:time}
 \vspace{-0.05in}
\begin{tabular}{ccrr}
\toprule  
 \multirow{2}{*}{Images}  & \multirow{2}{*}{Methods} &  \multicolumn{2}{c}{CPU time (min)} \\ 
 & & Analysis & Synthesis 
\\ \toprule 
\multirow{2}{*}{M31 (Fig. \ref{fig-m31} ) } &  Px-MALA &  $1307$ &  $944$   
\\ 
& MAP & .03 &  .02
\\ \midrule
\multirow{2}{*}{Cygnus A  (Fig. \ref{fig-others} ) } &  Px-MALA & $2274$  & $1762$  
\\ 
& MAP & .07  & .04
\\ \midrule
\multirow{2}{*}{W28  (Fig. \ref{fig-others} ) } &    Px-MALA & $1122 $    & $879$  
\\
&  MAP & .06 & .04
\\ \midrule
\multirow{2}{*}{3C288 (Fig. \ref{fig-others} ) } &  Px-MALA & $1144 $ &  $881$  
 \\ 
 &  MAP & .03 & .02
\\ \bottomrule
\end{tabular}
\end{center}
\end{table}

\subsection{Approximate HPD credible regions}
We compute the HPD credible regions for the four images considered. Precisely, we use formulas \eqref{eqn:gamma-a} and \eqref{eqn:gamma-s} to approximate the threshold or isocontour value $\gamma^\prime_\alpha$ defining the HPD regions for the analysis and synthesis models (recall that these are highly efficient approximations derived from the MAP estimates ${\vect x}_{\rm map}$ and ${\vect a}_{\rm map}$). Figure \ref{fig-hpd-cr} shows the threshold values obtained for each image and model, for $\alpha \in [0.01, 0.99]$; observe again that the results of the analysis and synthesis models are consistent with each other, as expected. 

{To assess the approximation error involved in using the approximations \eqref{eqn:gamma-a} and \eqref{eqn:gamma-s}, we also computed the exact HPD threshold values by using the Px-MALA MCMC algorithm ({\it cf.} \citealt[Figure 6]{CPM17}).} Recall than Px-MALA is several orders of magnitude more computationally expensive than MAP estimation (see Table \ref{tab:time}). This comparison revealed approximation errors of between $1\%$ and $5\%$ over all cases, which is in close agreement with the results reported in \cite{M16}.  These experiments confirm that the MAP-based approximations \eqref{eqn:gamma-a} and \eqref{eqn:gamma-s} deliver accurate estimates of the 
HPD credible regions with a dramatically lower computational cost.

\addtolength{\tabcolsep}{-\tabL}
{ \renewcommand{\arraystretch}{0.0}
\begin{figure*}
	\centering
	\begin{tabular}{cccc} 
		\includegraphics[trim={{.15\linewidth} {.07\linewidth} {.02\linewidth} {.073\linewidth}}, clip, width=0.24\linewidth, height = 0.21\linewidth]
		{./figs/M31_PMALA_mean_sample_ana}  \put(-128,30){\rotatebox{90}{Px-MALA}} &
		\includegraphics[trim={{.15\linewidth} {.07\linewidth} {.02\linewidth} {.073\linewidth}}, clip, width=0.24\linewidth, height = 0.21\linewidth]
		{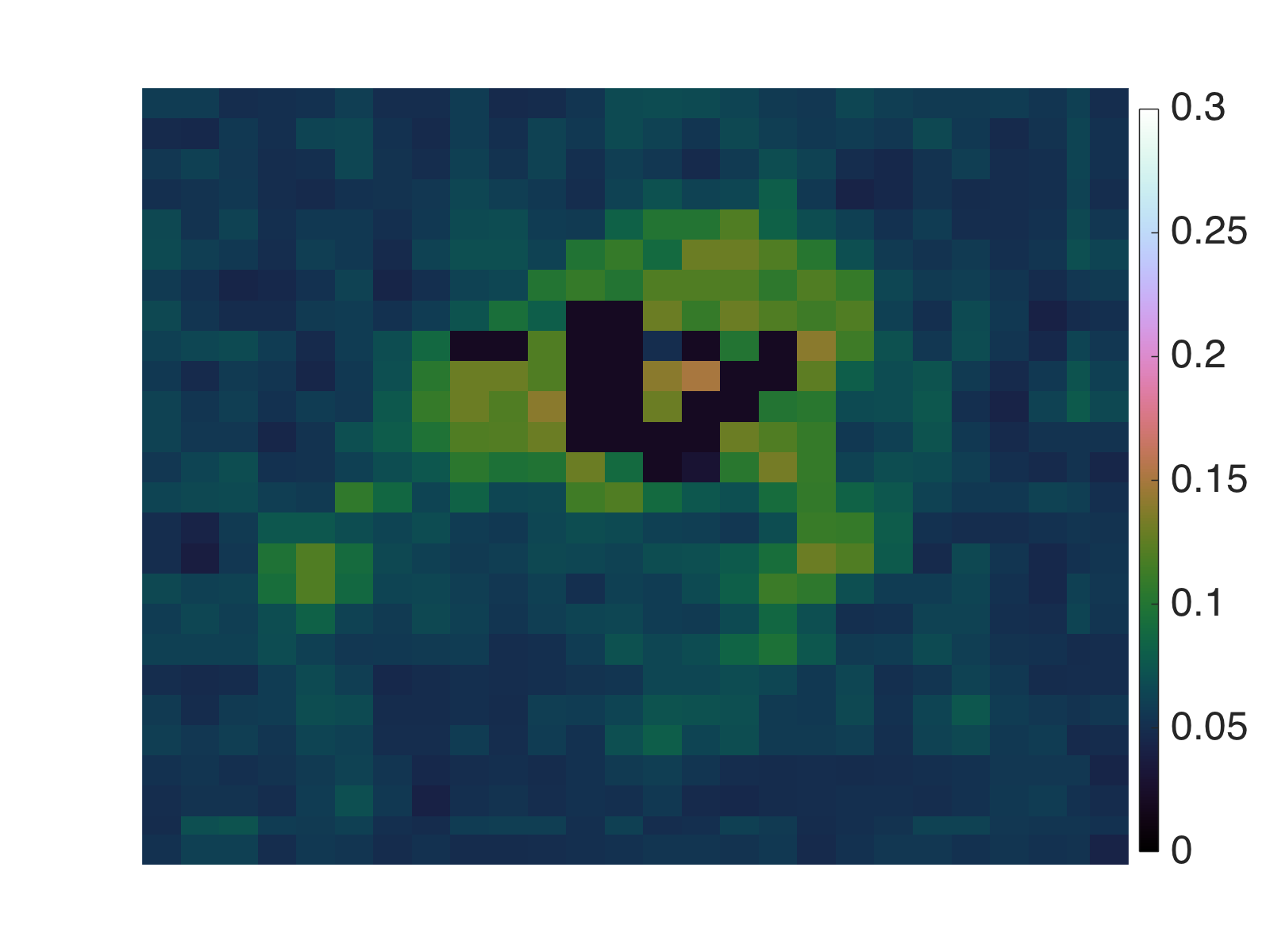}  &
		\includegraphics[trim={{.15\linewidth} {.07\linewidth} {.02\linewidth} {.073\linewidth}}, clip, width=0.24\linewidth, height = 0.21\linewidth]
		{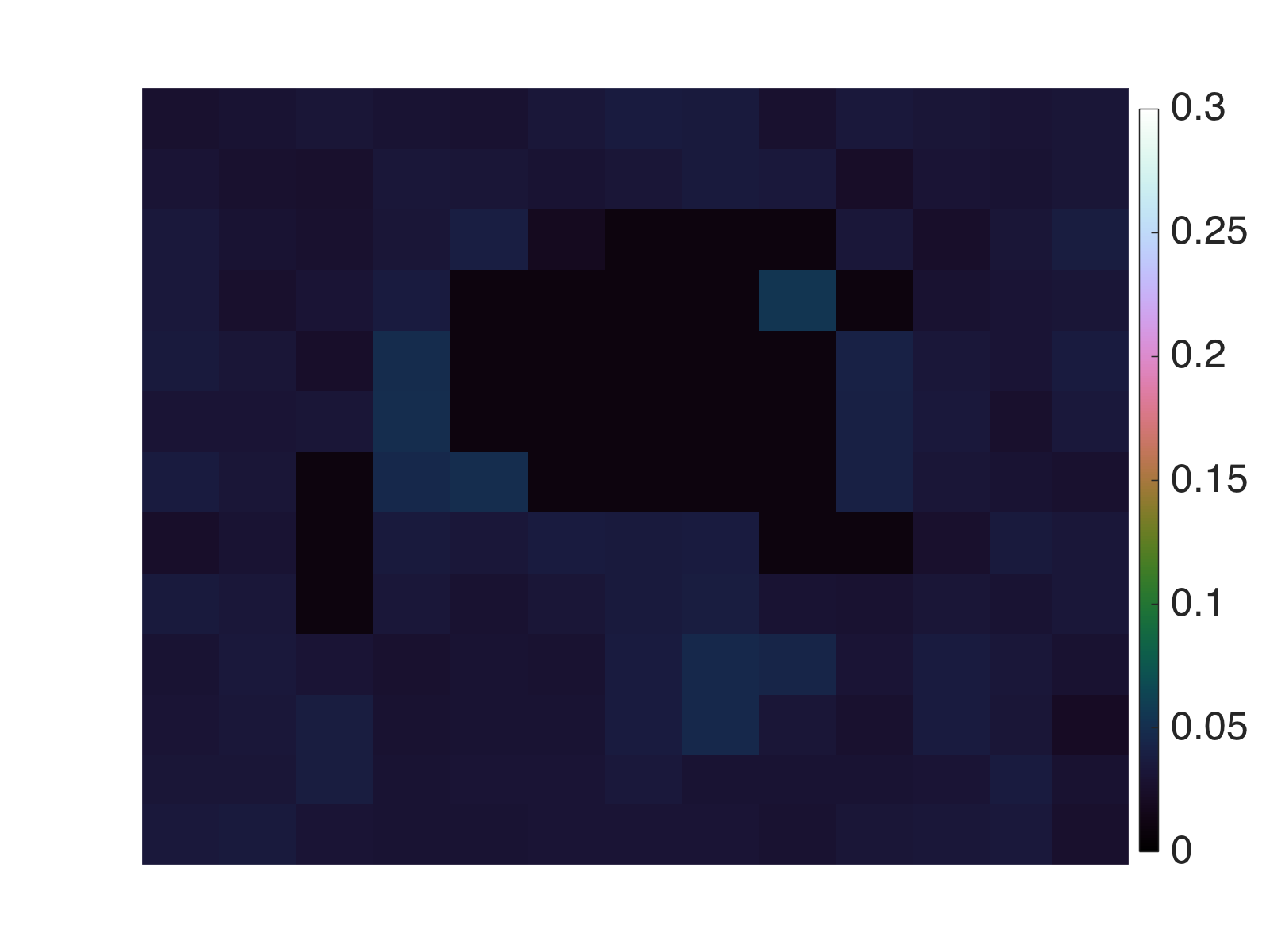} &
		\includegraphics[trim={{.15\linewidth} {.07\linewidth} {.02\linewidth} {.073\linewidth}}, clip, width=0.24\linewidth, height = 0.21\linewidth]
		{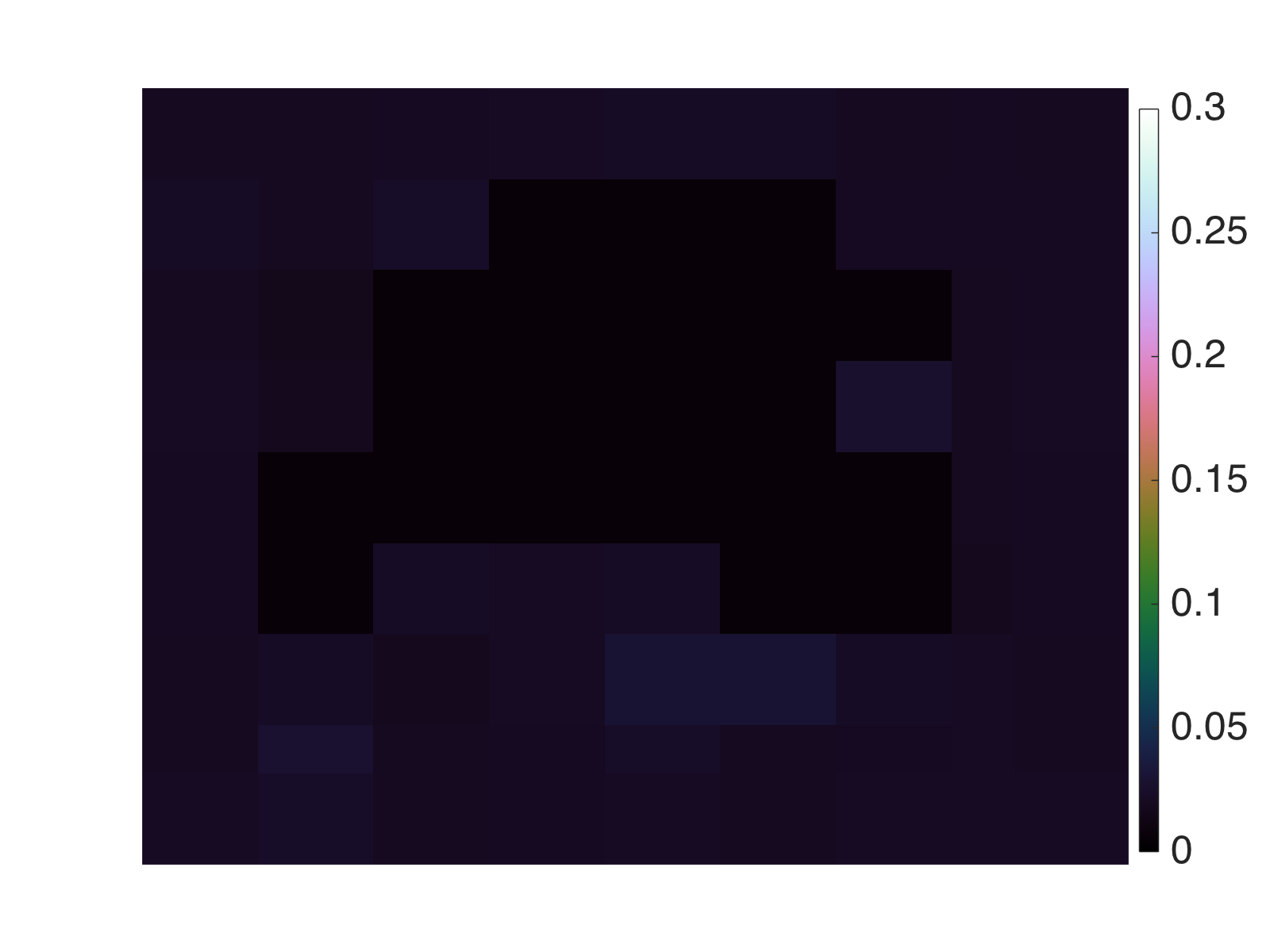}  
		\\	
		\includegraphics[trim={{.15\linewidth} {.07\linewidth} {.02\linewidth} {.073\linewidth}}, clip, width=0.24\linewidth, height = 0.21\linewidth]
		{./figs/M31_result_ana} \put(-128,40){\rotatebox{90}{MAP}}  &
		\includegraphics[trim={{.15\linewidth} {.07\linewidth} {.02\linewidth} {.073\linewidth}}, clip, width=0.24\linewidth, height = 0.21\linewidth]
		{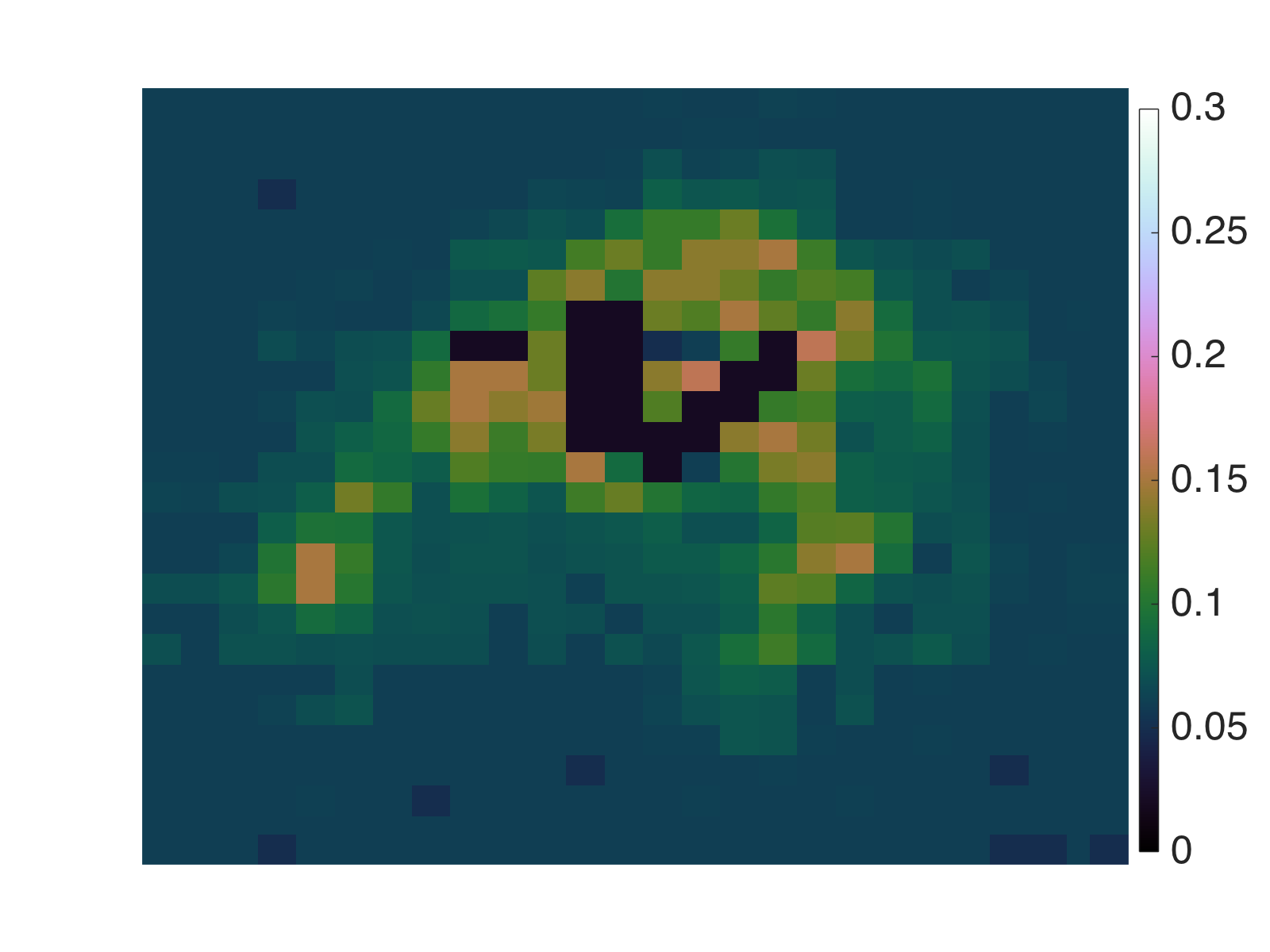} &
		\includegraphics[trim={{.15\linewidth} {.07\linewidth} {.02\linewidth} {.073\linewidth}}, clip, width=0.24\linewidth, height = 0.21\linewidth]
		{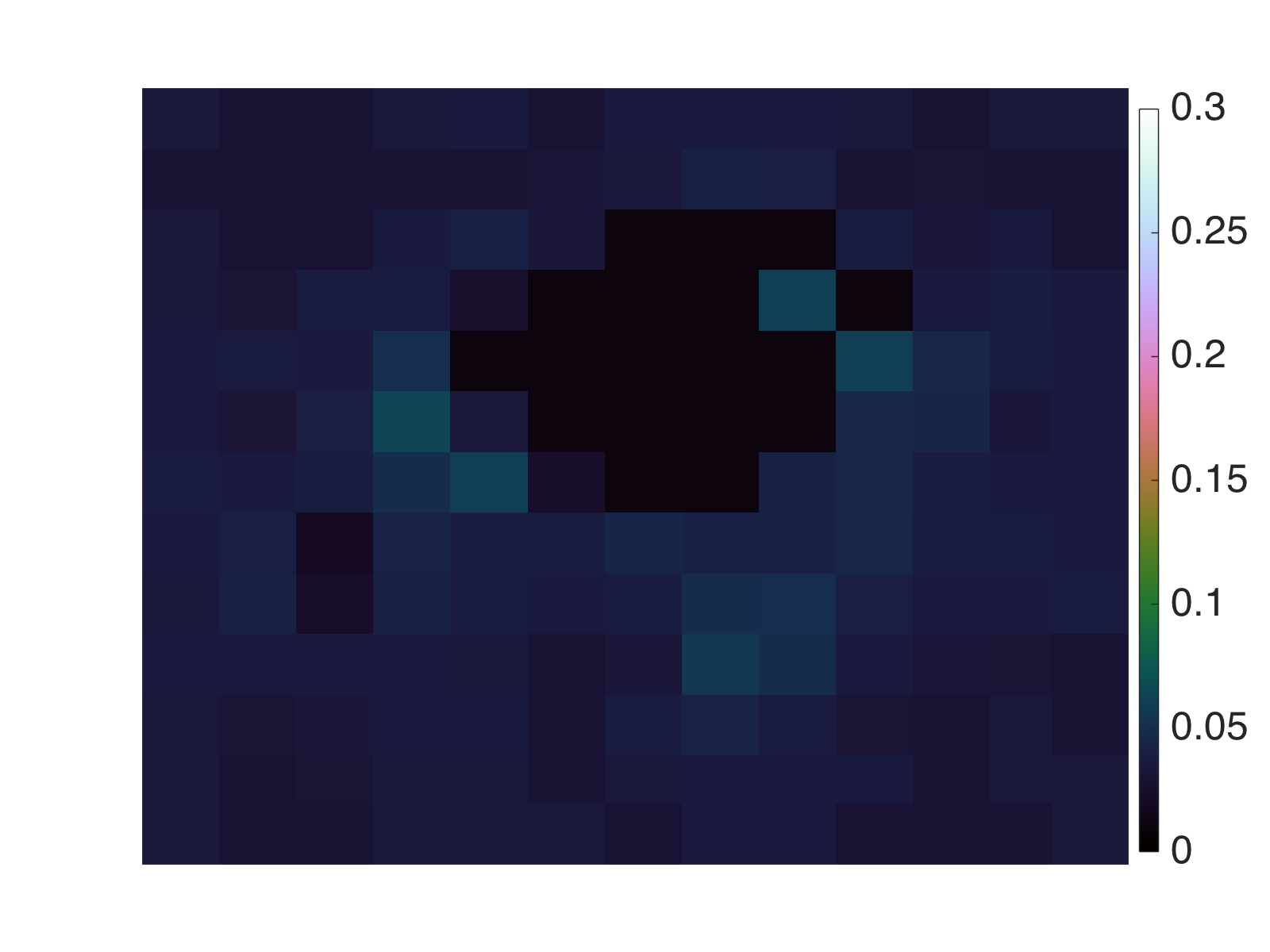} &
		\includegraphics[trim={{.15\linewidth} {.07\linewidth} {.02\linewidth} {.073\linewidth}}, clip, width=0.24\linewidth, height = 0.21\linewidth]
		{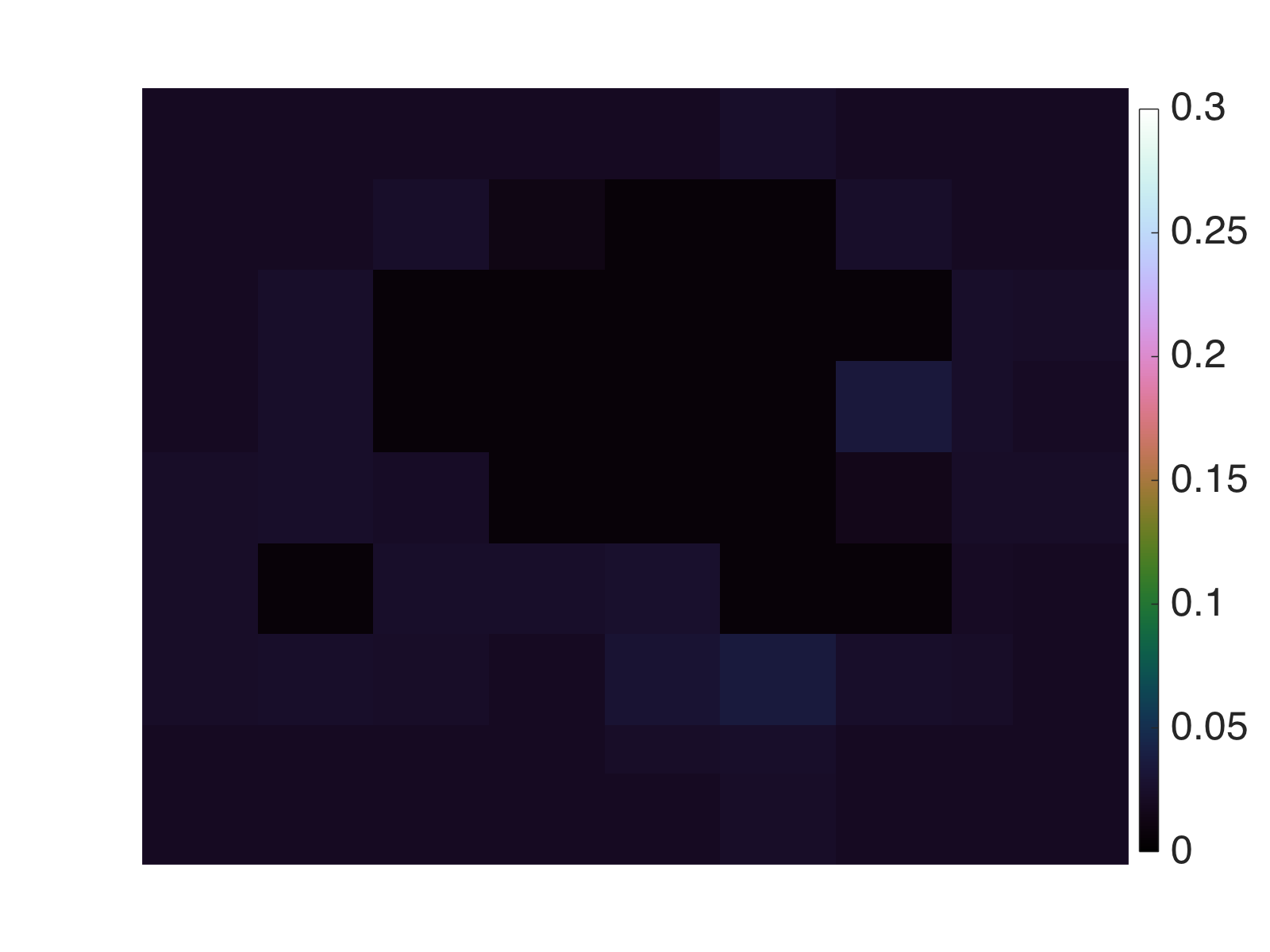} 
		\\ \vspace{0.05in}\\	
		\multirow{2}{*}{\small (a) point estimators } & {\small (b) local credible interval length } & {\small (c) local credible interval length }  & {\small (d) local credible interval length }
		 \vspace{0.03in} \\
		& {\small  grid size $10\times 10$ pixels } & {\small grid size $20\times 20$ pixels  } & {\small grid size $30\times 30$ pixels } 
        \end{tabular}
	\caption{Length of local credible intervals (99\% credible level), \textit{cf.} error bars, computed for M31 for the analysis model \eqref{eqn:ir-un-af}. 
	First column: (a) point estimators. 
	Second to fourth columns: (b)--(d) local credible intervals at grid sizes of $10\times 10$, $20\times 20$, and $30\times 30$ pixels, respectively.
	First row gives exact inferences computed with the MCMC method Px-MALA \citep{CPM17}. Second row gives MAP-based approximate 
	inferences computed by convex optimisation. Clearly, MAP-based approximations provide estimates of the length of local credible intervals (\textit{cf.} error bars)
	that are extremely consistent with the ones obtained by Px-MALA, while the MAP estimates can be computed several orders of magnitude more rapidly (Table \ref{tab:time}).  
  Moreover, the length of the approximate credible intervals computed by the MAP-based approach are theoretically conservative and can be seen to slightly overestimate the lengths computed by MCMC sampling.
	}
	\label{fig-all-ci-grid-m31-mean}
\end{figure*}
}
\addtolength{\tabcolsep}{\tabL}

\addtolength{\tabcolsep}{-1mm}
{ \renewcommand{\arraystretch}{0.0}
\begin{figure*}
	\centering
	\begin{tabular}{cccc}
		\includegraphics[trim={{.15\linewidth} {.07\linewidth} {.02\linewidth} {.073\linewidth}}, clip, width=0.24\linewidth, height = 0.13\linewidth]
		{./figs/CYN_PMALA_mean_sample_ana}  \put(-128,12){\rotatebox{90}{Px-MALA}}  &
		\includegraphics[trim={{.15\linewidth} {.07\linewidth} {.02\linewidth} {.073\linewidth}}, clip, width=0.24\linewidth, height = 0.13\linewidth]
		{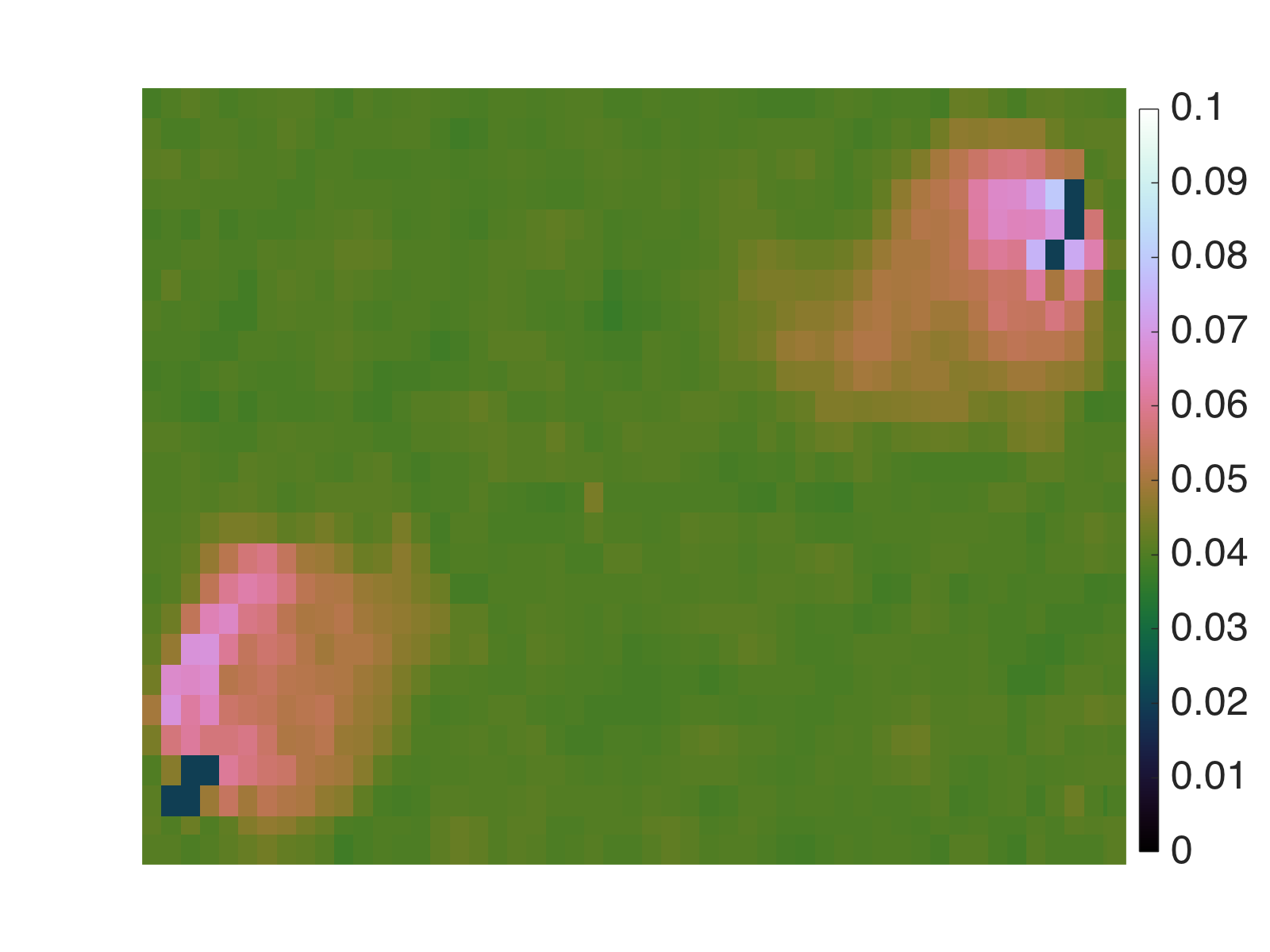} &
		\includegraphics[trim={{.15\linewidth} {.07\linewidth} {.02\linewidth} {.073\linewidth}}, clip, width=0.24\linewidth, height = 0.13\linewidth]
		{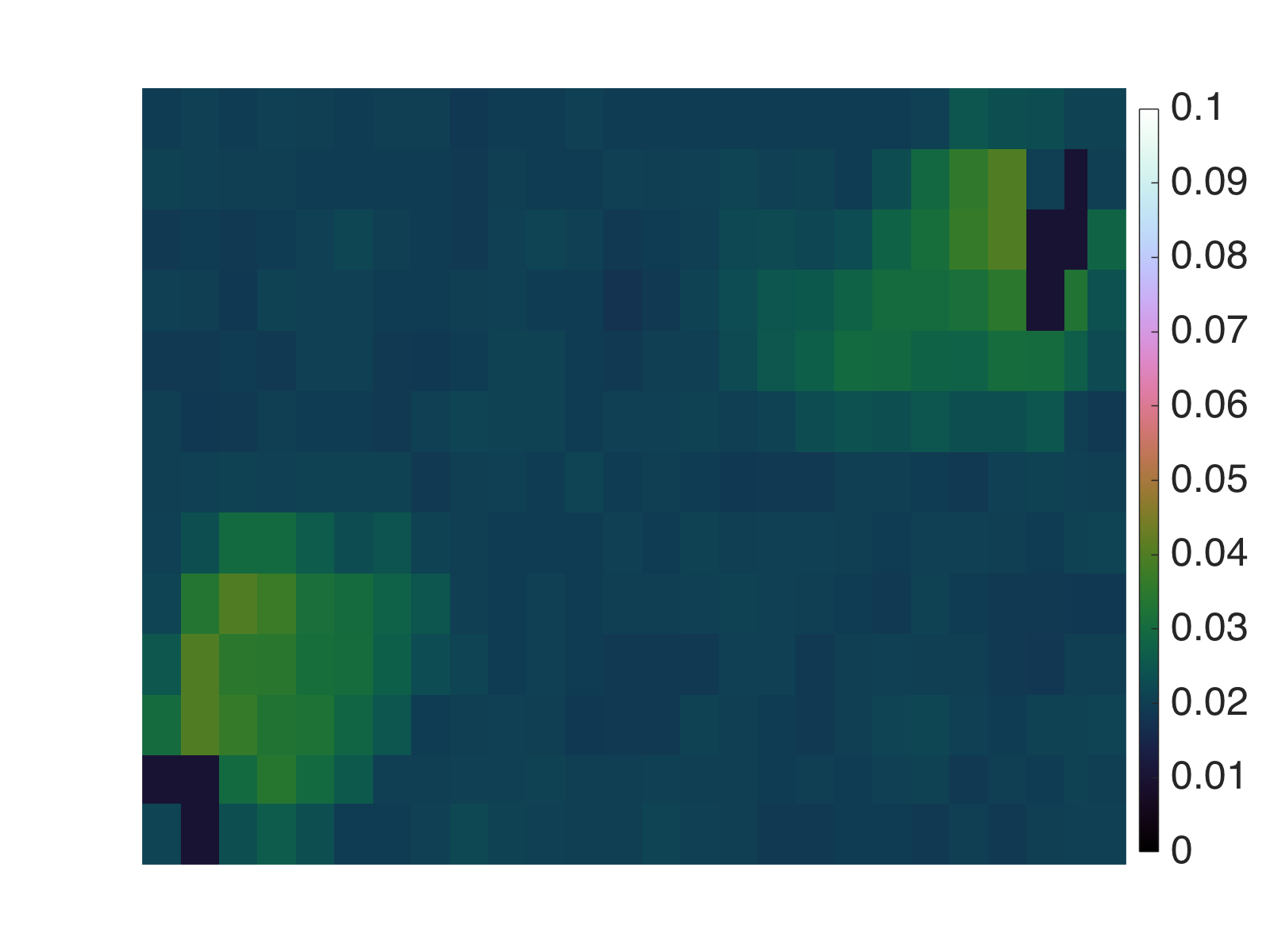} &
		\includegraphics[trim={{.15\linewidth} {.07\linewidth} {.02\linewidth} {.073\linewidth}}, clip, width=0.24\linewidth, height = 0.13\linewidth]
		{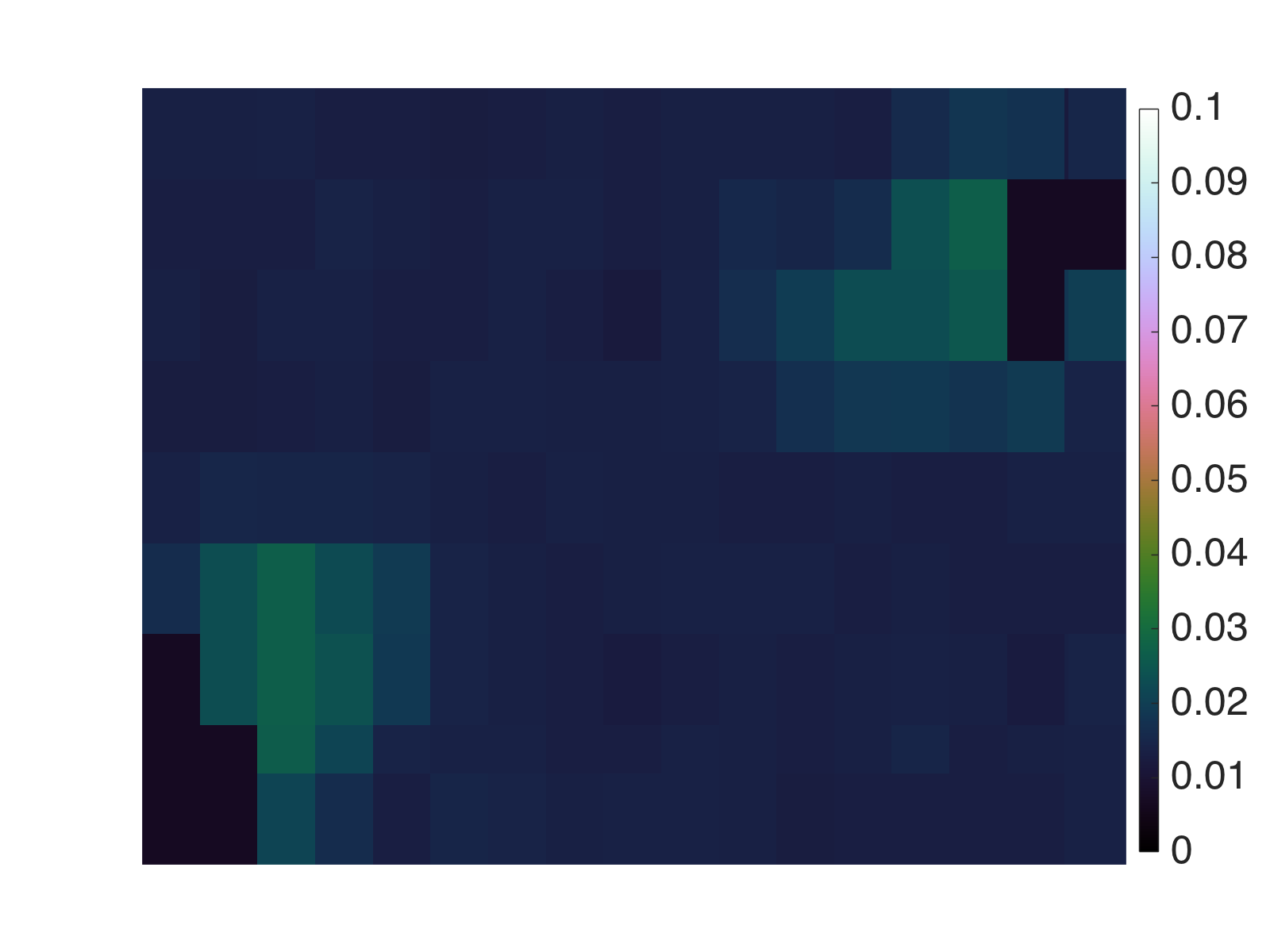}  
		\\	
		\includegraphics[trim={{.15\linewidth} {.07\linewidth} {.02\linewidth} {.073\linewidth}}, clip, width=0.24\linewidth, height = 0.13\linewidth]
		{./figs/CYN_result_ana} \put(-128,20){\rotatebox{90}{MAP}} &
		\includegraphics[trim={{.15\linewidth} {.07\linewidth} {.02\linewidth} {.073\linewidth}}, clip, width=0.24\linewidth, height = 0.13\linewidth]
		{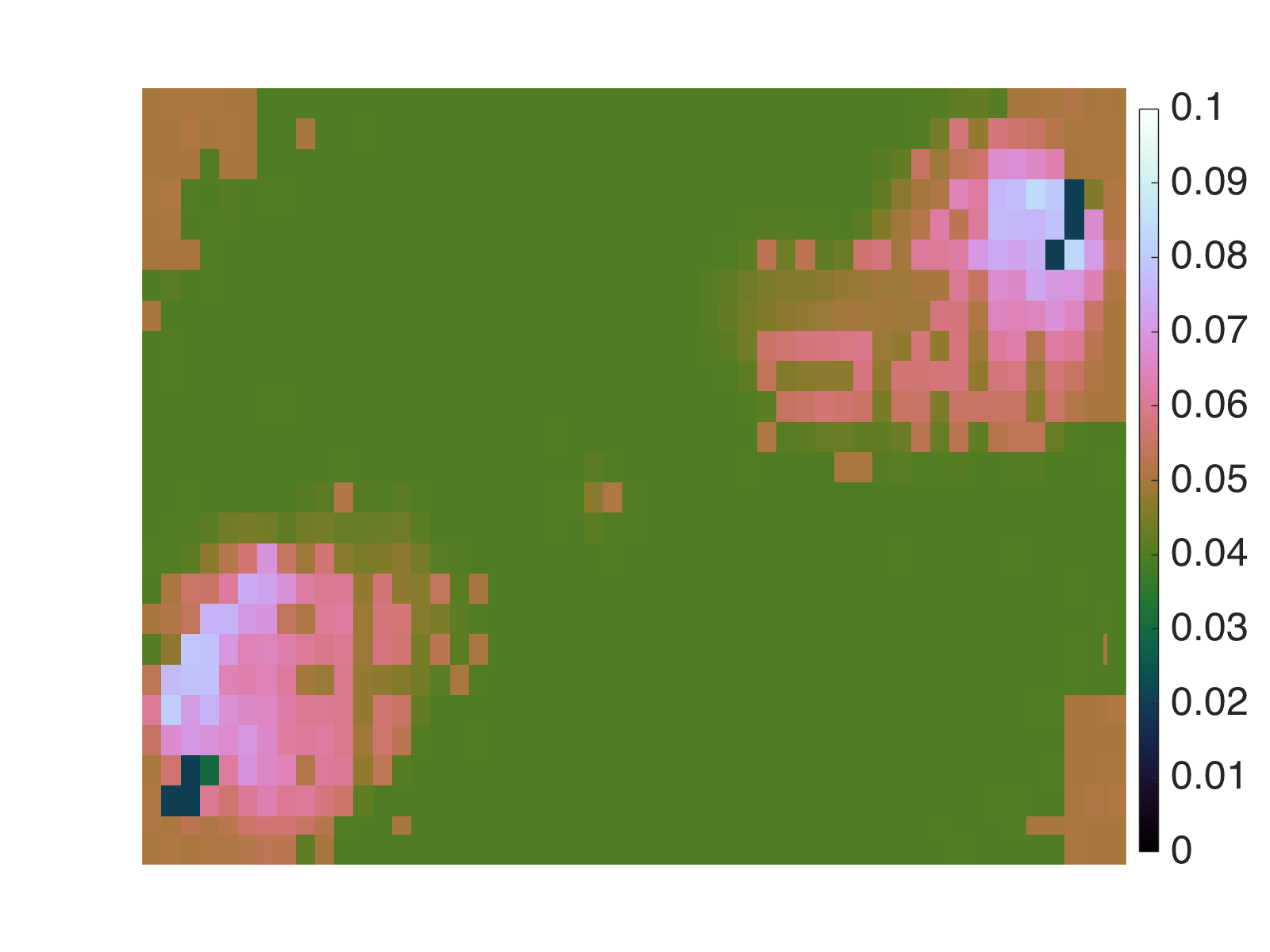} &
		\includegraphics[trim={{.15\linewidth} {.07\linewidth} {.02\linewidth} {.073\linewidth}}, clip, width=0.24\linewidth, height = 0.13\linewidth]
		{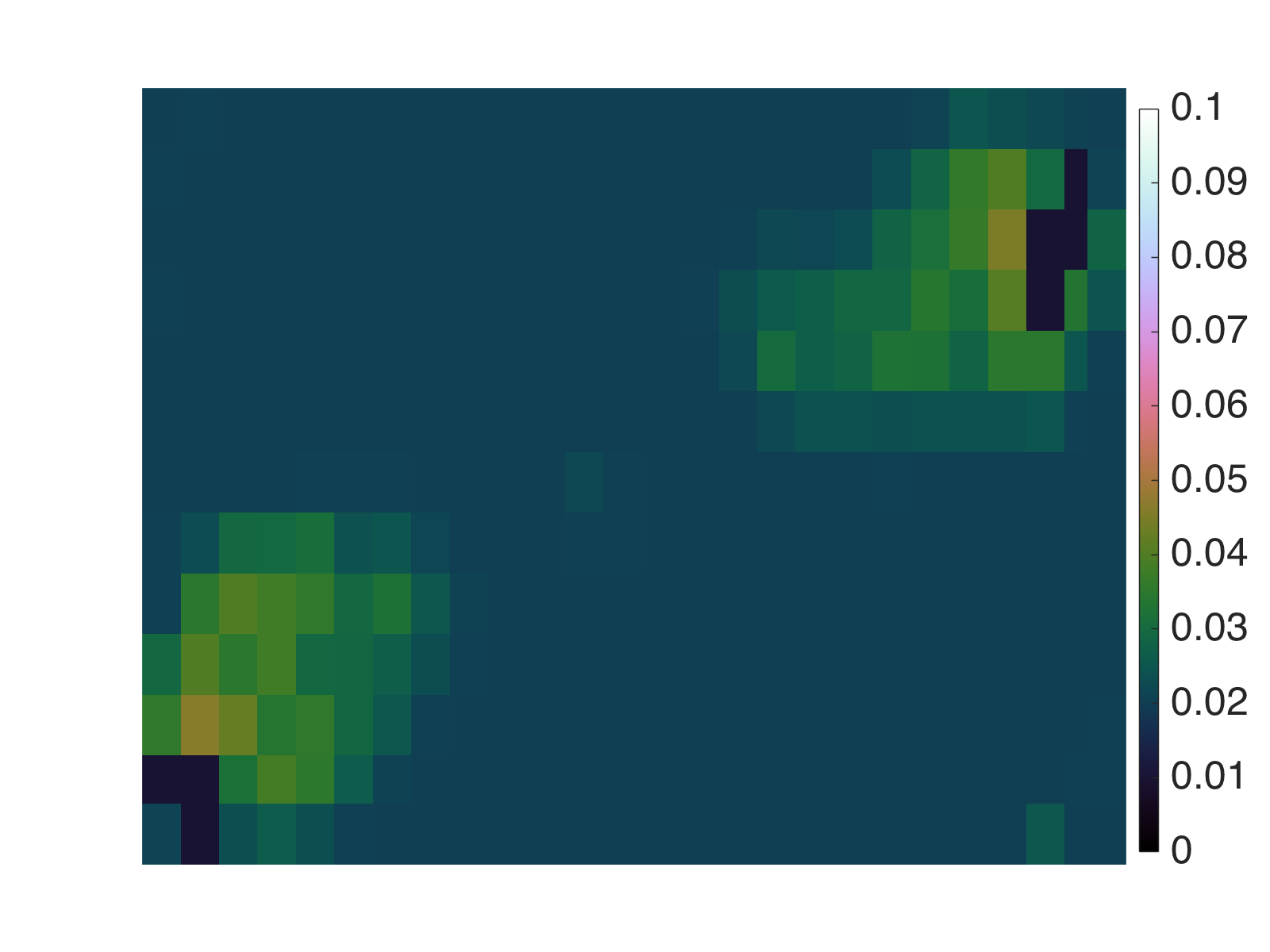} &
		\includegraphics[trim={{.15\linewidth} {.07\linewidth} {.02\linewidth} {.073\linewidth}}, clip, width=0.24\linewidth, height = 0.13\linewidth]
		{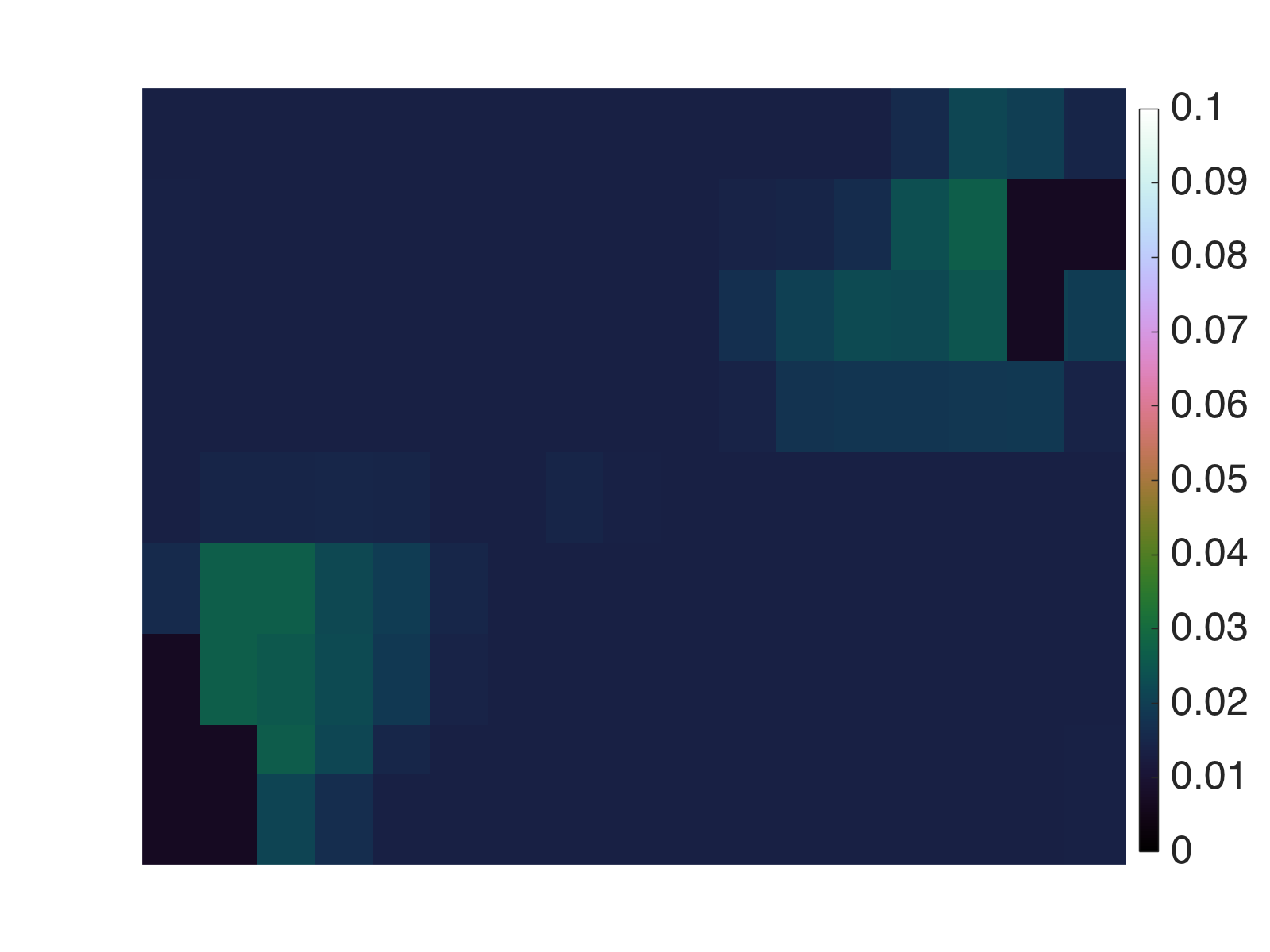} 
		\\ \vspace{0.05in}\\	
		\multirow{2}{*}{\small (a) point estimators } & {\small (b) local credible interval length } & {\small (c) local credible interval length }  & {\small (d) local credible interval length }
		 \vspace{0.03in} \\
		& {\small  grid size $10\times 10$ pixels } & {\small grid size $20\times 20$ pixels  } & {\small grid size $30\times 30$ pixels } 
        \end{tabular}
	\caption{Same as Figure \ref{fig-all-ci-grid-m31-mean} but for Cygnus A. 
	}
	\label{fig-all-ci-grid-cyn-mean}
\end{figure*}
}
\addtolength{\tabcolsep}{1mm}

\addtolength{\tabcolsep}{-\tabL}
{ \renewcommand{\arraystretch}{0.0}
\begin{figure*}
	\centering
	\begin{tabular}{cccc}
		\includegraphics[trim={{.15\linewidth} {.07\linewidth} {.02\linewidth} {.073\linewidth}}, clip, width=0.24\linewidth, height = 0.21\linewidth]
		{./figs/W28_PMALA_mean_sample_ana} \put(-128,30){\rotatebox{90}{Px-MALA}} &
		\includegraphics[trim={{.15\linewidth} {.07\linewidth} {.02\linewidth} {.073\linewidth}}, clip, width=0.24\linewidth, height = 0.21\linewidth]
		{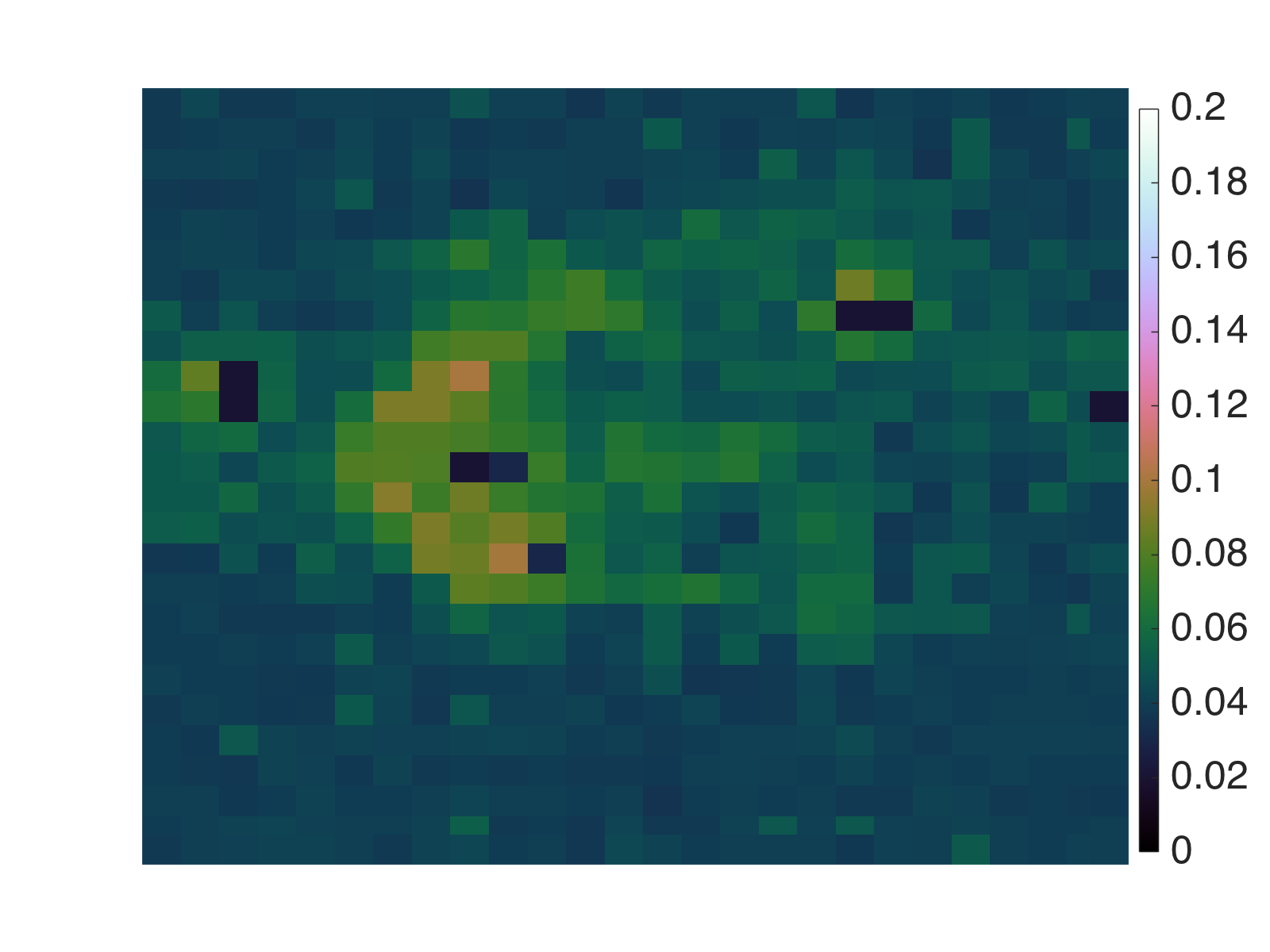} &
		\includegraphics[trim={{.15\linewidth} {.07\linewidth} {.02\linewidth} {.073\linewidth}}, clip, width=0.24\linewidth, height = 0.21\linewidth]
		{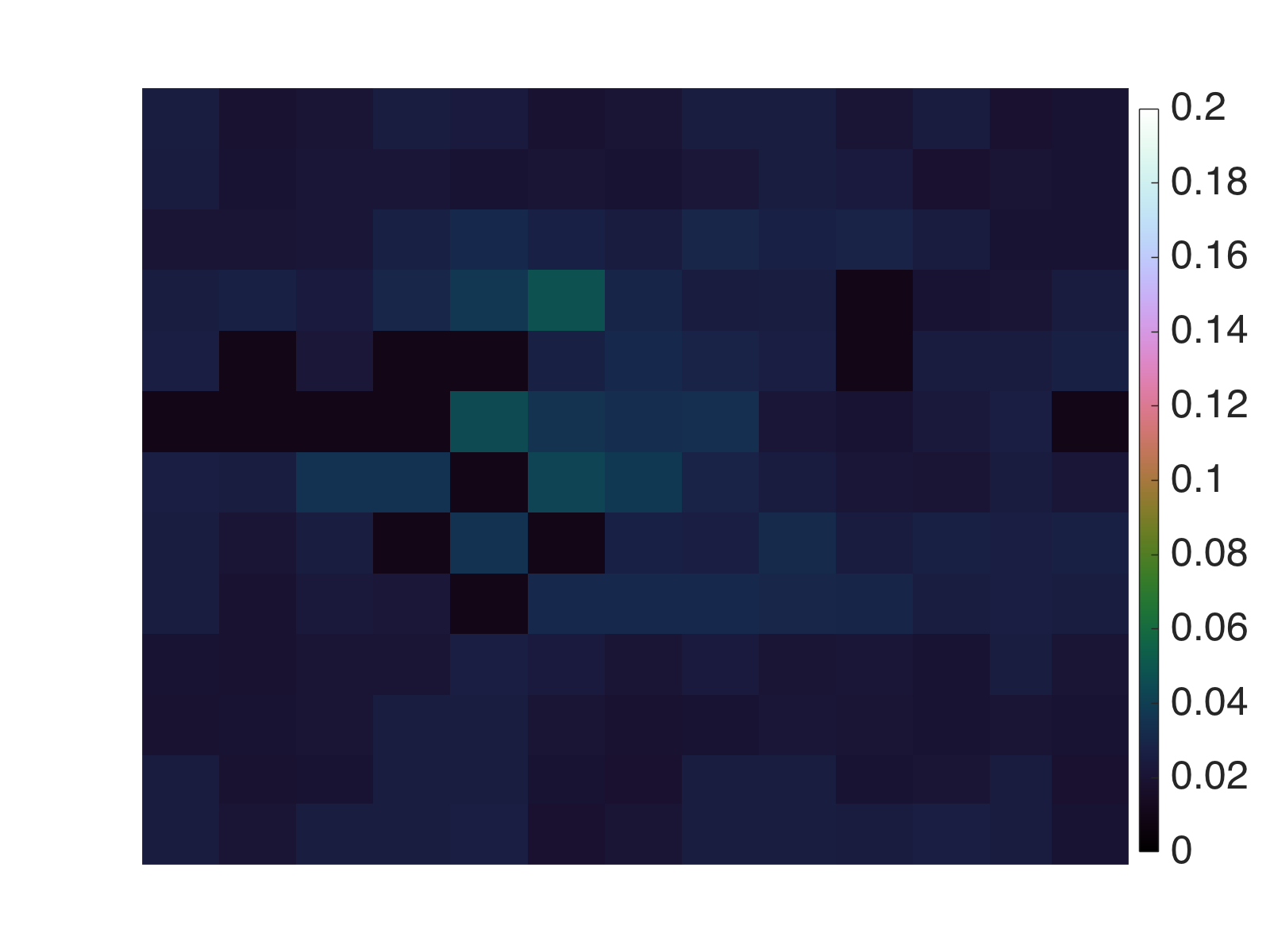} &
		\includegraphics[trim={{.15\linewidth} {.07\linewidth} {.02\linewidth} {.073\linewidth}}, clip, width=0.24\linewidth, height = 0.21\linewidth]
		{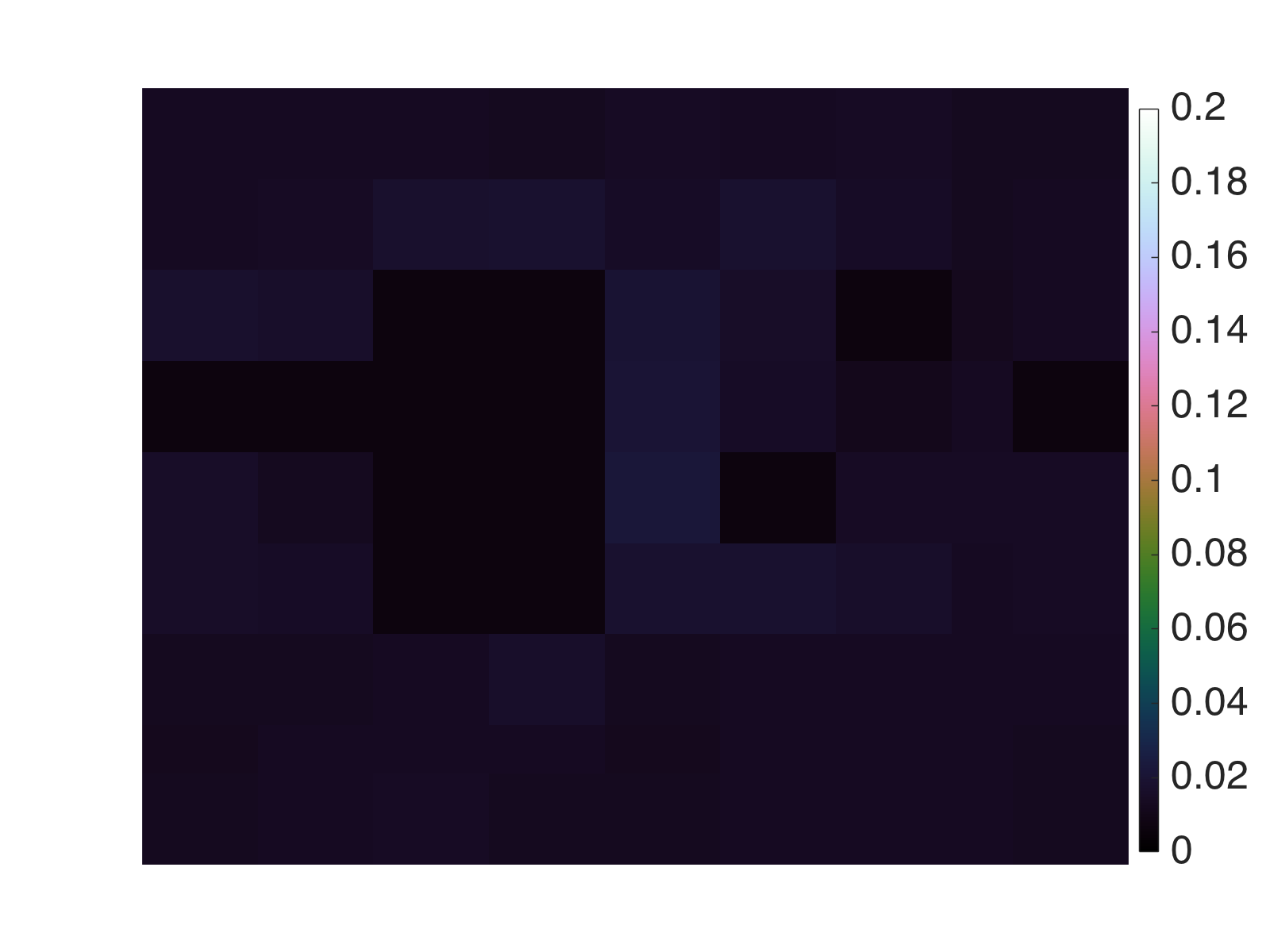} 
		\\
		\includegraphics[trim={{.15\linewidth} {.07\linewidth} {.02\linewidth} {.073\linewidth}}, clip, width=0.24\linewidth, height = 0.21\linewidth]
		{./figs/W28_result_ana} \put(-128,40){\rotatebox{90}{MAP}}  &
		\includegraphics[trim={{.15\linewidth} {.07\linewidth} {.02\linewidth} {.073\linewidth}}, clip, width=0.24\linewidth, height = 0.21\linewidth]
		{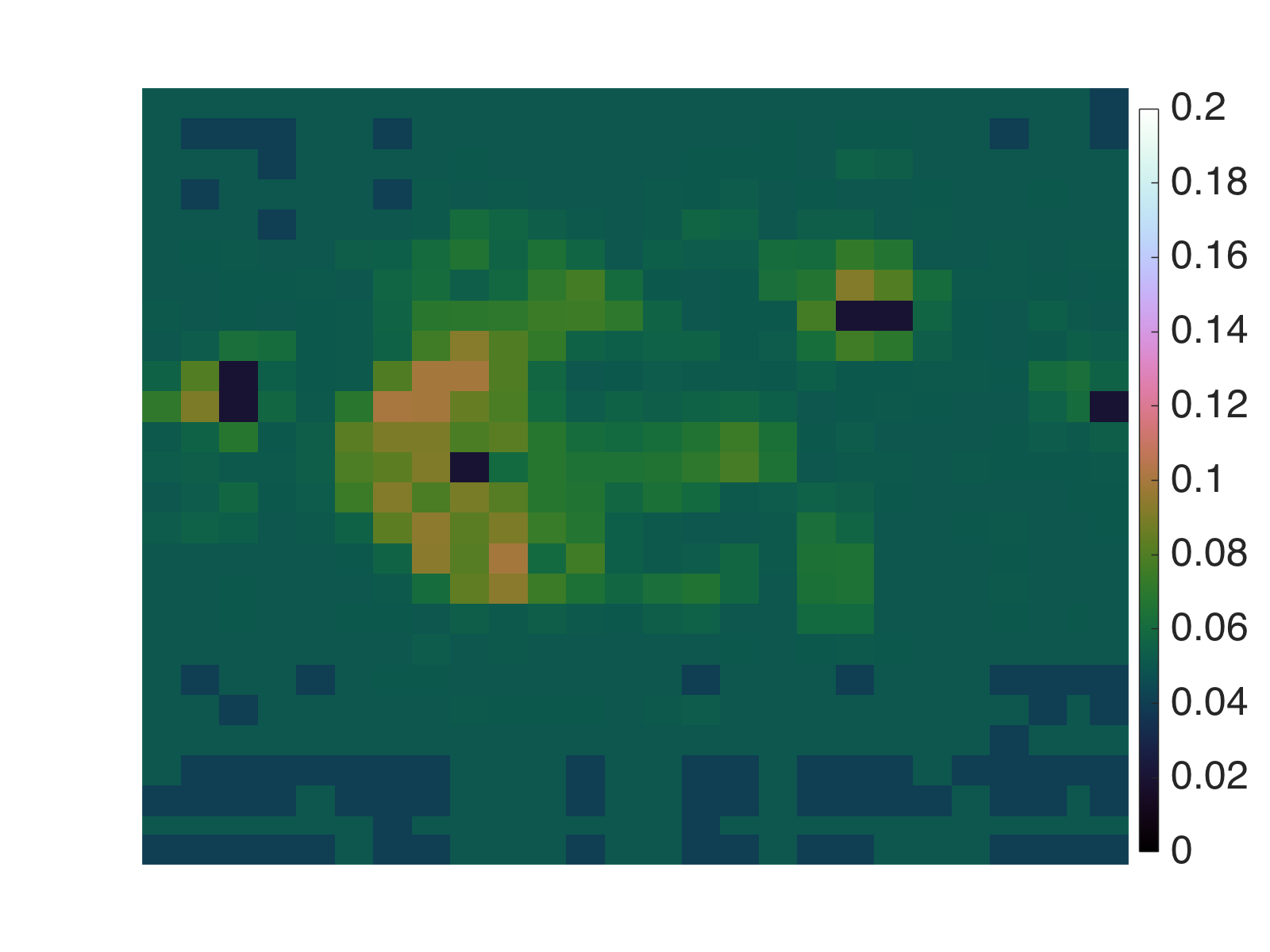} &
		\includegraphics[trim={{.15\linewidth} {.07\linewidth} {.02\linewidth} {.073\linewidth}}, clip, width=0.24\linewidth, height = 0.21\linewidth]
		{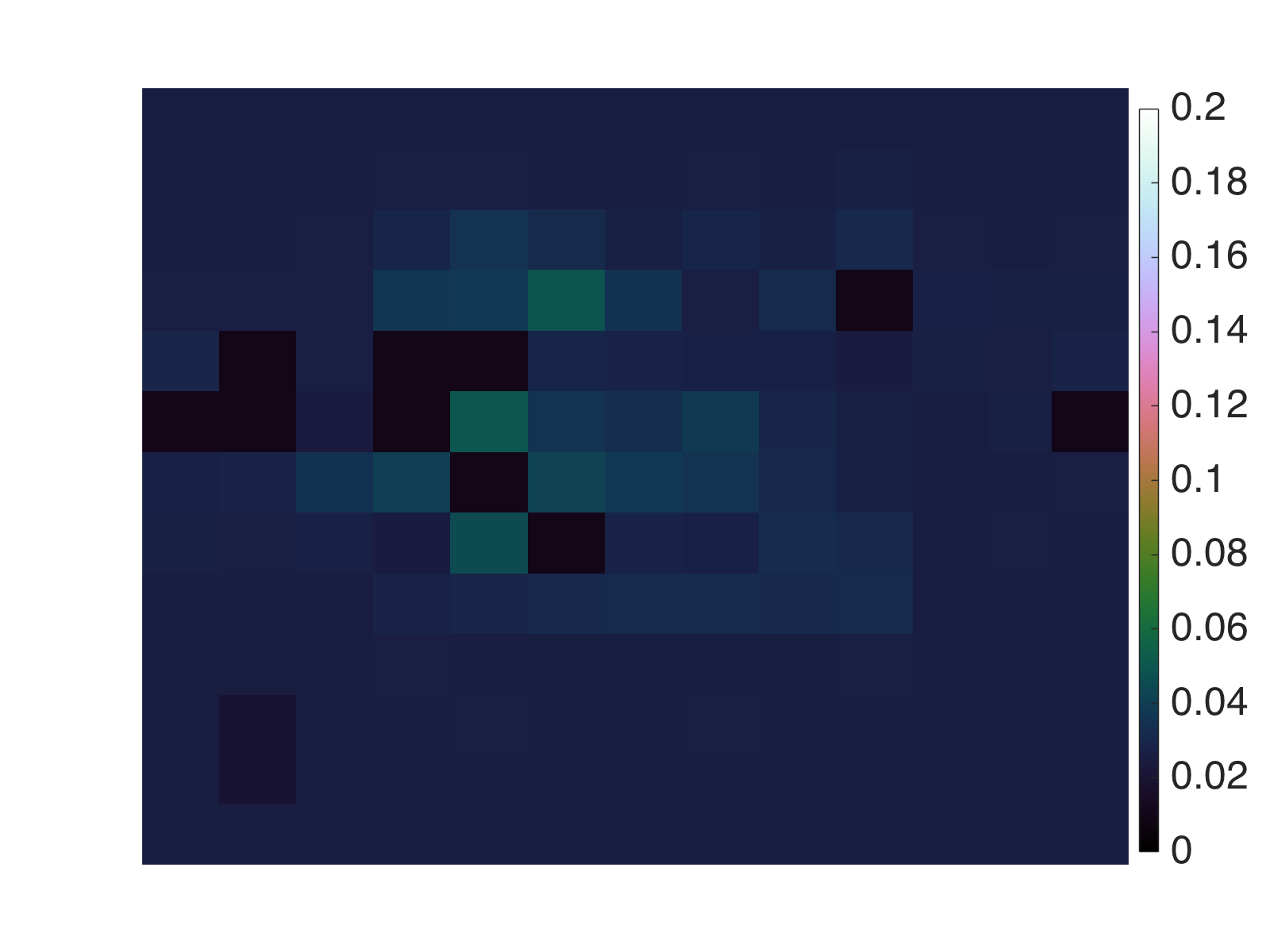} &
		\includegraphics[trim={{.15\linewidth} {.07\linewidth} {.02\linewidth} {.073\linewidth}}, clip, width=0.24\linewidth, height = 0.21\linewidth]
		{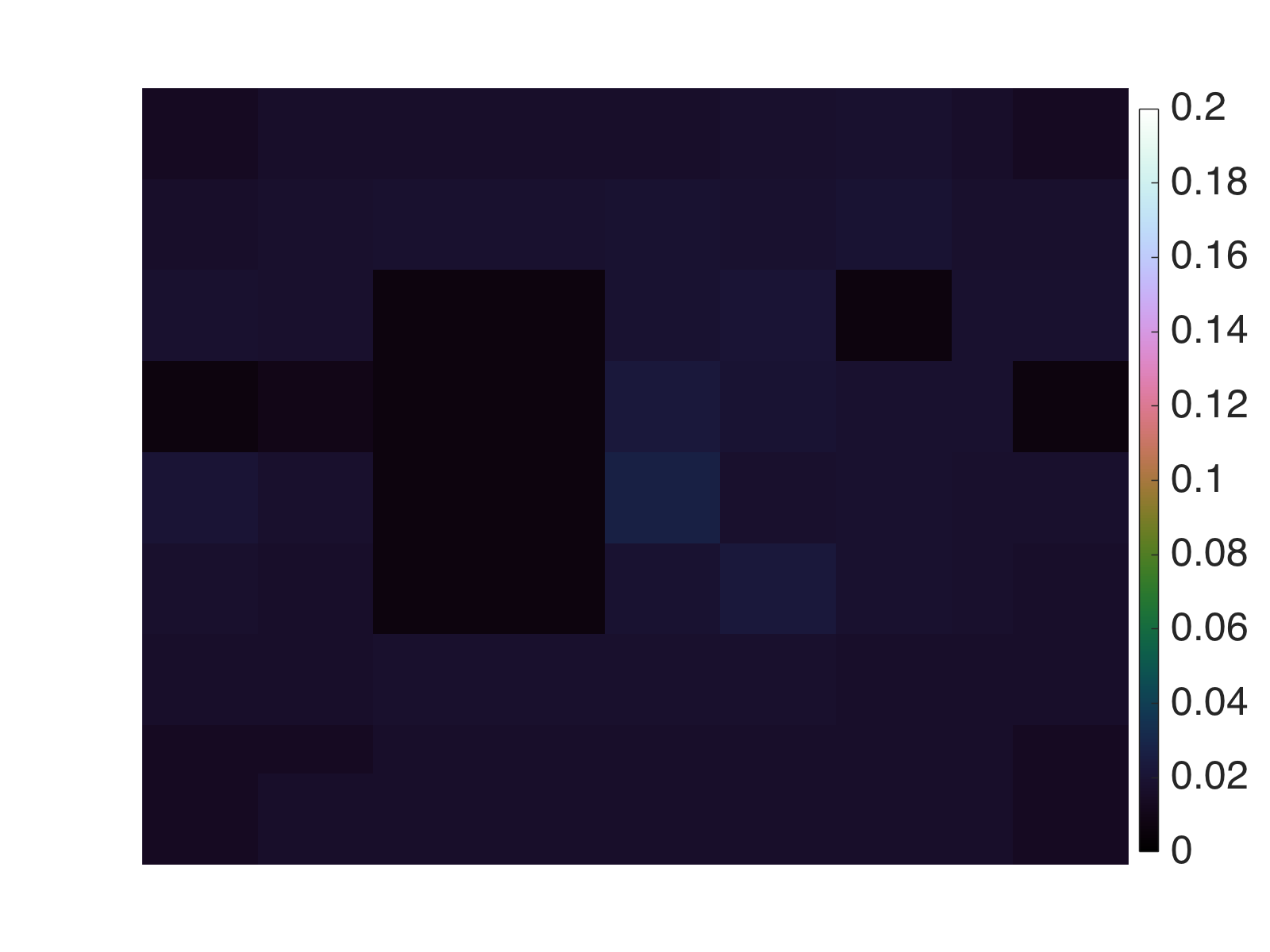} 
		\\ \vspace{0.05in}\\	
		\multirow{2}{*}{\small (a) point estimators } & {\small (b) local credible interval length} & {\small (c) local credible interval length}  & {\small (d) local credible interval length}
		 \vspace{0.03in} \\
		& {\small  grid size $10\times 10$ pixels } & {\small grid size $20\times 20$ pixels  } & {\small grid size $30\times 30$ pixels } 
        \end{tabular}
	\caption{Same as Figure \ref{fig-all-ci-grid-m31-mean} but for W28. 
	}
	\label{fig-all-ci-grid-w28-mean}
\end{figure*}
}
\addtolength{\tabcolsep}{\tabL}

\addtolength{\tabcolsep}{-\tabL}
{ \renewcommand{\arraystretch}{0.0}
\begin{figure*}
	\centering
	\begin{tabular}{cccc}
		\includegraphics[trim={{.15\linewidth} {.07\linewidth} {.02\linewidth} {.073\linewidth}}, clip, width=0.24\linewidth, height = 0.21\linewidth]
		{./figs/3C288_PMALA_mean_sample_ana}  \put(-128,30){\rotatebox{90}{Px-MALA}} &
		\includegraphics[trim={{.15\linewidth} {.07\linewidth} {.02\linewidth} {.073\linewidth}}, clip, width=0.24\linewidth, height = 0.21\linewidth]
		{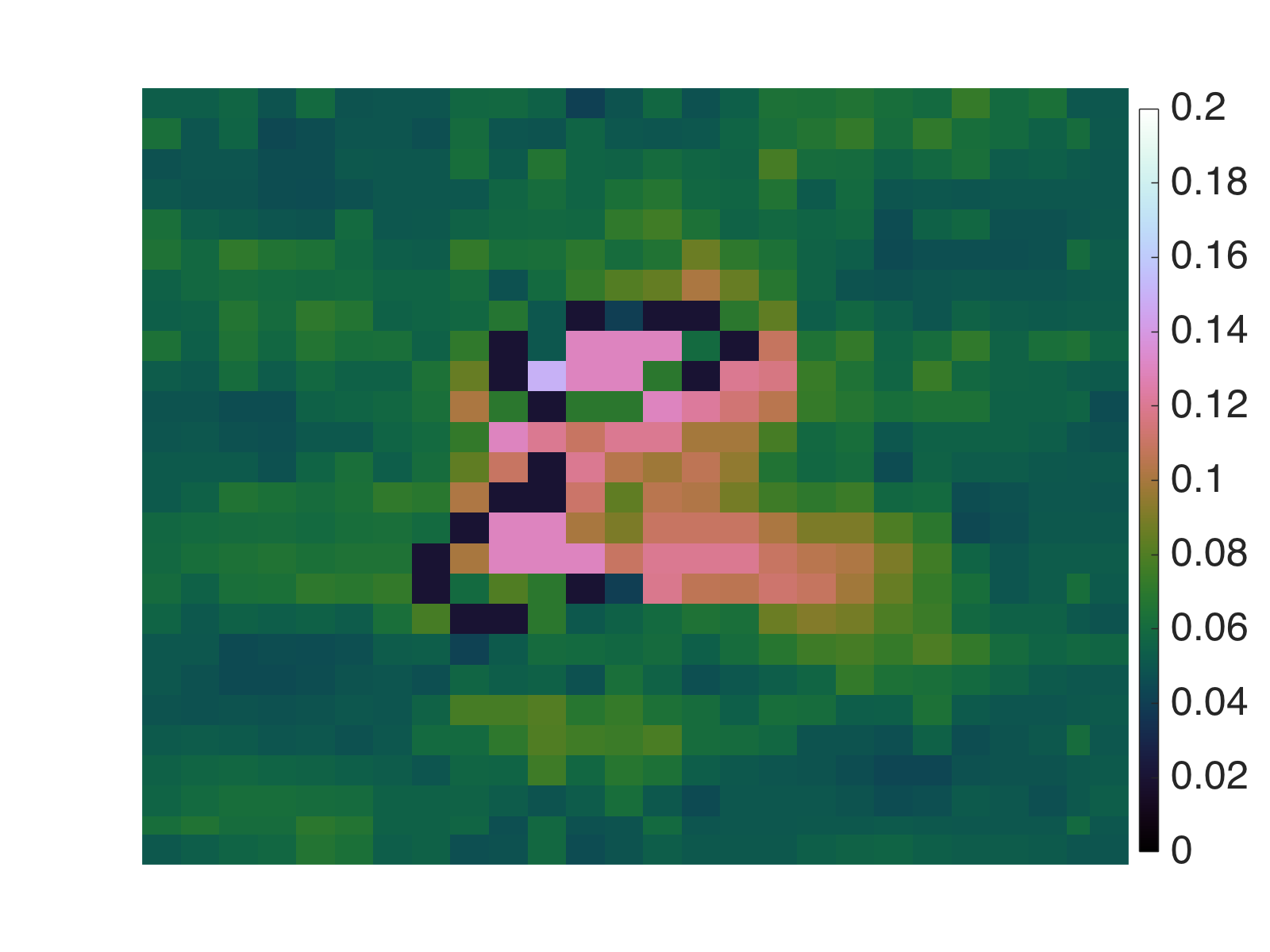} &
		\includegraphics[trim={{.15\linewidth} {.07\linewidth} {.02\linewidth} {.073\linewidth}}, clip, width=0.24\linewidth, height = 0.21\linewidth]
		{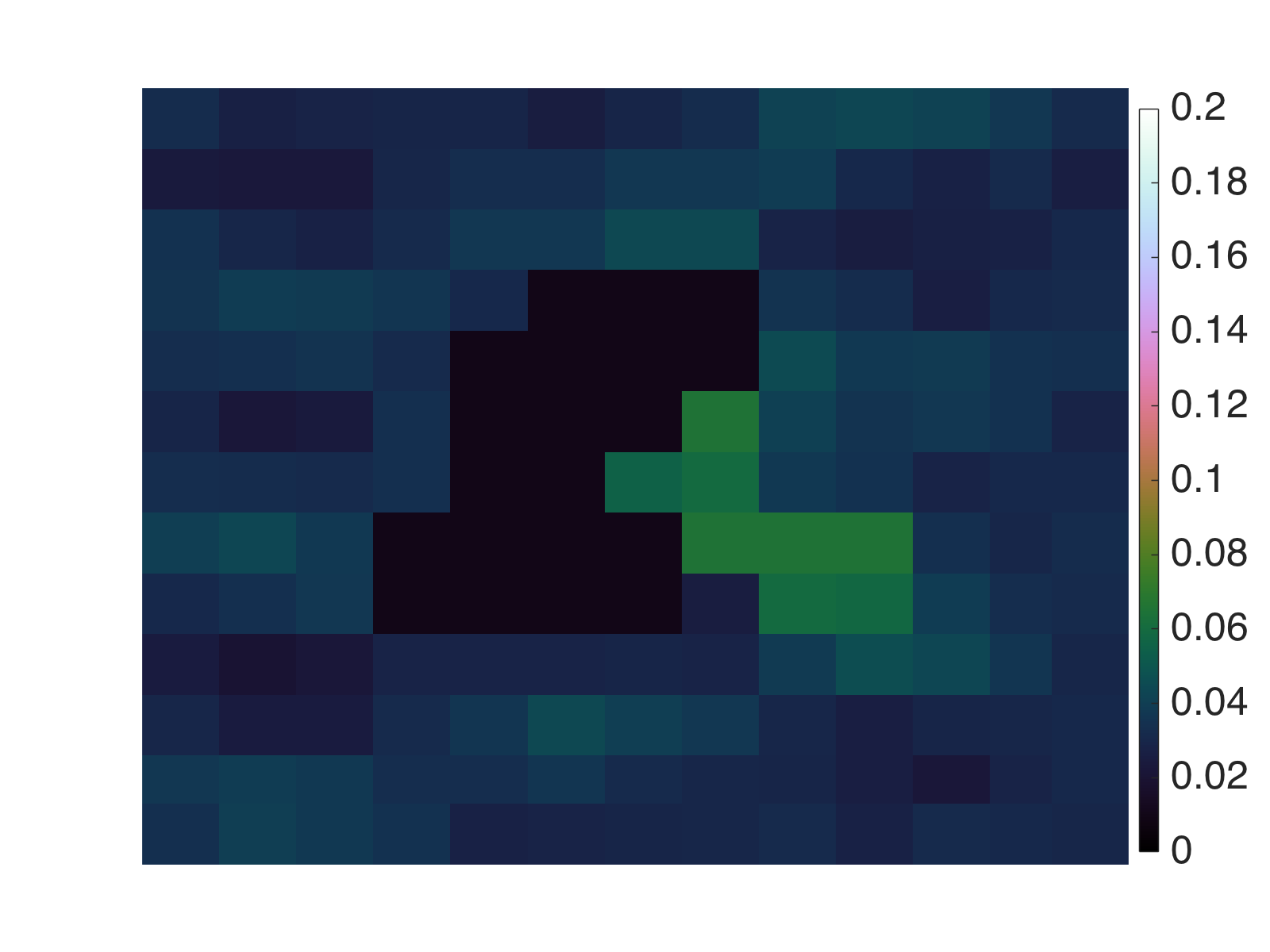} &
		\includegraphics[trim={{.15\linewidth} {.07\linewidth} {.02\linewidth} {.073\linewidth}}, clip, width=0.24\linewidth, height = 0.21\linewidth]
		{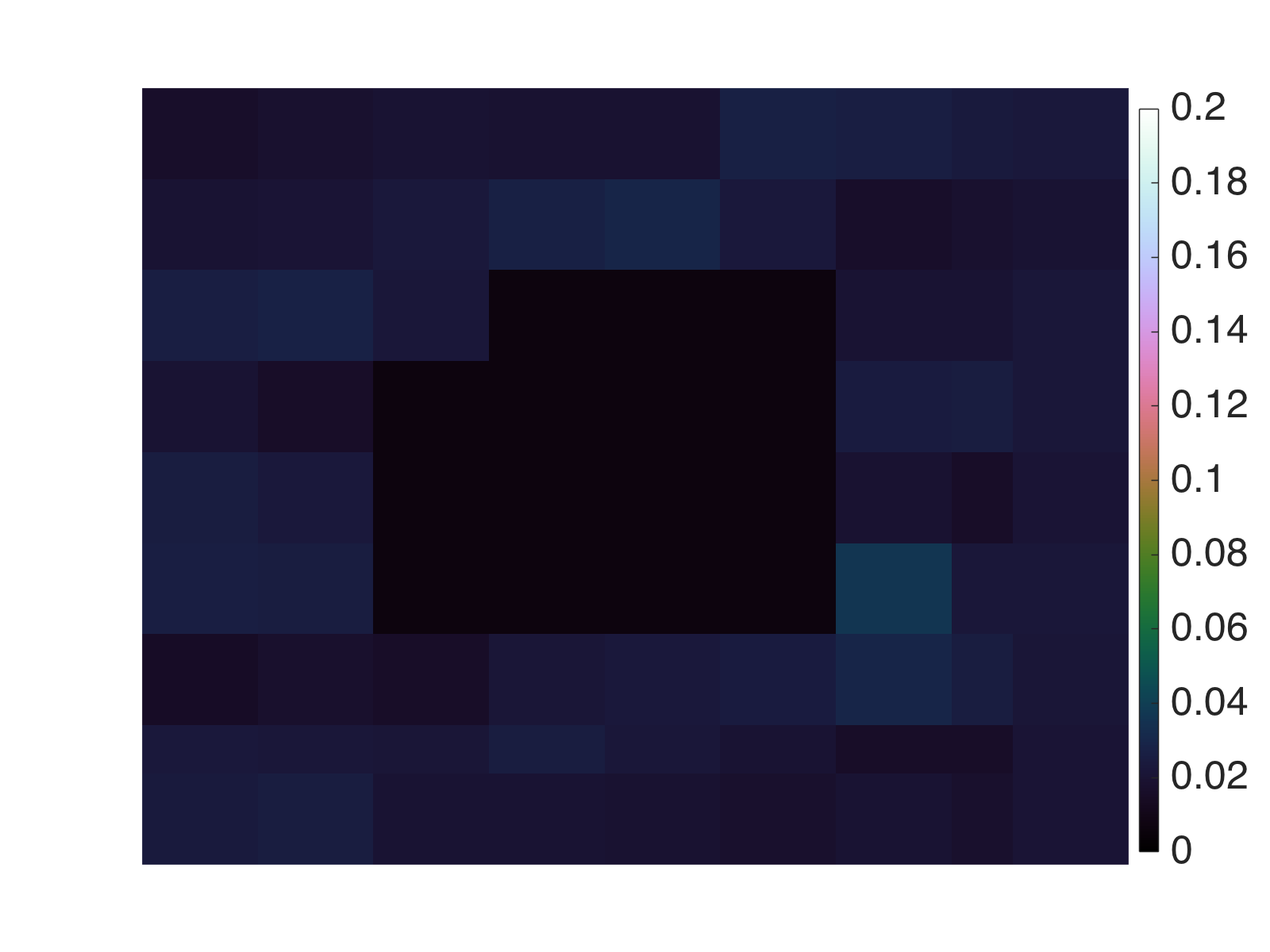} 
		\\	
		\includegraphics[trim={{.15\linewidth} {.07\linewidth} {.02\linewidth} {.073\linewidth}}, clip, width=0.24\linewidth, height = 0.21\linewidth]
		{./figs/3C288_result_ana} \put(-128,40){\rotatebox{90}{MAP}}  &
		\includegraphics[trim={{.15\linewidth} {.07\linewidth} {.02\linewidth} {.073\linewidth}}, clip, width=0.24\linewidth, height = 0.21\linewidth]
		{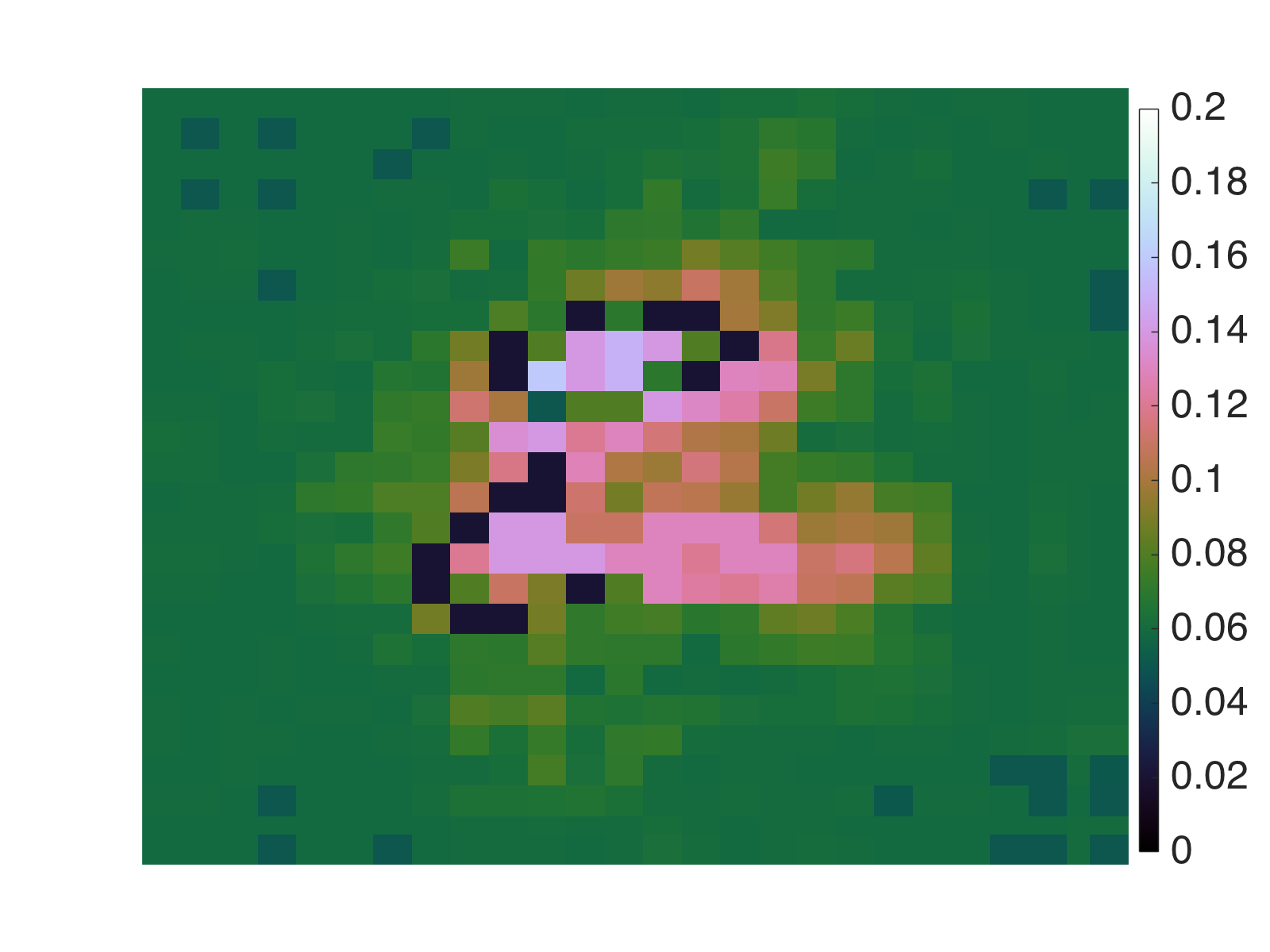} &
		\includegraphics[trim={{.15\linewidth} {.07\linewidth} {.02\linewidth} {.073\linewidth}}, clip, width=0.24\linewidth, height = 0.21\linewidth]
		{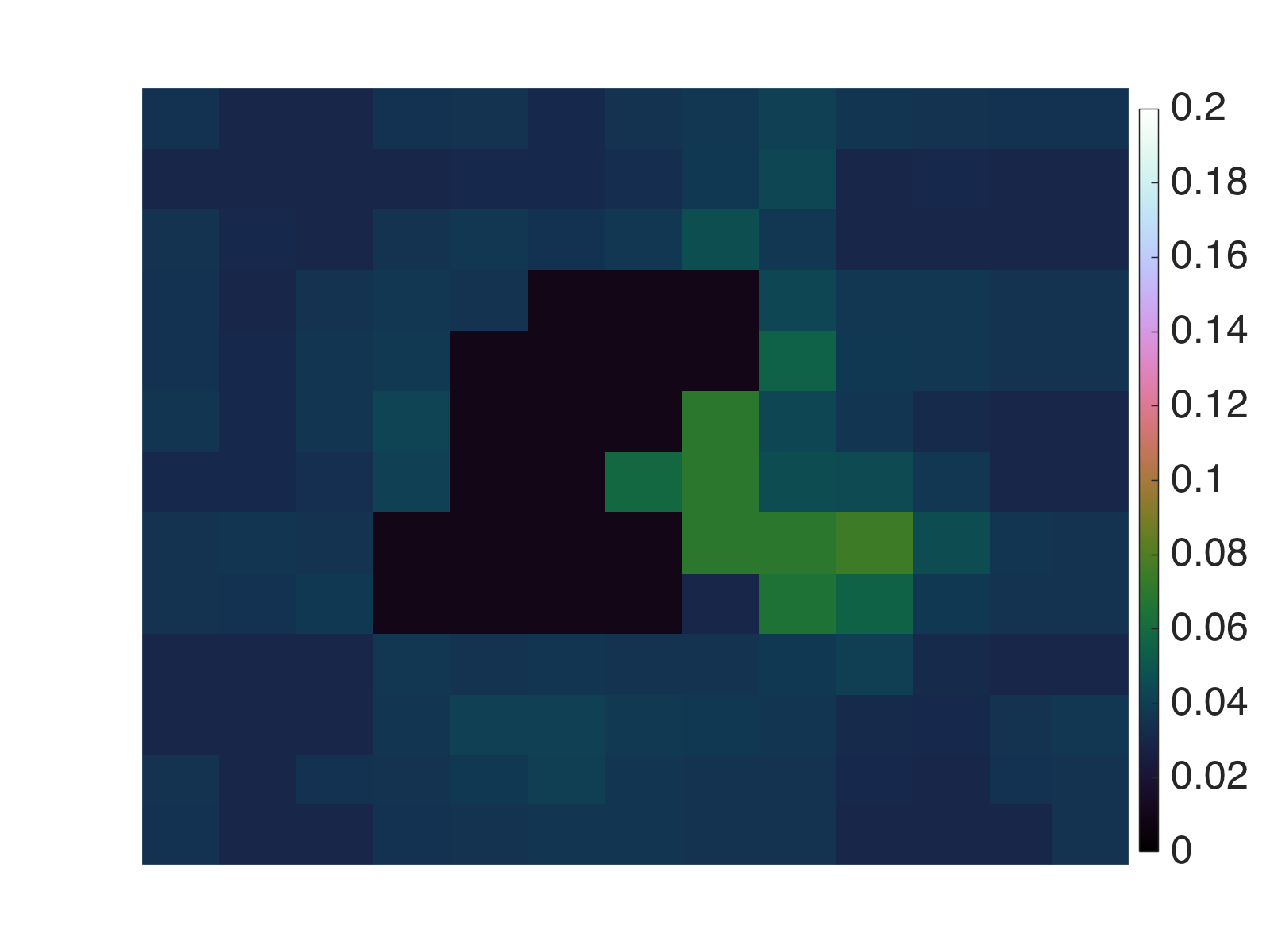} &
		\includegraphics[trim={{.15\linewidth} {.07\linewidth} {.02\linewidth} {.073\linewidth}}, clip, width=0.24\linewidth, height = 0.21\linewidth]
		{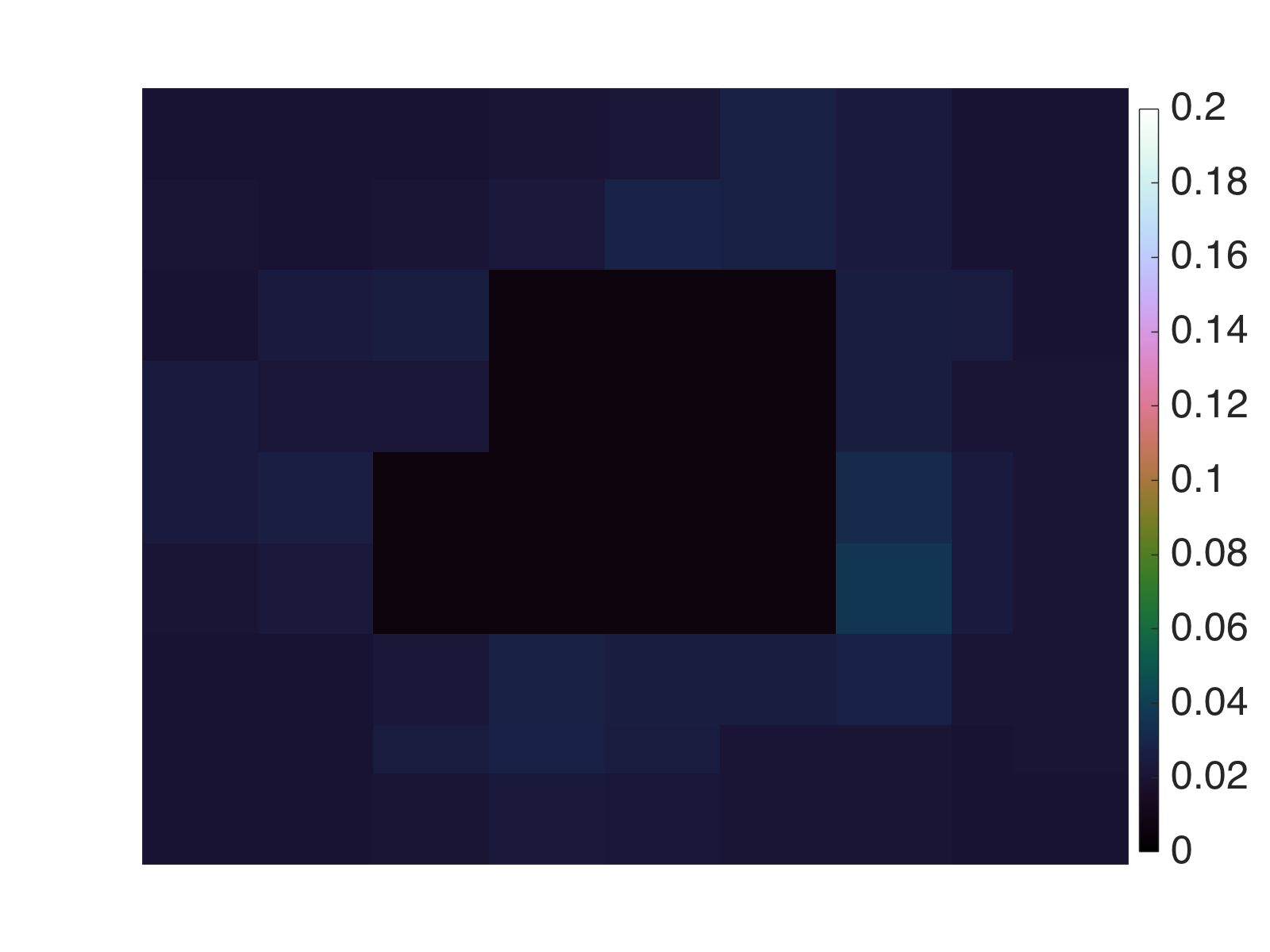} 
		\\ \vspace{0.05in}\\	
		\multirow{2}{*}{\small (a) point estimators } & {\small (b) local credible interval length } & {\small (c) local credible interval length }  & {\small (d) local credible interval length }
		 \vspace{0.03in} \\
		& {\small  grid size $10\times 10$ pixels } & {\small grid size $20\times 20$ pixels  } & {\small grid size $30\times 30$ pixels } 
        \end{tabular}
	\caption{Same as Figure \ref{fig-all-ci-grid-m31-mean} but for 3C288. 
	}
	\label{fig-all-ci-grid-3C288-mean}
\end{figure*}
}
\addtolength{\tabcolsep}{\tabL}

\subsection{Approximate local credible intervals}
We use the approximate HPD regions to calculate local credible intervals for image superpixels.
Precisely, Figures \ref{fig-all-ci-grid-m31-mean}--\ref{fig-all-ci-grid-3C288-mean} report the length of local credible intervals for 
the four test images for superpixel grid sizes of $10\times 10$, $20\times 20$, and $30\times 30$ pixels, computed w.r.t. 
the analysis model (the results for the synthesis model are very similar). {For comparison,  Figures \ref{fig-all-ci-grid-m31-mean}--\ref{fig-all-ci-grid-3C288-mean} also show the exact local credible estimates obtained by using the Px-MALA MCMC algorithm, which does not rely on the approximations \eqref{eqn:gamma-a} and \eqref{eqn:gamma-s}.}

We conclude the main observations as follows. Firstly, the results obtained with both approaches are extremely consistent with each other, 
indicating that the approximate credible intervals derived from the MAP estimation are very accurate. 
Secondly, the length of the approximate local credible intervals computed by MAP estimation are theoretically conservative and can be seen to slightly overestimate the lengths computed by MCMC sampling, and so are trustworthy.  
{Thirdly, note that (i) coarser scales have shorter credible intervals than narrower scales, and (ii) superpixels at object boundaries generally have longer credible intervals than superpixels in homogenous regions. These two observations are related to the fact that narrow scales are mainly sensitive to high spatial frequency information such as fine details and object boundaries that are difficult to accurately estimate, whereas coarser scales are also sensitive to lower frequencies and larger structures that are easier to estimate. More precisely, these two observations are a direct consequence of the fact that the sampling profile associated with the measurement operator $\bm{\mathsf{\Phi}}$ mainly covers low frequencies and has very few high-frequency measurements (see \citealt[Figure 2]{CPM17}). As a result, the likelihood $p(\vect{y}|\vect{x})$ has significantly less information about high-frequency image components, and this leads to higher uncertainty ({\it i.e.}, longer credible intervals) at fine scales, sharp details, and object boundaries.}

\begin{table*}
\begin{center}
  \caption{Hypothesis test results for test structures shown in Figure~\ref{fig-all-ci-hp} for M31, Cygnus A, W28, and 3C288.
   	Note that ${\gamma}_{\alpha}$ represents the isocontour defining the HPD credible region at credible level $(1-\alpha)$, where here $\alpha = 0.01$, 
   	${\vect x}^{*, {\rm sgt}}$ represents the surrogate generated from point estimator ${\vect x}^{*}$ (in particular, for Px-MALA ${\vect x}^{*}$ is the sample mean of the MCMC samples), 
   	and $(f+g)(\cdot)$ represents the objective function;  
   	symbols with labels \ $\bar{}$ \ and \ $\hat{}$ \ are related to 
   	the analysis model \eqref{eqn:ir-un-af} and the synthesis model \eqref{eqn:ir-un-sf}, respectively. 
   	Symbol \xmark \ indicates that the test area is artificial (and no strong statistical statement 
   	can be made as to the area), while \cmark \ indicates that the test area is physical. All values are in units $10^6$.
   	Clearly, both Px-MALA and MAP estimation give convincing and consistent hypothesis test results. Note that MAP estimation is dramatically more computationally efficient that Px-MALA (Table \ref{tab:time}).  	
   }      
 \label{tab:hp-test}
 \vspace{-0.05in}
\begin{tabular}{ccccHHIIc}
\toprule  
 \multirow{2}{*}{Images} &  Test  & Ground & \multirow{2}{*}{Method} &  \multicolumn{1}{c}{\multirow{2}{*}{$({\bar f} + {\bar g})(\bar{\vect x}^{*,{\rm sgt}})$ } }
 	&  \multicolumn{1}{c}{Isocontour} &  \multicolumn{1}{c}{\multirow{2}{*}{$({\hat f} + {\hat g})(\bm{\mathsf{\Psi}}^\dagger \hat{\vect x}^{*,{\rm sgt}})$ } }
 	& \multicolumn{1}{c}{Isocontour} & Hypothesis \\
	& areas & truth &  & \multicolumn{1}{c}{}  & \multicolumn{1}{c}{$\bar{\gamma}_{0.01}$} 
	&   \multicolumn{1}{c}{} & \multicolumn{1}{c}{$\hat{\gamma}_{0.01}$} & test
\\ \toprule
\multirow{2}{*}{M31 (Fig. \ref{fig-all-ci-hp} ) } & \multirow{2}{*}{1}  &  \multirow{2}{*}{ \cmark}  &
Px-MALA &  $\bf 2.44$  & $2.34$ & $\bf 2.43$   & $2.34 $ & \cmark
\\ 
& & & MAP & $\bf 2.29$  & $2.26$ & $\bf 2.29$  & $2.26$ &  \cmark
\\ \midrule
\multirow{2}{*}{Cygnus A  (Fig. \ref{fig-all-ci-hp} ) } &  \multirow{2}{*}{1} &  \multirow{2}{*}{ \cmark}  &
Px-MALA & $ 1.17$   & $ \bf 1.26$ & $ 1.18$  & $ \bf 1.27$ &  \xmark
\\ 
& & & MAP & $1.02$  & $\bf 1.14$ & $1.02$ & $\bf 1.14$ & \xmark
\\ \midrule
\multirow{2}{*}{W28  (Fig. \ref{fig-all-ci-hp} ) } &   \multirow{2}{*}{1}  &   \multirow{2}{*}{ \cmark}  &
Px-MALA & $\bf 3.38$   & $1.84$ &  $ \bf 3.37$   & $1.85$   &  \cmark
\\ 
& & &  MAP& $\bf 3.47$  & $1.89$ & $\bf 3.47$  & $1.89$ & \cmark
\\ \midrule
\multirow{4}{*}{3C288 (Fig. \ref{fig-all-ci-hp} ) } &  \multirow{2}{*}{1}  &   \multirow{2}{*}{\cmark}  &
Px-MALA & $\bf 3.27$  & $2.02$  &  $\bf 3.25$ & $2.01$  &  \cmark
 \\ 
 & & & MAP & $\bf 3.11$  & $  1.91$ & $\bf 3.11$ & $  1.91$ &  \cmark     \\  \cdashline{2-9}
  &  \multirow{2}{*}{2} &  \multirow{2}{*}{ \xmark}  &
Px-MALA & $1.971$  & $\bf 2.027$ & $1.954$ & $\bf 2.010$ & \xmark
 \\ 
   & & & MAP & $1.844$   & $ \bf 1.912$ & $1.844$  & $ \bf 1.912$ & \xmark
\\ \bottomrule
\end{tabular}
\end{center}
\end{table*}

\addtolength{\tabcolsep}{-\tabL}
{ \renewcommand{\arraystretch}{0.0}
\begin{figure}
	\centering
	\begin{tabular}{cc}
		\includegraphics[trim={{.15\linewidth} {.07\linewidth} {.02\linewidth} {.155\linewidth}}, clip, width=0.49\linewidth, height = 0.43\linewidth]
		{./figs/M31_result_ana}  \put(-120,40){\rotatebox{90}{ M31}} 
		\put(-100,36){\yellow{\framebox(12,12){ }}}  \put(-100,36){\yellow{\bf 1}} 
		 &		 
		\includegraphics[trim={{.15\linewidth} {.07\linewidth} {.02\linewidth} {.155\linewidth}}, clip, width=0.49\linewidth, height = 0.43\linewidth]
		{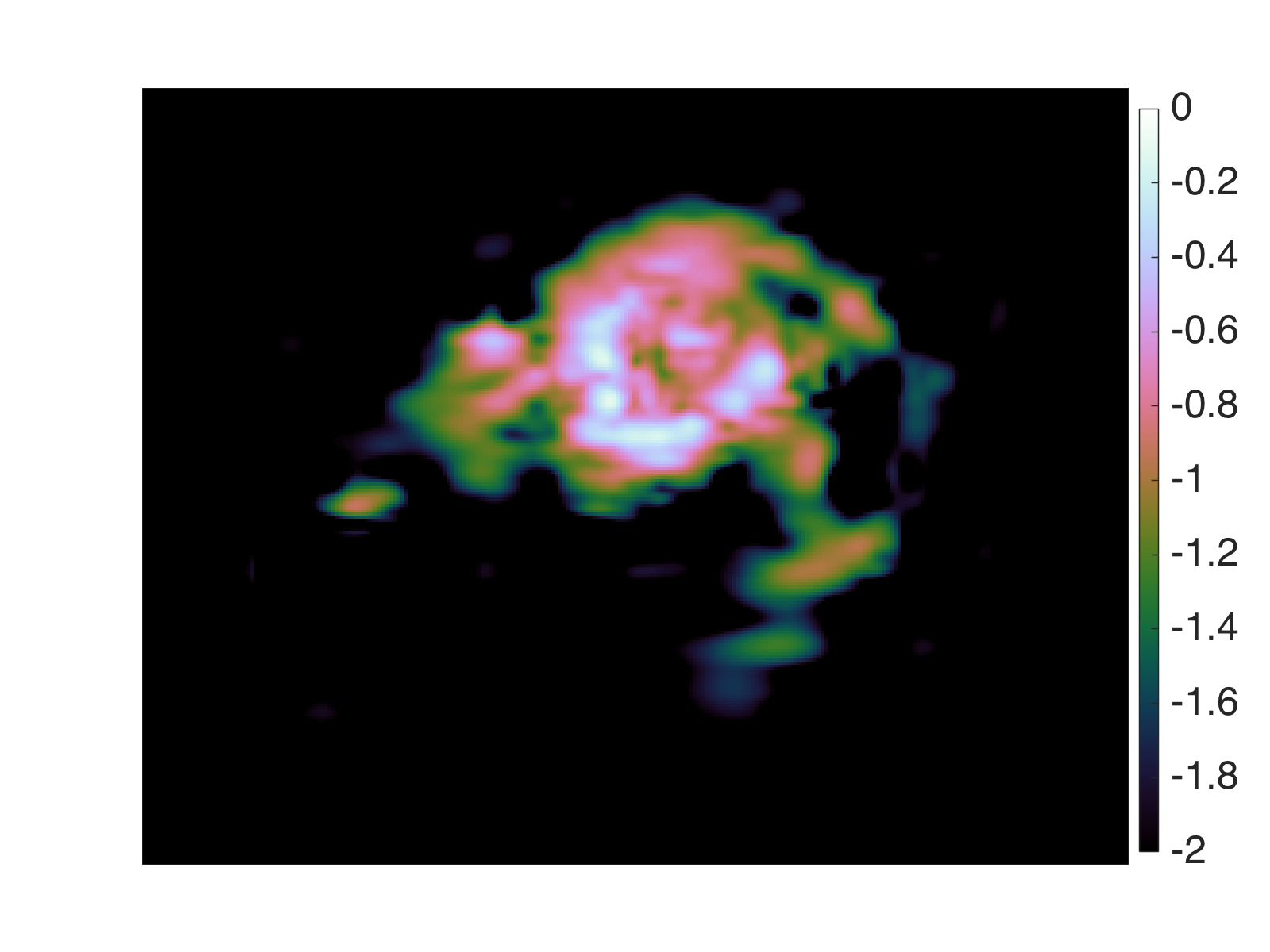}	 
		\\  
		\includegraphics[trim={{.15\linewidth} {.07\linewidth} {.02\linewidth} {.155\linewidth}}, clip, width=0.49\linewidth, height = 0.24\linewidth]
		{./figs/CYN_result_ana}   \put(-120,11){\rotatebox{90}{ Cygnus A}}
		\put(-69,23){\yellow{\framebox(10,10){ }}}  \put(-69,23){\yellow{\bf 1}} 
		&		
		\includegraphics[trim={{.15\linewidth} {.07\linewidth} {.02\linewidth} {.155\linewidth}}, clip, width=0.49\linewidth, height = 0.24\linewidth]
		{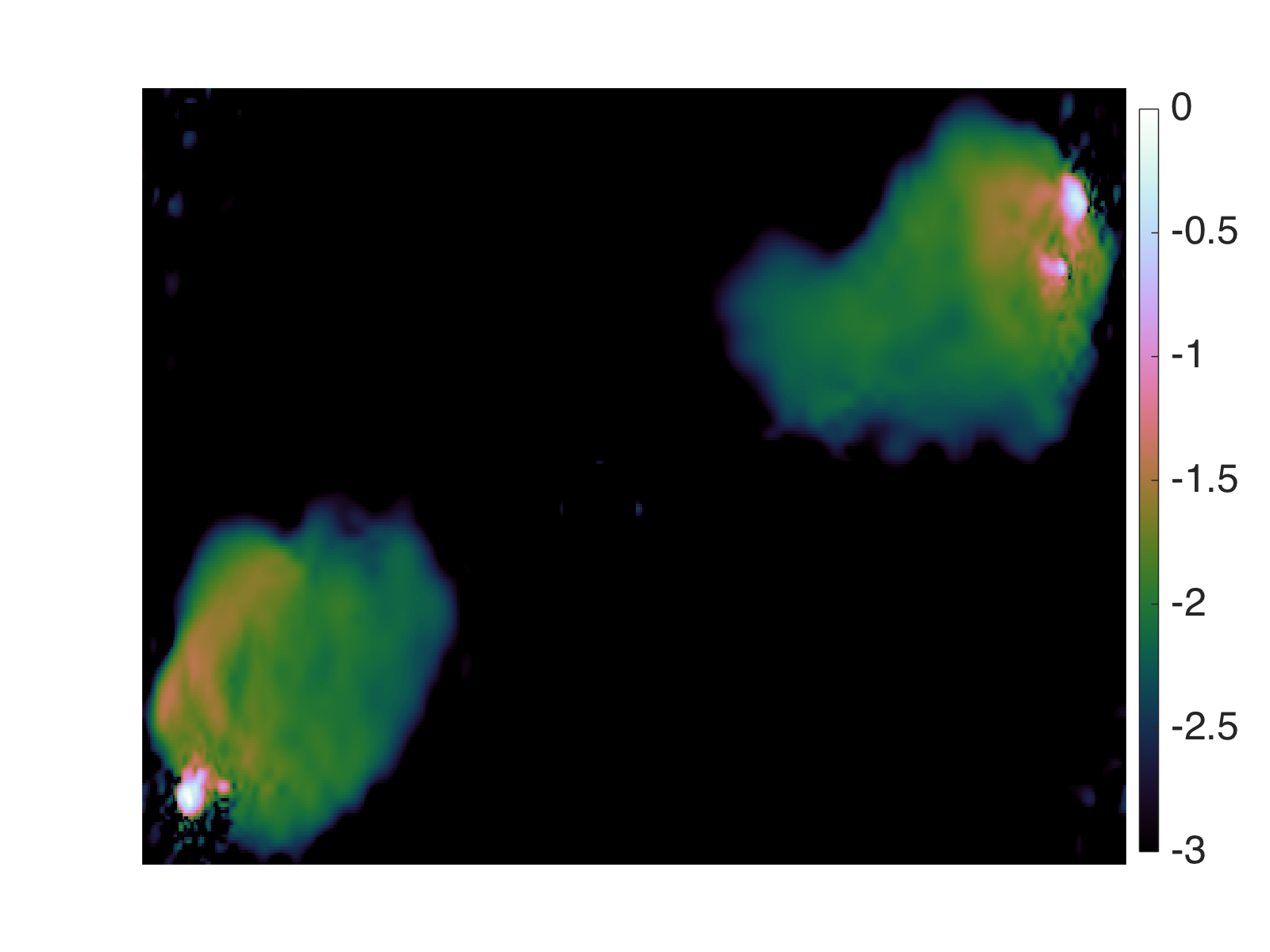} 	 
		 \\		 
		 \includegraphics[trim={{.15\linewidth} {.07\linewidth} {.02\linewidth} {.155\linewidth}}, clip, width=0.49\linewidth, height = 0.43\linewidth]
		{./figs/W28_result_ana} \put(-120,40){\rotatebox{90}{ W28}}
		\put(-109,59){\yellow{\framebox(10,10){ }}}  \put(-109,59){\yellow{\bf 1}}
		&
		\includegraphics[trim={{.15\linewidth} {.07\linewidth} {.02\linewidth} {.155\linewidth}}, clip, width=0.49\linewidth, height = 0.43\linewidth]
		{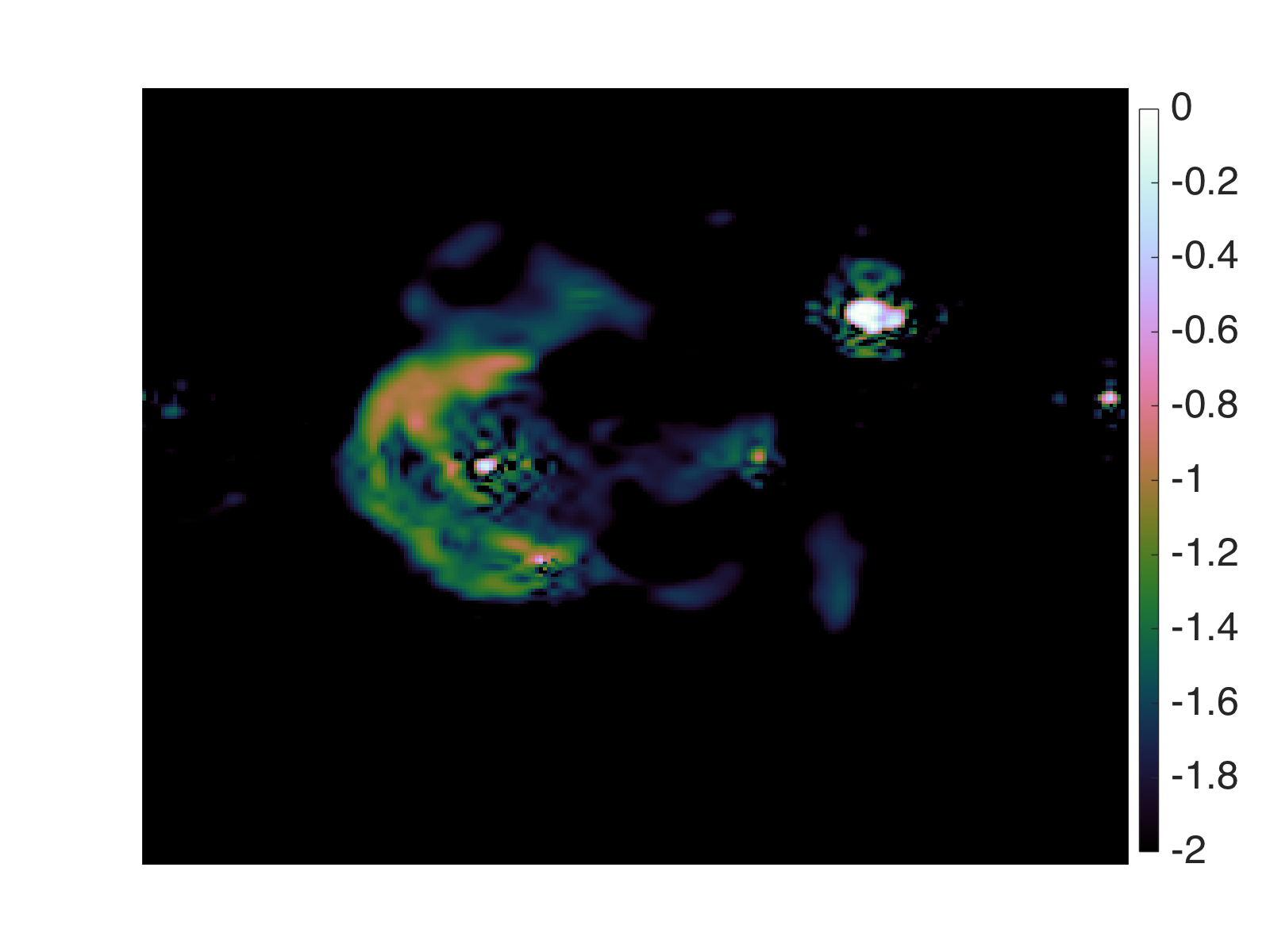}
		\\
		 \includegraphics[trim={{.15\linewidth} {.07\linewidth} {.02\linewidth} {.155\linewidth}}, clip, width=0.49\linewidth, height = 0.43\linewidth]
		{./figs/3C288_result_ana}   \put(-120,38){\rotatebox{90}{ 3C288}}
		\put(-76,49){\yellow{\framebox(9,9){ }}}  \put(-76,49){\yellow{\bf 1}}
		 \put(-54,84){\yellow{\framebox(12,12){ }}}  \put(-54,84){\yellow{\bf 2}}
		&
		\includegraphics[trim={{.15\linewidth} {.07\linewidth} {.02\linewidth} {.155\linewidth}}, clip, width=0.49\linewidth, height = 0.43\linewidth]
		{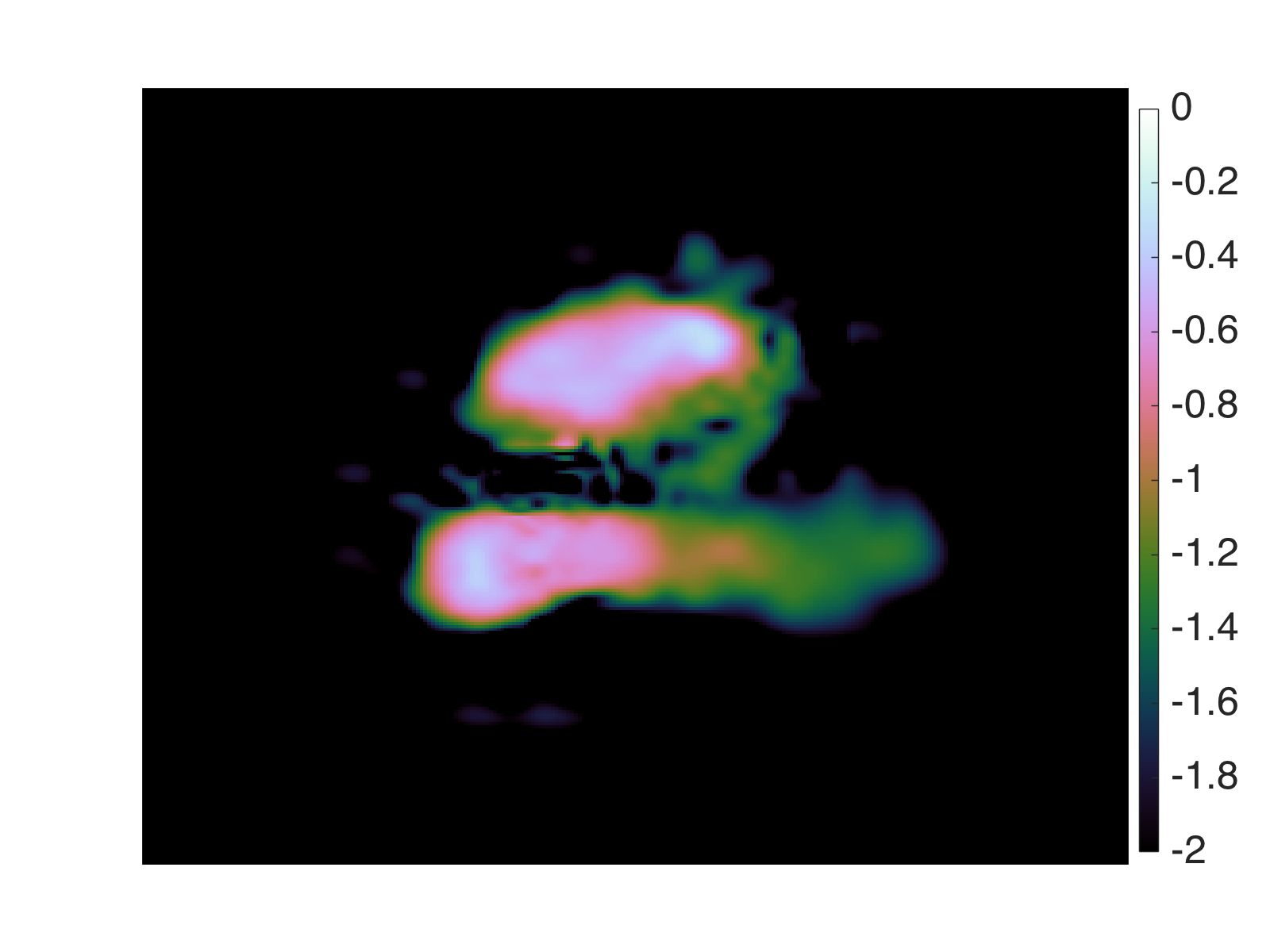}  \\
		{\small (a) MAP point estimators } & {\small (b) inpainted surrogate }
        \end{tabular}  
  \caption{Hypothesis testing {of image structure} for M31, Cygnus A, W28, and 3C288. 
  The five structures depicted in yellow are considered, all of which are physical ({\it i.e.} present in the ground truth images), except for structure 2 in 3C288, which is a reconstruction artefact.
	First column (a): point estimators obtained by MAP estimation for the analysis model \eqref{eqn:ir-un-af} (shown 
	in ${\tt log}_{10}$ scale). Second column (b): segmented-inpainted surrogate test images with information 
	in the yellow rectangular areas removed and replaced by inpainted background  (shown 
	in ${\tt log}_{10}$ scale). 
	Hypothesis testing is then performed to
	test whether the structure considered is physical by checking whether the surrogate test images shown in (b) fall outside of the HPD credible regions.  Results of these hypothesis tests are specified in Table~\ref{tab:hp-test}.
	Note that for the case shown in the last row the structures within areas 1 and 2 are tested independently.
	}
	\label{fig-all-ci-hp}
\end{figure}
}
\addtolength{\tabcolsep}{\tabL}

\addtolength{\tabcolsep}{-\tabL}
{ \renewcommand{\arraystretch}{0.0}
\begin{figure}
	\centering
	\begin{tabular}{cc}
		\includegraphics[trim={{.15\linewidth} {.07\linewidth} {.02\linewidth} {.155\linewidth}}, clip, width=0.49\linewidth, height = 0.43\linewidth]
		{./figs/M31_result_ana} 
		\put(-61.5,65){\yellow{\framebox(11,11){ }}}  
		 &		 
		\includegraphics[trim={{.15\linewidth} {.07\linewidth} {.02\linewidth} {.155\linewidth}}, clip, width=0.49\linewidth, height = 0.43\linewidth]
		{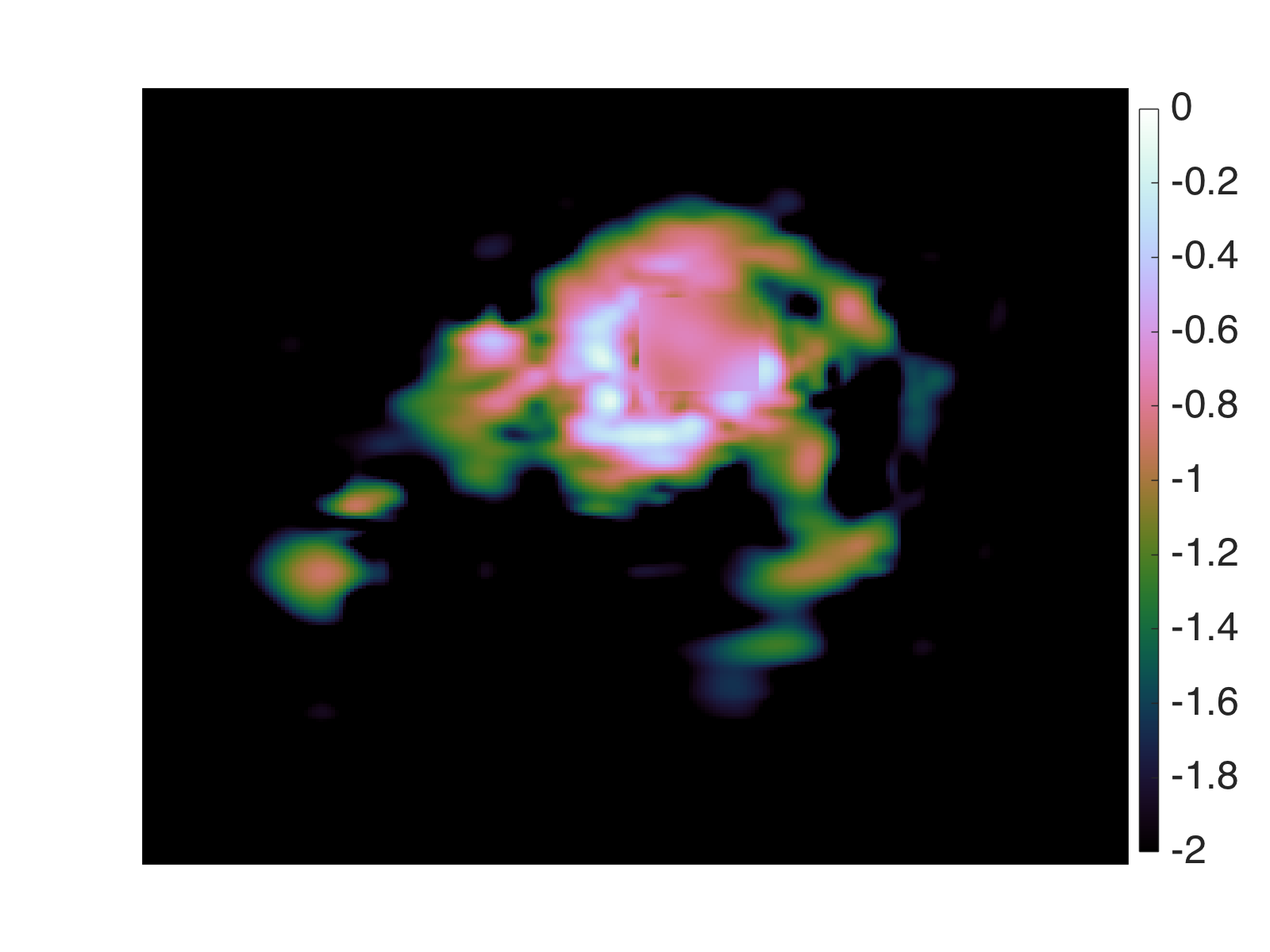}	  \\
		{\small (a) MAP point estimator } & {\small (b) smoothed surrogate }
        \end{tabular}  
  \caption{{Hypothesis testing {of image sub-structure} for M31 (both images are shown in ${\tt log}_{10}$ scale). 
  	The area depicted in yellow is considered, where the sub-structure presented in it is physical in the ground truth image.
	First column (a): point estimator obtained by MAP estimation for the analysis model \eqref{eqn:ir-un-af}. 
	Second column (b): smoothed surrogate test image with information in the yellow rectangular area smoothed (a MATLAB built-in function 
	 {\tt imgaussfilt} using Gaussian filtering with standard deviation 6 is applied).
	Hypothesis testing is then performed to
	test whether the sub-structure in the area considered is physical by checking whether the surrogate test image shown in (b) falls outside of the HPD credible regions.  
	The null hypothesis is rejected and the sub-structure of interested is correctly classified as physical and not a reconstruction artefact. }
	}
	\label{fig-m31-smooth-hp}
\end{figure}
}
\addtolength{\tabcolsep}{\tabL}

\subsection{Hypothesis testing of image structure}
We conclude our experimental results by demonstrating our methodology for testing structure in reconstructed images. We consider the same images and 
structures of interest as in \cite{CPM17}, shown in the yellow rectangular areas in the first column of Figure \ref{fig-all-ci-hp}. 
All of these structures are physical (\emph{i.e.} present in the ground truth images), except for structure 2 in 3C288 which is a reconstruction artefact.

{Recall that the methodology proceeds as follows. First, we construct a carefully designed surrogate image ${\boldsymbol{x}}^{*, {\rm sgt}}$ by modifying the MAP estimator $\boldsymbol{x}_{\rm map}$ to remove the structure of interest via segmentation-inpaiting, computed using formula \eqref{eqn:inpaint} (notice that this modification produces a surrogate that is in agreement with the prior distribution).
Each structure is assessed individually. Second, we check if ${\boldsymbol{x}}^{*, {\rm sgt}} \notin {C}^{\prime, {\rm map}}_{\alpha}$ (\emph{i.e.} if $f({\boldsymbol{x}}^{*, {\rm sgt}}) + g({\boldsymbol{x}}^{*, {\rm sgt}}) > {\gamma}^\prime_\alpha$) to determine whether the surrogate falls within the conservative HPD credible region or not. The resulting surrogate images are displayed in the second column of Figure~\ref{fig-all-ci-hp}. If the fact of removing the structure from $\boldsymbol{x}_{\rm map}$, which is at the centre of ${C}^{\prime, {\rm map}}_{\alpha}$, produces a surrogate that is outside ${C}^{\prime, {\rm map}}_{\alpha}$, this indicates that the likelihood is in clear disagreement with that modification. In that case we conclude that there exists significant evidence in the observed data in favour of the structure considered. Otherwise, we conclude that we fail to establish that there is significant evidence in favour of that structure}. We emphasise at this point that conclusions are generally not highly sensitive to the exact value of $\alpha$; here we report results for $\alpha = 0.01$ related to a 99\% credible level.

The results of these tests are shown in Table \ref{tab:hp-test}. For comparison, we also include the results obtained with the reference 
method Px-MALA \citep{CPM17}. Again, the two methods produce excellent results that are consistent with each other. 
From Table \ref{tab:hp-test}, we observe that the methods have correctly classified the three main physical structures of 
M31, W28, and 3C288, and correctly identified the minor structure of 3C288 as a potential reconstruction artefact. 
Moreover, the methods have found that it is not possible to make a strong statistical statement about the small 
physical structure in image Cygnus A, which is difficult because it is only a few pixels in size, isolated, and significantly 
weaker in intensity than the other structures in the image.

{To test the performance of hypothesis testing in terms of assessing sub-structure within areas of interest, we 
consider sub-structure in an area in M31 (see Figure~\ref{fig-m31-smooth-hp}). We find that the surrogate test image shown in Figure~\ref{fig-m31-smooth-hp}~(b) 
falls outside of the HPD credible region (the objective of the surrogate is $2.38 \times 10^6$, which is larger than the HPD isocontour of $\bar{\gamma}_{0.01} = 2.26 \times 10^6$)
according to the analysis model (the hypothesis testing result regarding the synthesis model is the same). 
Therefore, the sub-structure shown in the specified area in Figure~\ref{fig-m31-smooth-hp}~(a) is correctly classified as physical at a high credible level.}

Before closing this section, we emphasise again that the methods presented in this article deliver a variety of forms of 
uncertainty quantification with a very low computational cost.
While these new forms of uncertainty quantification can also be achieved 
by using state-of-the-art proximal MCMC methods, such as Px-MALA and MYULA, as presented in the companion article
\cite{CPM17}, MCMC techniques cannot scale to massive data sizes. Nevertheless, they are useful for medium-scale 
problems and provide accurate benchmarks for the highly efficient methods presented herein, which will scale very 
well to the emerging big-data era of radio astronomy.

\section{Conclusions}\label{sec:con}
Uncertainty quantification is an important missing component in RI imaging that will only become increasingly important as the big-data era of radio interferometry emerges.  No existing RI imaging techniques that are used in practice ({\it e.g.} CLEAN, MEM or CS approaches) provide uncertainty quantification.  
In this article, as an alternative to MCMC methods, such as Px-MALA and MYULA that were presented in \cite{CPM17}, 
we present new uncertainty quantification methods  
MAP estimation by convex optimisation. The proposed uncertainty quantification methods exhibit extremely fast computation speeds and allow uncertainty quantification to be performed practically and in a manner that will scale to the emerging big-data era of RI imaging.

Our proposed methods, which inherit the advantages of convex optimisation methods,
are much more efficient than proximal MCMC methods that explore the entire posterior distribution of the image. Note, however, that the methods proposed here give an approximation of HPD credible regions and, consequently, the additional forms of uncertainty quantification that are built on the approximate HPD credible regions are also approximate.  Nevertheless, we show these approximations are very accurate.  Moreover, the approximations are conservative so that uncertainties are not underestimated.  In contrast, proximal MCMC methods can theoretically provide HPD credible regions and other forms of uncertainty quantification that are more accurate.
Therefore, the proposed fast MAP-based methods and the proximal MCMC methods complement each other, rather than being mutually exclusive. We anticipate that when it comes to the big-data era, we will use predominantly fast uncertainty quantification methods such as those based on MAP estimation, and reserve MCMC methods for benchmarking and detailed comparison.

A variety of forms of uncertainty quantification for MAP estimation were constructed, including HPD credible regions, local credible intervals (\textit{cf.} error bars) 
for individual pixels and superpixels, and tests for image structure. Our methods were evaluated on four test images that are representative in RI imaging. 
These experiments demonstrated that our MAP-based methods exhibit excellent performance and can reconstruct images with sharp detail.  Moreover, they simultaneously underpin highly accurate approximate techniques to quantify uncertainties. In terms of computation time, MAP techniques were found to be approximately $10^5$ times faster than state-of-the-art proximal MCMC methods, even when MAP estimation is run on a standard laptop and proximal MCMC methods on a high-performance workstation.  Moreover, they lead to algorithmic structures that can be highly distributed and parallelised.

In the near future, we plan to apply the uncertainty quantification techniques presented in this article to RI observations acquired by a variety of different telescopes and to make the methods publicly available.  The methods will be implemented in the existing {PURIFY}\footnote{\url{https://github.com/basp-group/purify}} package for RI imaging.  Furthermore, novel algorithms will be developed to implement our methods with improved computational efficiency and to highly distribute and parallelise computations and data.
{We will also investigate optimal techniques for setting the regularisation parameter in a hierarchical Bayesian framework, applying the strategies developed by \citet{MBF15}.}

It is our hope that uncertainty quantification, \textit{e.g.} in the form of recovering error bars (Bayesian credible intervals) and hypothesis testing of image structure and sub-structure, will become an important standard component in RI imaging for statistically principled and robust scientific inquiry. For the first time, we propose techniques for the practical quantification of uncertainties in RI imaging.  These techniques can be applied not only to observations made by existing telescopes but also to the emerging big-data era of radio astronomy.

\section*{Acknowledgements}
This work is supported by the UK Engineering and Physical Sciences Research Council (EPSRC) by grant EP/M011089/1, and
Science and Technology Facilities Council (STFC) ST/M00113X/1.
{We also thank the editor and the anonymous reviewer for their constructive comments, which have significantly improved this manuscript.}

\appendix

\section{{Convex optimisation methods for MAP estimation}}
\label{sec:appendix}
Forward-backward splitting algorithms solve optimisation problems of the form 
\begin{equation} \label{eqn:fb-gen}
\argmin_{\vect x \in \mathbb{R}^{N}} (f+g)(\vect x) ,
\end{equation}
by using a splitting of $ (f+g)(\vect x)$.
We consider the setting where $f \notin \mathcal{C}^1$ is proper, convex and lower semi-continuous (l.s.c.)  
and $g \in \mathcal{C}^1$ is l.s.c. convex 
and $\beta_{\rm Lip}$-Lipchitz differentiable, {\it i.e.},
\begin{equation}
\|\nabla g(\hat{\vect z}) - \nabla g(\bar{\vect z})\| \le \beta_{\rm Lip} \|\hat{\vect z} - \bar{\vect z} \|, 
	\ \  \forall (\hat{\vect z}, \bar{\vect z}) \in \mathbb{C}^{N} \times \mathbb{C}^{N}.
\end{equation}
Precisely, forward-backward algorithms solve \eqref{eqn:fb-gen} by using the iteration
\begin{equation} \label{eqn:fb-i}
{\vect x}^{(i+1)} = {\rm prox}_{ \lambda^{(i)} f} ({\vect x}^{(i)} -  \lambda^{(i)} \nabla g({\vect x}^{(i)})),  
\end{equation}
where $ \lambda^{(i)}$ is the step size in a suitable bounded interval \citep[see, \textit{e.g.},][]{CP10}.
The
 {\it proximity operator} of $\lambda f$ is defined as \citep{M65}
\begin{equation} \label{eqn:prox-ope}
{\rm prox}_{ \lambda f} (\vect{z}) \equiv \argmin_{{\vect u}\in \mathbb{R}^N} \left \{   f(\vect{u}) + \|{\vect u} - {\vect z}\|^2/2\lambda \right \}.
\end{equation}
It is worth mentioning that when $f$ is associated with the $\ell_1$ norm, then computing \eqref{eqn:prox-ope}
goes to the so-called pointwise soft-thresholding of $\vect{z}$, {\it i.e.}, 
${\rm soft}_{\lambda}({\vect z}) = \big({\rm soft}_{\lambda}({z}_1), {\rm soft}_{\lambda}({z}_2), \cdots \big)$ defined by
\begin{equation} \label{eqn:soft-t}
{\rm soft}_{\lambda}({z}_j) =
\begin{cases}
 {z}_j (|{z}_j|  -  \lambda)/|{z}_j|   & {\rm if} \  |{z}_j| > \lambda, \\
 0 & {\rm otherwise},
\end{cases} 
\end{equation}
for every component $z_j$. 

There are several refinements of \eqref{eqn:fb-i} with better convergence properties. For example, using relaxation leads to the iteration 
\begin{equation} \label{eqn:fb-i-e1}
{\vect x}^{(i+1)} = (1-\beta^{(i)}) {\vect x}^{(i)} + \beta^{(i)}  \tilde{\vect x}^{(i+1)},
\end{equation}
where $\tilde{\vect x}^{(i+1)}$ is computed by \eqref{eqn:fb-i}, $\beta^{(i)}$ is a sequence of relaxation parameters, 
$ \lambda^{(i)} \in (\epsilon, 2/\beta_{\rm Lip} - \epsilon)$, $ \beta^{(i)} \in (\epsilon, 1)$, 
and $\epsilon\in (0,  \min \{1, 1/\beta_{\rm Lip}\} )$ \citep{CW05}; or with $ \lambda^{(i)} = 1/\beta_{\rm Lip}$, 
$ \beta^{(i)} \in (\epsilon, 3/2 - \epsilon)$, and $\epsilon\in (0, 3/4)$ \citep{BC11}.  
Furthermore, algorithmic structures that allow computations to be highly distributed and parallelised  \citep[\textit{e.g.}][]{car14,OCRMTPW16} and computed in an online manner \citep{CPraM17} can also be developed to assist in scaling to big-data.

\bibliographystyle{mnras}
\bibliography{refs_xhcai}

\label{lastpage}

\end{document}

%% file: fig_uq_diag_part2.tex

\begin{tikzpicture}[>=stealth,every node/.style={shape=rectangle,draw,rounded corners},]
     \node (c0){};
    \node (c1) [fill=red!30, right of = c0, node distance=1cm]{Observed visibilities in RI imaging: $\vect y$};
    \node (c2) [fill=blue!20, below left of = c0, node distance=2cm, text width=2.5cm,align=center]{MAP image estimation: ${\vect x}_{\rm map}$ };
    \node (c3) [fill=green!10, right of = c2, node distance=5cm, text width=2.5cm,align=center]{Approximate HPD credible regions: $C^{\prime}_{\alpha}$  };
    \node (c5) [fill=green!10, below = 0.75cm of c3,text width=2.8cm,align=center]{Approximate local credible intervals: $({\vect \xi}_-, {\vect \xi}_+)$ };
    \node (c4) [fill=green!10, below = 0.75cm of c5,text width=2.5cm,align=center]{Hypothesis testing};

    \draw[->] (c1) -- (c2);
    \draw[->] (c1) -- (c3);
    \draw[->] (c2) -- (c3);
    \draw[->] (c3.east) to [out=-60,in=60] (c4.east);
    \draw[->] (c2) -- (c4.west);
    \draw[->] (c3) -- (c5);
    \draw[->] (c2) -- (c5);
\end{tikzpicture}